\documentclass[a4paper,fleqn]{cas-dc}

\usepackage[numbers]{natbib}

\begin{document}
\let\WriteBookmarks\relax
\def\floatpagepagefraction{1}
\def\textpagefraction{.001}
\shorttitle{Gravitational emissions and light curves of quasi-periodic orbits in Schwarzschild spacetime embedded in a Dehnen-type dark matter halo}
\shortauthors{S. J. Tan et~al.}

\title [mode = title]{Gravitational emissions and light curves of quasi-periodic orbits in Schwarzschild spacetime embedded in a Dehnen-type dark matter halo}                      

\author[1]{Shijie Tan}
\credit{Software, Writing -- original draft}

\affiliation[1]{organization={School of Mathematics and Physics, University of South China},
                city={Hengyang},
                postcode={421001}, 
                state={Hunan},
                country={People's Republic of China}}

\author[1,2]{Chunhua Jiang}
\credit{Writing -- review $\&$ editing}

\affiliation[2]{organization={Hunan Provincial Key Laboratory of Mathematical Modeling and Scientific Computing, University of South China},
                city={Hengyang},
                postcode={421001}, 
                state={Hunan}, 
                country={People's Republic of China}}

\author[1,2]{Dan Li}
\credit{Writing -- review $\&$ editing}

\author[1,2]{Shiyang Hu}[orcid=0000-0001-8177-9762]
\cormark[1]
\ead{husy_arcturus@163.com}
\credit{Conceptualization, Methodology, Software, Supervision, Writing -- review $\&$ editing}

\author[3]{Chen Deng}
\credit{Methodology, Writing -- review $\&$ editing}

\affiliation[3]{organization={School of Astronomy and Space Science, Nanjing University},
                city={Nanjing},
                postcode={210023}, 
                state={Jiangsu}, 
                country={People's Republic of China}}

\author[1,2]{Wenbin Lin}
\credit{Writing -- review $\&$ editing}

\cortext[cor1]{Corresponding author}

\begin{abstract}
Timelike orbits in curved spacetimes encode intrinsic information about the background geometry and serve as critical probes for investigating gravitational theories and source distributions. In this study, we investigate strictly closed timelike orbits within a Schwarzschild spacetime embedded in a Dehnen-type dark matter halo. By solving the geodesic equations, we identify various configurations of these closed orbits and simulate their corresponding gravitational waves and electromagnetic light curves. Our findings reveal that the morphology of closed orbits is primarily governed by the ratio of the azimuthal period to the radial period. Notably, dark matter halo parameters such as the core scale and density parameters exert a significant amplification effect on the orbital scale, which further induces a discernible phase lag in the gravitational wave signals. Furthermore, within a specific parameter space, we discover a linear relationship between the number of peaks in the light curves and the number of orbital leaves. From a theoretical perspective, these findings reveal the multimessenger signatures of closed orbits, which may provide a potential theoretical foundation for establishing a connection between orbital dynamics and the surrounding dark matter environment.
\end{abstract}


\maketitle

\section{Introduction}
\label{sec:intro}
General relativity (GR) has remained the cornerstone of modern gravitational physics since its inception, consistently demonstrating remarkable agreement with experimental and observational tests across a vast range of scales. Despite its successes, mounting observational evidence suggests that GR may not provide a complete description of gravity across all scales. The most prominent challenges arise from galactic rotation curves \cite{1974Natur.250..309E,1978ApJ...225L.107R}, gravitational lensing in galaxy clusters \cite{2004ApJ...604..596C,2006ApJ...648L.109C,2020ApJ...898...81C,2021PhRvD.103d4045M}, and large-scale structure formation; these observations collectively indicate the existence of unseen mass, commonly referred to as dark matter, which has become widely accepted within the scientific community as observational evidence continues to accumulate \cite{2020A&A...641A...6P,2023A&A...679A..31D,2025PhRvL.135p1005M,2025JCAP...11..080T,2025NatAs...9.1714P,2026NatAs.tmp...38C}.

In the standard cosmological framework, the cold dark matter (CDM) model has achieved remarkable success in explaining the large-scale structure of the Universe \cite{1984ApJ...285L..45B}. A cornerstone of this paradigm is the formation of extended dark matter halos, which provide the essential gravitational scaffolding for galaxy formation and evolution. To accurately characterize the spatial mass distribution within these halos, various density profiles have been proposed and extensively applied, including the Navarro-Frenk-White (NFW) \cite{1996ApJ...462..563N}, Einasto \cite{1989A&A...223...89E}, and Burkert \cite{1995ApJ...447L..25B} models. Among these, the Dehnen-type family of density profiles \cite{1993MNRAS.265..250D} stands out due to its exceptional analytical flexibility and broad compatibility. Fundamentally, the spatial distribution in this model is governed not only by two core physical quantities, namely a scale radius $r_{\textrm{s}}$ and a scale density $\rho_{\textrm{s}}$ which are of significant observational interest and serve as key parameters for characterizing dark matter halos, but also by a set of dimensionless shape parameters typically denoted as $(\alpha,\beta,\gamma)$. This parameterization allows the model to encompass an exceptionally wide range of halo morphologies within a single analytical framework, making it capable of reproducing almost all standard dark matter density profiles through appropriate parameter adjustments. Specifically, $\gamma$ and $\beta$ dictate the inner and outer logarithmic density slopes, respectively, while $\alpha$ controls the sharpness of the transition between these two asymptotic regimes. On physical grounds, a particularly well-motivated choice is to fix $\alpha$ and $\beta$ to $1$ and $4$, respectively; the condition $\alpha=1$ ensures a smooth and physically natural transition between the inner and outer density regimes, while $\beta=4$ guarantees that the total dark matter halo mass remains finite, avoiding the logarithmic divergence that plagues profiles with shallower outer slopes such as NFW. Furthermore, setting $\gamma=0$ endows the dark matter halo with a relatively flat central core, avoiding the central cusp that would arise for $\gamma > 0$ and directly addressing the longstanding core-cusp tension between CDM predictions and observations of low surface brightness galaxies. The resulting $(\alpha,\beta,\gamma)=(1,4,0)$ Dehnen model thus offers a promising and physically plausible description of realistic dark matter halo distributions. Building on this perspective, Gohain et al. \cite{2024PDU....4601683G} derived the Schwarzschild spacetime embedded in a $(1,4,0)$ Dehnen dark matter halo and investigated the influence of the halo parameters on black hole phase transitions and light deflection. However, the geometric narrative of this spacetime remains incomplete: the impact of the dark matter halo parameters on the motion of matter and the associated observational signatures within this geometry are still unclear, warranting further systematic investigation.

To investigate the properties of black hole spacetimes, the dynamical behavior of timelike particles serves as a well-established starting point. Phenomena such as particle acceleration mechanisms \cite{2010PhRvL.104b1101J,2010PhRvD..82j3005W,2016PhRvD..93h4025G}, chaotic effects \cite{2009ApJ...693..472T,2014ApJ...787..117K,2019EPJC...79..479P,2019EPJP..134...96L,2021ApJS..257...40H,2021EPJC...81..785S,2021ApJ...909...22W,2022ApJ...925..158H,2023EPJC...83..828Z,2025EPJC...85..770X}, and scattering processes \cite{2016EPJC...76...32S}, all of which are closely connected to high-energy astrophysics, provide powerful probes for unveiling the nature of spacetime geometry. Notably, a technique was developed in \cite{2008PhRvD..77j3005L} for searching for rational orbits in the equatorial plane of curved spacetimes by exploiting the periodicity of particle orbits in both the radial and angular directions. This method enables the identification of strictly closed orbits with distinct topological structures; such an approach not only lays the foundation for classifying particle motions but also paves the way for uncovering potential correlations between orbital characteristics and spacetime properties. Consequently, the search for and simulation of distinctive periodic and quasi-periodic orbits within various gravitational backgrounds has become a highly active research topic \cite{Misra:2010pu,Wei:2019zdf,Liu:2018vea,Deng:2020yfm,Deng:2020hxw,Lin:2021noq,Wang:2022tfo,Gao:2020wjz,Lin:2022wda,Zhang:2022zox,Gao:2021arw,Lin:2023rmo,QiQi:2024dwc}. Furthermore, after obtaining these timelike orbits, researchers have employed analytic kludge gravitational waveform models \cite{2007PhRvD..75b4005B} to simulate the corresponding gravitational wave signals \cite{Tu:2023xab,Meng:2024cnq,Li:2024tld,Wang:2025hla,Lu:2025xlp,Lu:2025cxx,Ahmed:2025azu,Deng:2025wzz,Alloqulov:2025dqi,Yang:2024lmj,Alloqulov:2025ucf,Alloqulov:2025bxh,Zhao:2024exh,Shabbir:2025kqh,Haroon:2025rzx,Zhang:2025wni,Junior:2024tmi,Zare:2025aek,Huang:2025vpi,Hua:2026kvw,Li:2025sfe,Huang:2024oli,Huang:2025czx,Jiang:2024cpe}. These efforts establish the functional dependence among signal features, orbital characteristics, and spacetime properties, exploring the possibility of decoding such relations with future space-based gravitational wave interferometers.

Timelike particles in curved spacetimes can be regarded as moving luminous sources whose electromagnetic radiation is observable in the form of light curves. Consequently, light curves---much like gravitational wave signals---encode orbital configurations and facilitate the extraction of spacetime parameters \cite{Huang:2024wpj,Zhu:2024vxw}. It is therefore highly desirable to integrate these two messengers, namely gravitational and electromagnetic radiation, to investigate timelike orbits and unveil the underlying spacetime properties. Following this approach, we numerically simulate various timelike orbits around a Schwarzschild black hole embedded in a $(1,4,0)$ Dehnen-type dark matter halo. Subsequently, we generate both the gravitational wave signals and the light curves associated with these orbits. Our primary objective is to identify the imprints of orbital characteristics and dark matter halo parameters within these two messengers, thereby revealing the influence of dark matter halo on the surrounding geometry. Notably, while gravitational wave-based studies of timelike orbits have received significant attention, the application of light curves for this purpose remains relatively underexplored. This work thus presents a novel theoretical approach to probing spacetimes modified by dark matter through the electromagnetic signatures of orbital motion.

The remainder of this paper is organized as follows. In Sec. II, we introduce the metric of a Schwarzschild spacetime embedded in a Dehnen-type dark matter halo and derive the geodesic equations for timelike particles. Sec. III is dedicated to the study of the effective potential for timelike particles within the target spacetime. In Sec. IV, we present the strictly closed rational orbits and the numerical scheme employed to obtain them, with particular emphasis on their initial conditions. Sec. V provides our primary results, including closed rational orbits of various configurations, irrational orbits, and their corresponding gravitational wave signals and light curves; we further discuss the potential correlations among orbital morphologies, multi-messenger signatures, and spacetime parameters. Finally, we summarize our findings and present our conclusions. Throughout this paper, we adopt geometric units in which the black hole mass $M$, the gravitational constant $G$, and the speed of light $c$ are all set to unity ($M=G=c=1$). These physical units are explicitly restored only when evaluating the observational significance of the gravitational wave signals and light curves.
\section{Metric and geodesic equations of the Schwarzschild black hole embedded in a Dehnen-type dark matter halo}
\label{sec:metric}
Following Gohain et al. \cite{2024PDU....4601683G}, a spherically symmetric black hole embedded in a Dehnen-type $(1,4,0)$ dark matter halo is described by the line element
\begin{equation}
\label{eq:metric}
\textrm{d}s^{2}
= -f(r)\,\textrm{d}t^{2}
+ \frac{\textrm{d}r^{2}}{f(r)}
+ r^{2}\bigl(\textrm{d}\theta^{2} + \sin^{2}\theta\,\textrm{d}\phi^{2}\bigr),
\end{equation}
where the metric function 
\begin{equation}
\label{eq:f}
f(r)
 = 1 - \frac{2}{r}
   - \frac{4\pi (r_{\textrm{s}} + 2r)\,r_{\textrm{s}}^{3}\rho_{\textrm{s}}}{3(r_{\textrm{s}} + r)^{2}}
\end{equation}
incorporates the gravitational contribution from the dark matter halo, with $r_{\textrm{s}}$ and $\rho_{\textrm{s}}$ denoting the scale radius and scale density, respectively. Notably, in the limit where either $r_{\textrm{s}}$ or $\rho_{\textrm{s}}$ vanishes, $f(r)$ recovers the standard Schwarzschild form. It is also observed that $r_{\textrm{s}}$ and $\rho_{\textrm{s}}$ influence the spacetime geometry through distinct scaling behaviors: the metric's dependence on the scale density is linear, whereas its dependence on the scale radius exhibits a more complex, non-linear character. Furthermore, it must be emphasized that the dark matter halo parameters in geometric units are not assigned arbitrarily; rather, they must be strictly consistent with empirical observations. In a parallel study, utilizing the shadow observations of M87$^{*}$ and the Galactic Center black hole Sgr A$^{*}$, we constrained the magnitudes of $r_{\textrm{s}}$ and $\rho_{\textrm{s}}$ to the order of $10^{-1}$ \cite{Hu:2025lyp}, which is highly consistent with the results reported in \cite{2025JCAP...03..054J}. Similarly, based on the quasi-periodic oscillations of X-ray sources, other researchers have also restricted the dark matter halo parameters for the Dehnen family models to the $10^{-1}$ order of magnitude \cite{2025EPJC...85.1193X}. Consequently, in all subsequent calculations, we assign values to the dark matter halo parameters strictly within the interval of [0, 1].

The Lagrangian for a test particle moving along geodesics in this background is given by
\begin{equation}
\label{eq:Lagrangian}
2\mathcal{L}
 = -f(r)\,\dot t^{2}
   + \frac{\dot r^{2}}{f(r)}
   + r^{2}\left(\dot\theta^{2} + \sin^{2}\theta\,\dot\phi^{2}\right),
\end{equation}
where the overdot denotes differentiation with respect to an affine parameter $\tau$. Owing to the spherical symmetry of the spacetime, the motion can be confined to the equatorial plane $\theta=\pi/2$ without loss of generality. From the Euler--Lagrange equations, the canonical momenta $p_{\mu}$ are defined as
\begin{align}
p_t &= \frac{\partial\mathcal{L}}{\partial\dot t}
     = -\,f(r)\,\dot t \equiv -E,
\label{eq:pt} \\
p_r &= \frac{\partial\mathcal{L}}{\partial\dot r}
     = \frac{\dot r}{f(r)},
\label{eq:pr} \\
p_\phi &= \frac{\partial\mathcal{L}}{\partial\dot\phi}
       = r^{2}\dot\phi \equiv L.
\label{eq:pphi}
\end{align}
Here, $E$ and $L$ represent the conserved energy and angular momentum per unit rest mass of the particle, respectively. According to the Legendre transformation, the Hamiltonian governing the particle motion is expressed as 
\begin{equation}
\mathcal{H}
= p_{t}\dot t + p_{r}\dot r + p_{\phi}\dot\phi - \mathcal{L},
\end{equation}
which satisfies the normalization condition
\begin{equation}
\label{eq:H-delta}
2\mathcal{H}
= g_{\mu\nu}\dot x^\mu\dot x^\nu
= \delta,
\qquad
\delta =
\begin{cases}
-1, & \text{timelike particle},\\[1mm]
0,  & \text{lightlike particle}.
\end{cases}
\end{equation}
Consequently, $2\mathcal{H}=2\mathcal{L}=\delta$ constitutes a constant of motion. Given the initial generalized coordinates $x^{\mu}=\left(t,r,\theta,\phi\right)$ and the conjugate momenta $p_{\mu}=\left(p_{t},p_{r},p_{\theta},p_{\phi}\right)$, the trajectory can be numerically integrated using the Hamilton's canonical equations:
\begin{eqnarray}\label{eq:can}
\dot{x^{\mu}} = \frac{\partial\mathcal{H}}{\partial p_{\mu}}, \quad \dot{p_{\mu}} = - \frac{\partial\mathcal{H}}{\partial x^{\mu}}.
\end{eqnarray}
\section{Effective potential for timelike particles}
In general relativity, particle orbits are typically classified into circular and precessing orbits. Determining the existence and stability of these orbits requires an analysis of the effective potential. For timelike particles, the effective potential $V_{\textrm{eff}}(r)$ is conventionally defined via the energy conservation relation:
\begin{equation}\label{veff1}
\dot{r}^{2} + V_{\text{\textrm{eff}}}(r) = E^{2}.
\end{equation}
By substituting the metric function \eqref{eq:f} into the Lagrangian \eqref{eq:Lagrangian}, we obtain:
\begin{eqnarray}\label{veff2}
V_{\text{eff}}(r) ={}& f(r)\left(1 + \frac{L^{2}}{r^{2}}\right) \nonumber \\
  ={}& \left( 1 - \frac{2}{r}
  - \frac{4\pi (r_{\textrm{s}} + 2r)\,r_{\textrm{s}}^{3}\rho_{\textrm{s}}}{3(r_{\textrm{s}} + r)^{2}} \right) \left(1 + \frac{L^{2}}{r^{2}}\right).
\end{eqnarray}
The effective potential is governed by the radial coordinate $r$, the specific angular momentum $L$, and the dark matter halo parameters $r_{\textrm{s}}$ and $\rho_{\textrm{s}}$, as illustrated in Fig. 1. Across most of the parameter space, $V_{\textrm{eff}}(r)$ initially increases with $r$, reaches a local maximum, decreases to a local minimum, and finally approaches unity as $r \rightarrow +\infty$, consistent with the asymptotically flat nature of the spacetime. The maximum and minimum correspond to unstable and stable circular orbits, respectively, while the region within the potential well supports precessing orbits. Potential curves that lack both a peak and a trough, such as the red curve in Fig. 1(a), indicate the absence of bound orbits; in such cases, particles will inevitably be captured by the black hole.
\begin{figure*}
\center{
\includegraphics[width=5cm]{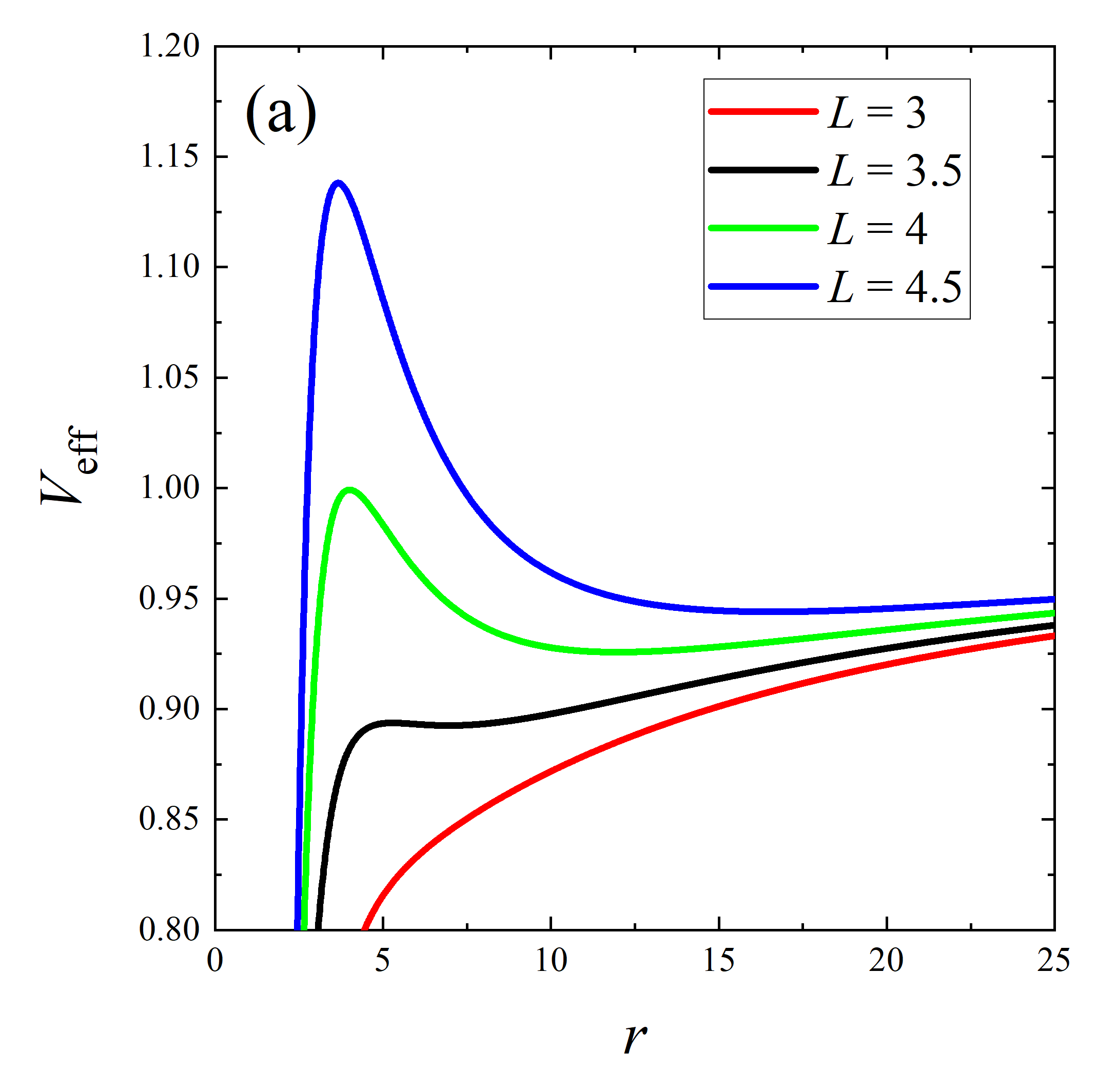}
\includegraphics[width=5cm]{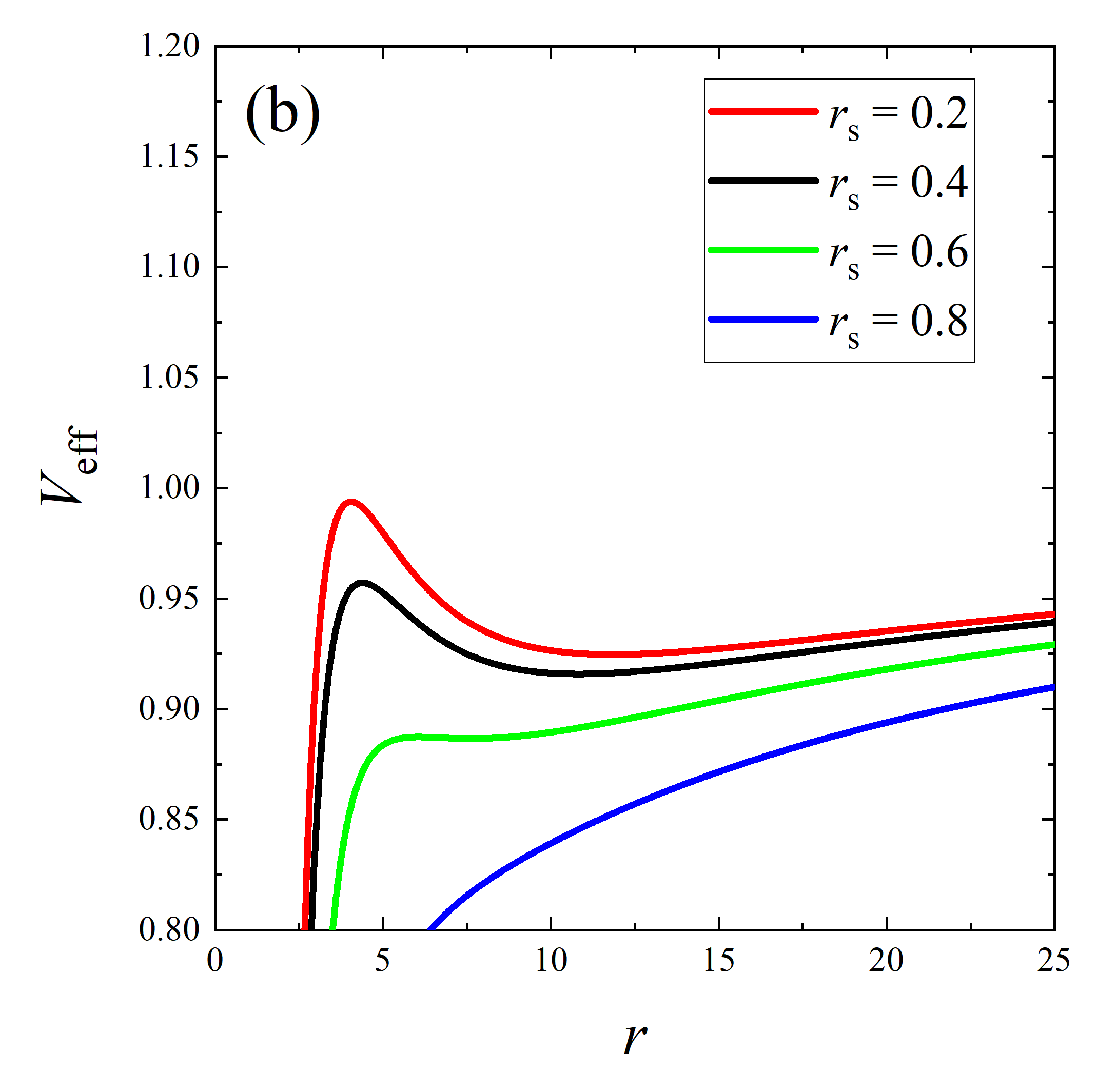}
\includegraphics[width=5cm]{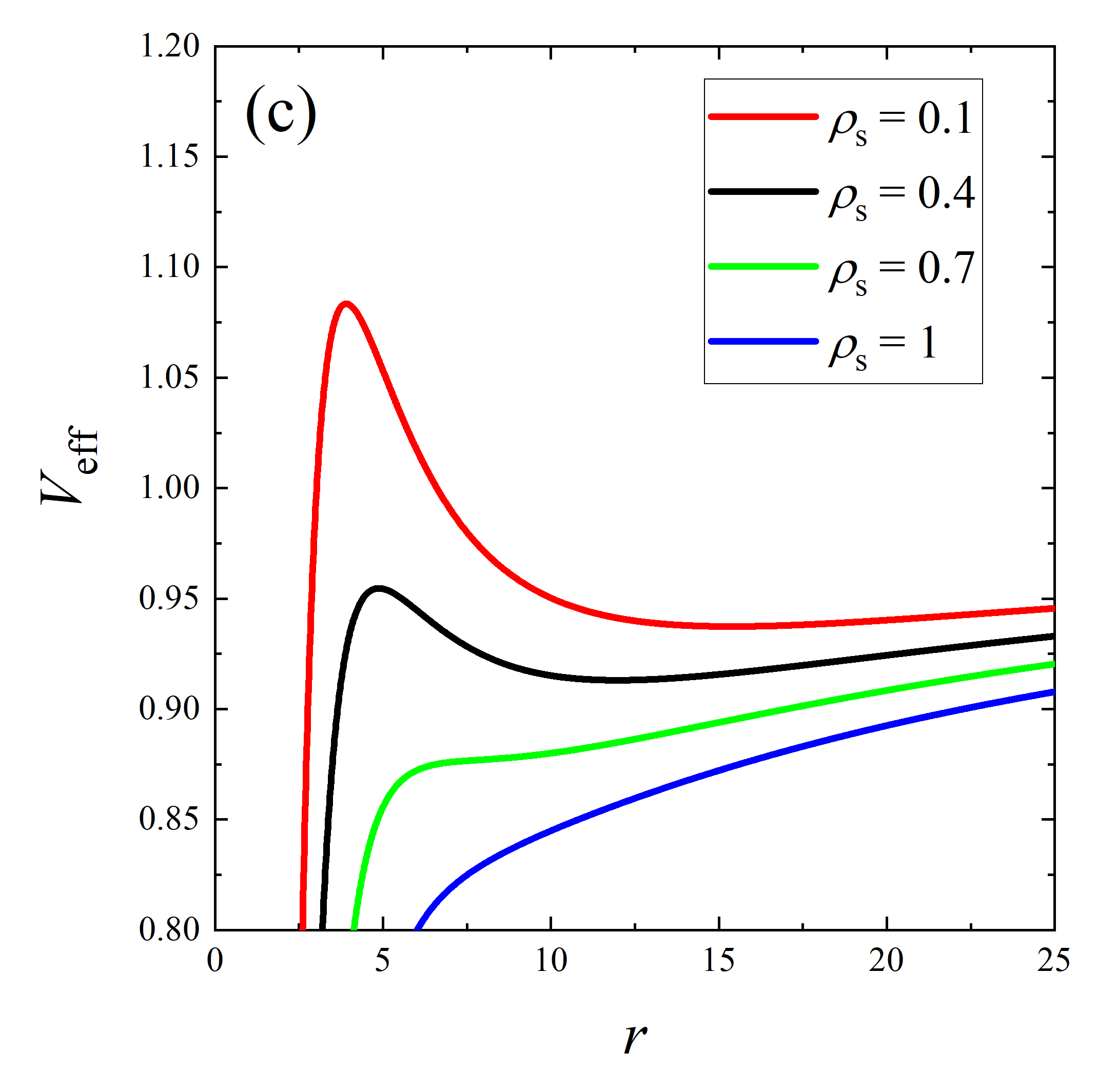}
\caption{Evolution of the effective potential $V_{\textrm{eff}}(r)$ for different parameter spaces. From left to right, the panels display the effects of varying $L$ (with $(r_{\textrm{s}},\rho_{\textrm{s}})=(0.1,0.2)$ fixed), $r_{\textrm{s}}$ (with $(\rho_{\textrm{s}},L)=(0.2,4)$ fixed), and $\rho_{\textrm{s}}$ (with $(r_{\textrm{s}},L)=(0.5,4.5)$ fixed).}}\label{fig1}
\end{figure*}

We find that increasing the specific angular momentum $L$ significantly expands the potential well, which suggests a broader range for precessing orbits. Simultaneously, the potential minimum shifts toward larger radii, implying an increase in the radius of stable circular orbits. In contrast, the influence of dark matter halo parameters is markedly different. As shown in Figs. 1(b) and 1(c), increasing either $r_{\textrm{s}}$ or $\rho_{\textrm{s}}$ gradually diminishes the regions available for bound particles; specifically, the potential curves shift downward and the well shrinks until it eventually vanishes. Furthermore, we observe that for a fixed angular momentum, higher dark matter halo parameters cause the radii of circular orbits to shift inward. The change in the effective potential morphology is fundamentally due to the fact that larger dark matter halo parameters amplify the central gravitational field. This amplification forces the stable orbital range of timelike particles to shrink, thereby increasing their tendency to fall into the black hole.

The radii of stable circular orbits, which correspond to the minima of the effective potential, are determined by the conditions:
\begin{eqnarray}
\frac{\partial V_{\textrm{eff}}}{\partial r} = 0, \label{veffdr} \\
\frac{\partial^{2} V_{\textrm{eff}}}{\partial r^{2}} > 0.
\end{eqnarray}
For any given parameter set $(r_{\textrm{s}},\rho_{\textrm{s}})$, there exists a stable circular orbit with the minimum possible radius, known as the innermost stable circular orbit (ISCO). The ISCO parameters must satisfy the simultaneous conditions: 
\begin{eqnarray}
\frac{\partial V_{\textrm{eff}}}{\partial r} = \frac{\partial^{2} V_{\textrm{eff}}}{\partial r^{2}} = 0.
\end{eqnarray}
By substituting the explicit expression for the effective potential \eqref{veff2}, these conditions yield the following algebraic equations:
\begin{eqnarray}
&0=6\,r^{2}(r+r_{\textrm{s}})^{3}\,(r^{2}+3L^{2})
 + 8\pi r_{\textrm{s}}^{3}\rho_{\textrm{s}}\,r^{5}(L^{2}+r^{2}) \nonumber \\
&\quad+8\pi L^{2} r^{3} r_{\textrm{s}}^{3}\rho_{\textrm{s}}\,(2r+r_{\textrm{s}})(r+r_{\textrm{s}}) \nonumber \\
& - 6 L^{2} r^{3}(r+r_{\textrm{s}})^{3},
\end{eqnarray}
and
\begin{eqnarray}
0 ={}&(L^{2}+r^{2})\Bigl(12 r^{2}(r+r_{\textrm{s}})^{4}
      - 32\pi r_{\textrm{s}}^{3}\rho_{\textrm{s}}\, r^{5}(r+r_{\textrm{s}}) \nonumber \\
     &{} + 24\pi r_{\textrm{s}}^{3}\rho_{\textrm{s}}\, r^{5}(2r+r_{\textrm{s}}) \Bigr) \nonumber \\
&{}+ 36 L^{2} r^{2}(r+r_{\textrm{s}})^{4}
     - 18 L^{2} r^{3}(r+r_{\textrm{s}})^{4} \nonumber \\
&{}+ 24\pi L^{2} r^{3} r_{\textrm{s}}^{3}\rho_{\textrm{s}}(2r+r_{\textrm{s}})(r+r_{\textrm{s}})^{2} \nonumber \\
&{}+ 24 L^{2} r^{2}(r+r_{\textrm{s}})^{4}
     - 32\pi L^{2} r^{4} r_{\textrm{s}}^{3}\rho_{\textrm{s}}(r+r_{\textrm{s}})^{2} \nonumber \\
&{}+ 32\pi L^{2} r^{4} r_{\textrm{s}}^{3}\rho_{\textrm{s}}(2r+r_{\textrm{s}})(r+r_{\textrm{s}}).
\end{eqnarray}
By solving these equations numerically, we obtain the ISCO radius $r_{\textrm{isco}}$ and the corresponding specific angular momentum $L_{\textrm{isco}}$ for various dark matter halo parameters. Furthermore, the specific energy $E_{\textrm{isco}}$ can be directly calculated using the relation \eqref{veff1} under the circular orbit condition $\dot{r}=0$.

Fig. 2 illustrates the distribution of $r_{\textrm{isco}}$ within the two-dimensional parameter space $(r_{\textrm{s}},\rho_{\textrm{s}})$. It is evident that increasing either $r_{\textrm{s}}$ or $\rho_{\textrm{s}}$ enlarges the ISCO radius, with the scale parameter $r_{\textrm{s}}$ exerting a more pronounced influence. This phenomenon stems from the fact that the dark matter halo effectively enhances the gravitational field intensity near the central black hole.
\begin{figure*}
\center{
\includegraphics[width=5cm]{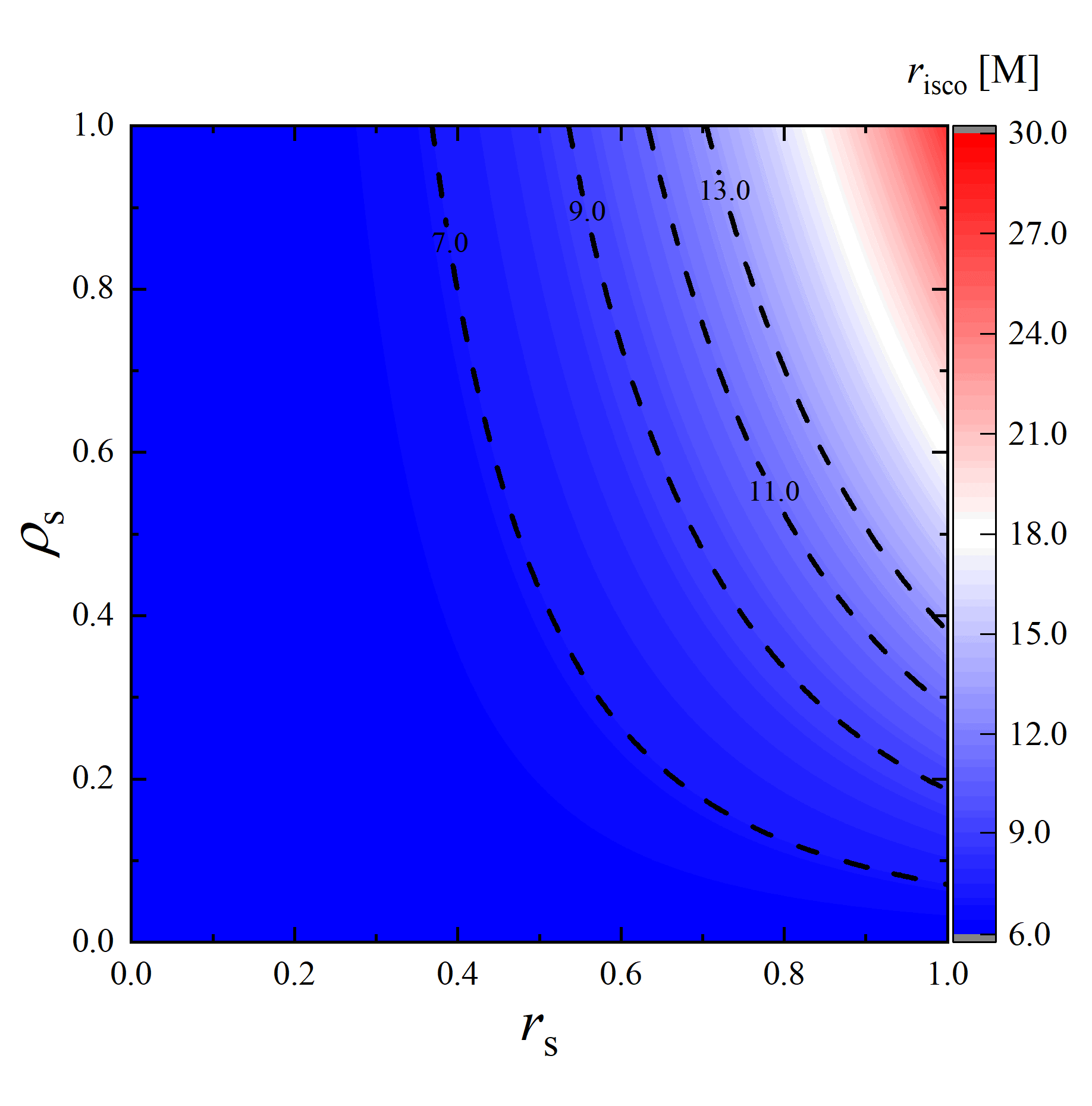}
\caption{Distribution of the ISCO radius $r_{\textrm{isco}}$ as a function of the dark matter halo parameters $r_{\textrm{s}}$ and $\rho_{\textrm{s}}$. The ISCO radius increases with both parameters, though the impact of the scale parameter is notably more significant.}}\label{fig2}
\end{figure*}

The radii of unstable circular orbits, which correspond to the local maxima of the effective potential, are determined by the conditions:
\begin{eqnarray}
\frac{\partial V_{\textrm{eff}}}{\partial r} = 0, \\
\frac{\partial^{2} V_{\textrm{eff}}}{\partial r^{2}} < 0.
\end{eqnarray}
In this context, we focus on the marginally bound orbit (MBO), which is defined by the specific energy condition $V_{\textrm{eff}}=E^{2}=1$. By solving this condition alongside the extremum condition $\partial V_{\textrm{eff}}/\partial r = 0$, one can obtain the MBO radius $r_{\textrm{mbo}}$, the corresponding specific angular momentum $L_{\textrm{mbo}}$, and the specific energy $E_{\textrm{mbo}}$.

When a particle is released near the minimum of the effective potential, specifically within the potential well, it undergoes quasi-periodic motion. The radial extent of this motion can be determined from Eq. \eqref{veff1}. Fig. 3 illustrates the dependence of $\dot{r}^{2}$ on $r$ across various parameter spaces. All curves exhibit a consistent trend: as $r$ increase, $\dot{r}^{2}$ drops sharply from an initially large value, reaches a minimum, rebounds to a maximum, and then gradually declines. In this process, if a curve intersects the horizontal blue dashed line ($\dot{r}^{2}=0$) at three points, it indicates the existence of a stable bound region for the particle. In such cases, the two larger roots correspond to the periastron and apastron of the orbit, as illustrated by the green, red, and pink curves in Fig. 3(a). It is worth noting that while the smallest root of $\dot{r}^{2}=0$ is mathematically valid, as Eq. \eqref{veff1} typically yields a cubic equation in $r$, this root lacks physical significance for the orbits considered here.
\begin{figure*}
\center{
\includegraphics[width=5cm]{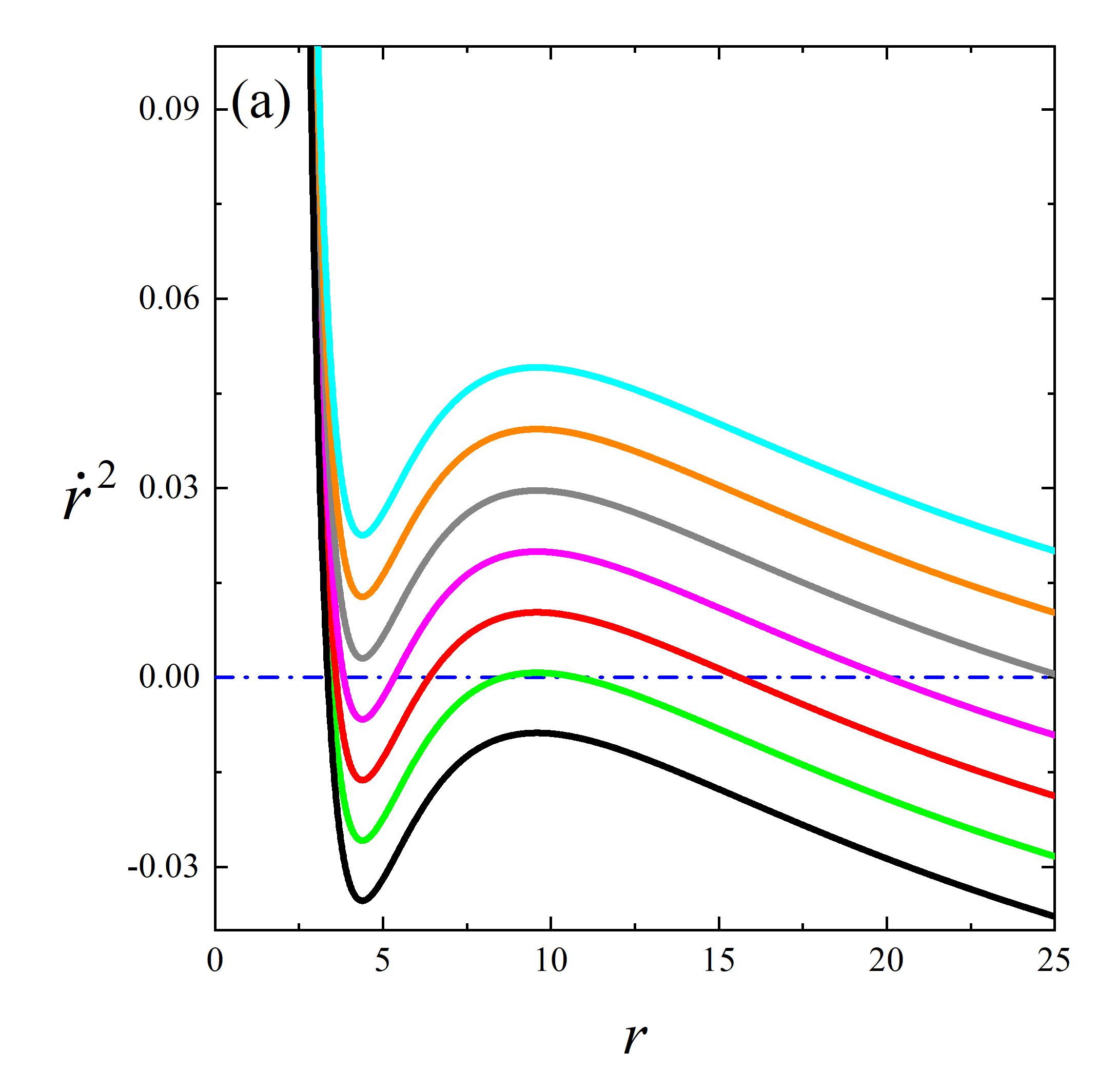}
\includegraphics[width=5cm]{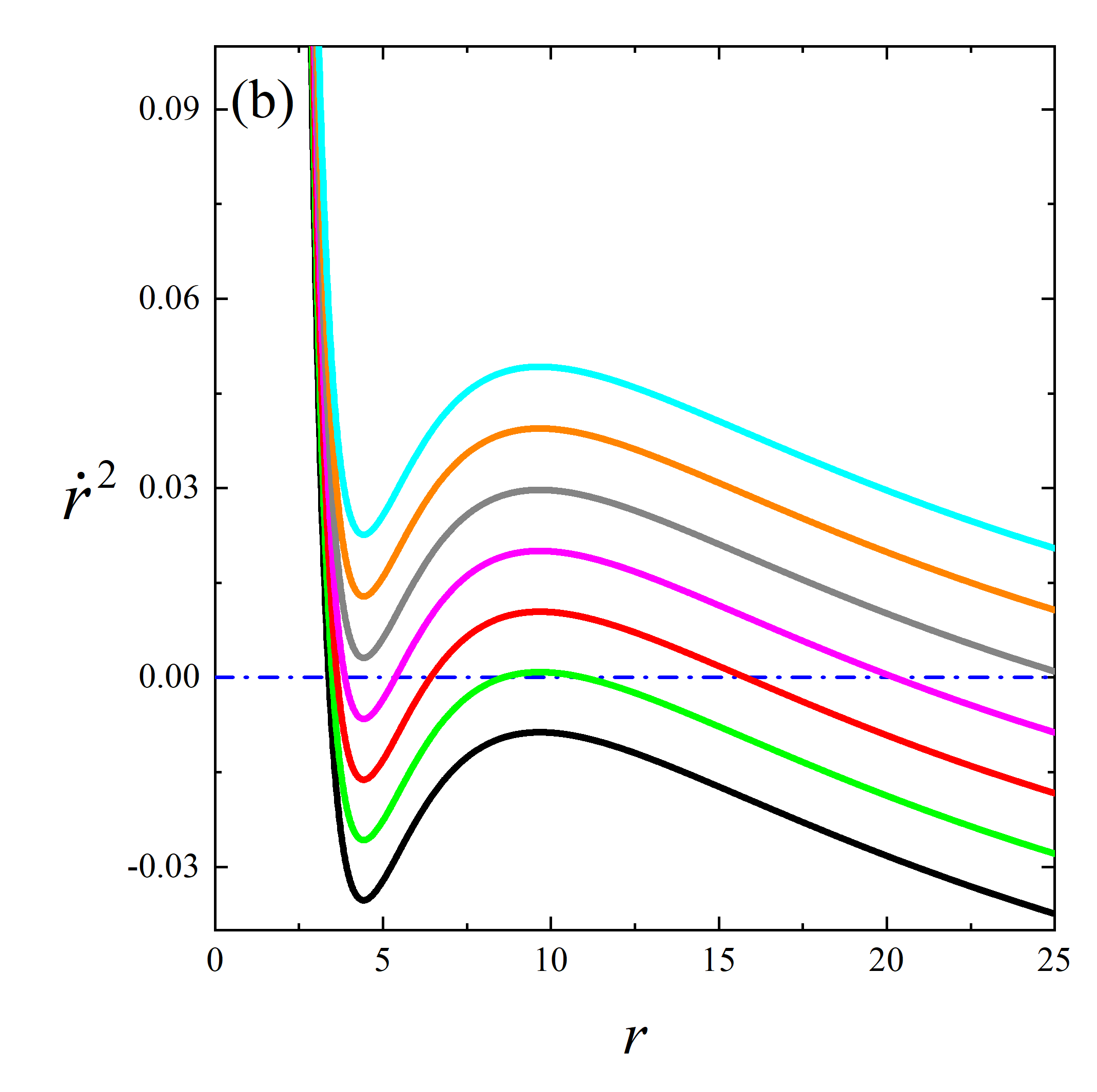}
\includegraphics[width=5cm]{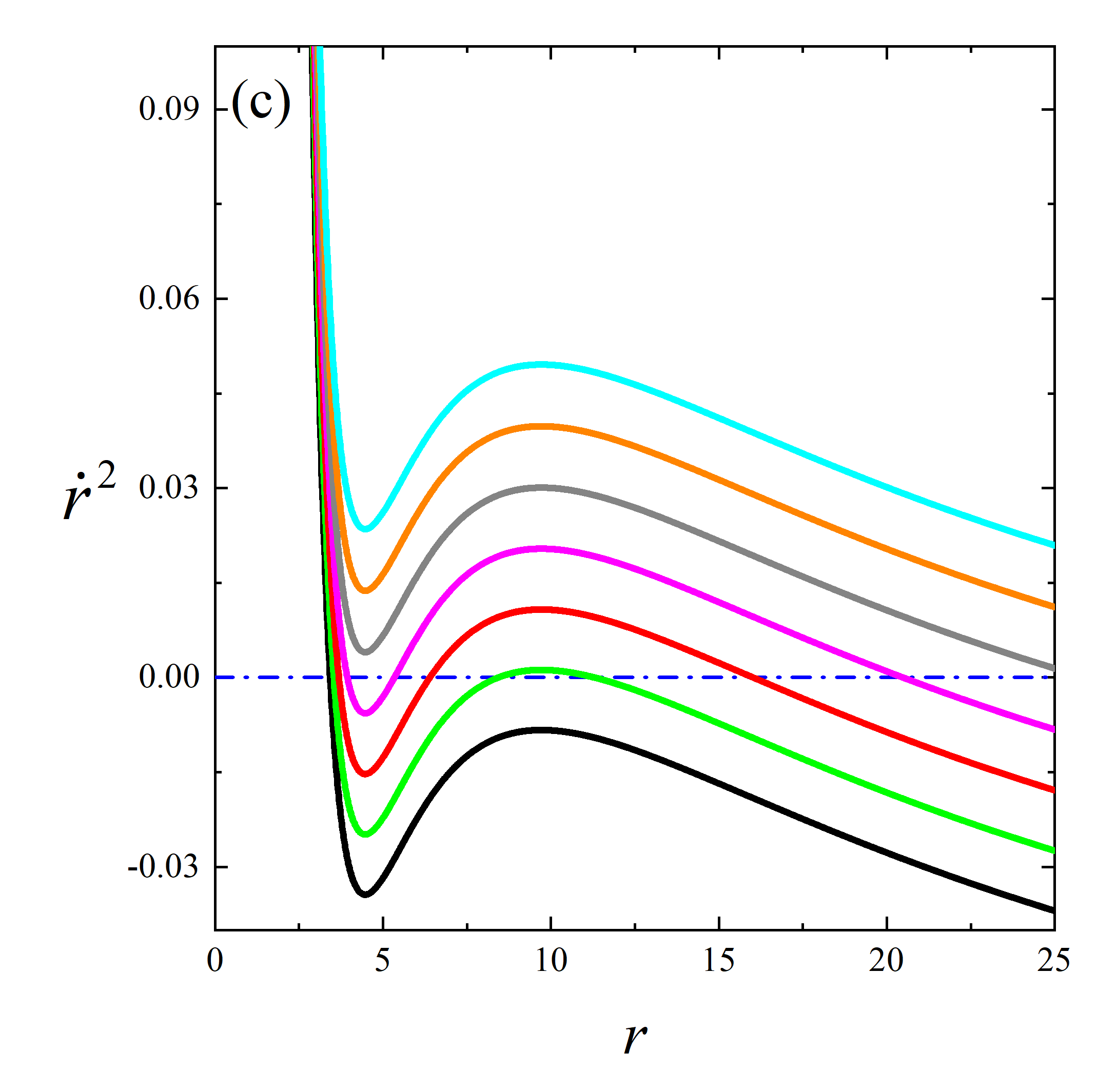}
\includegraphics[width=5cm]{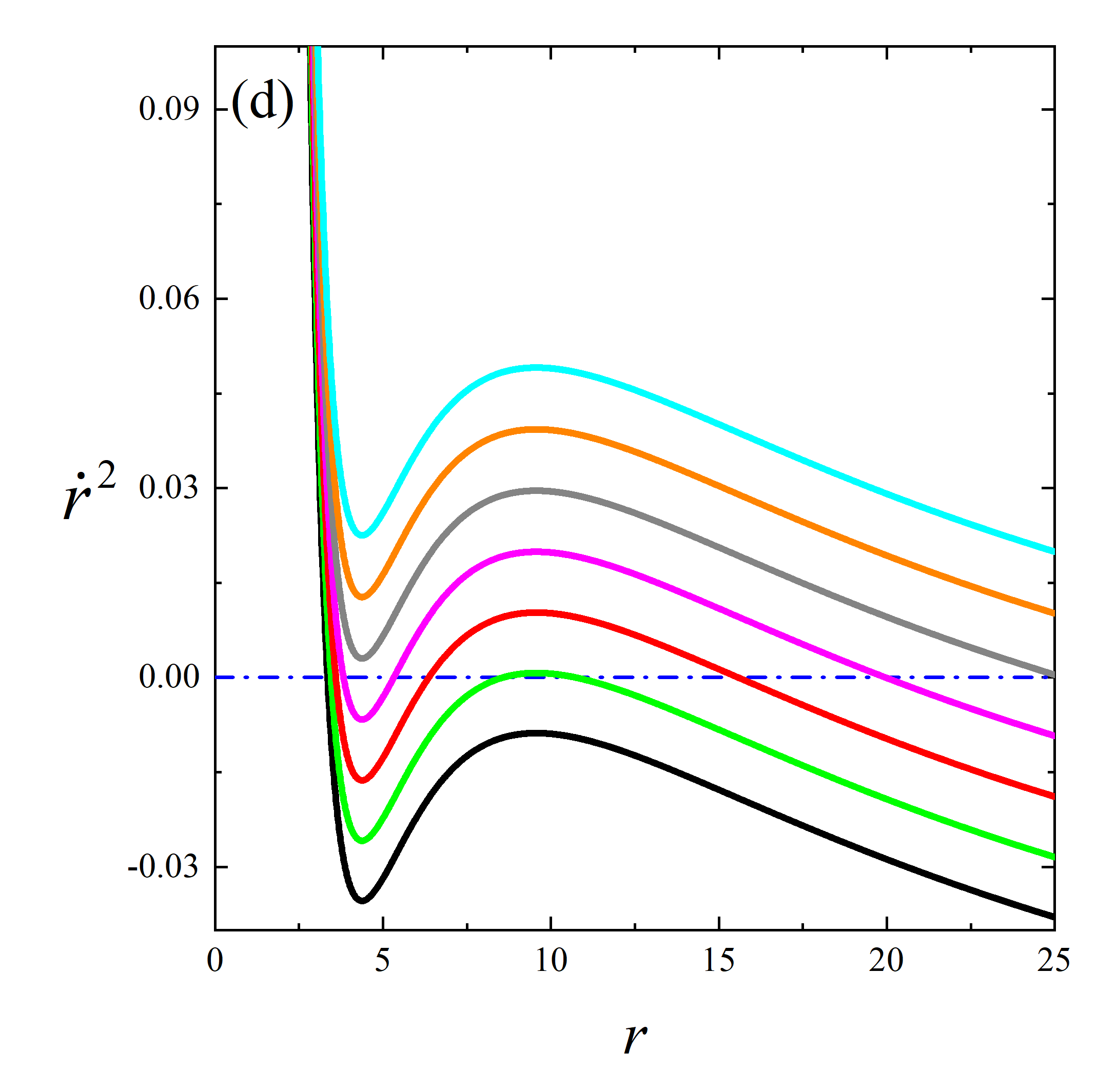}
\includegraphics[width=5cm]{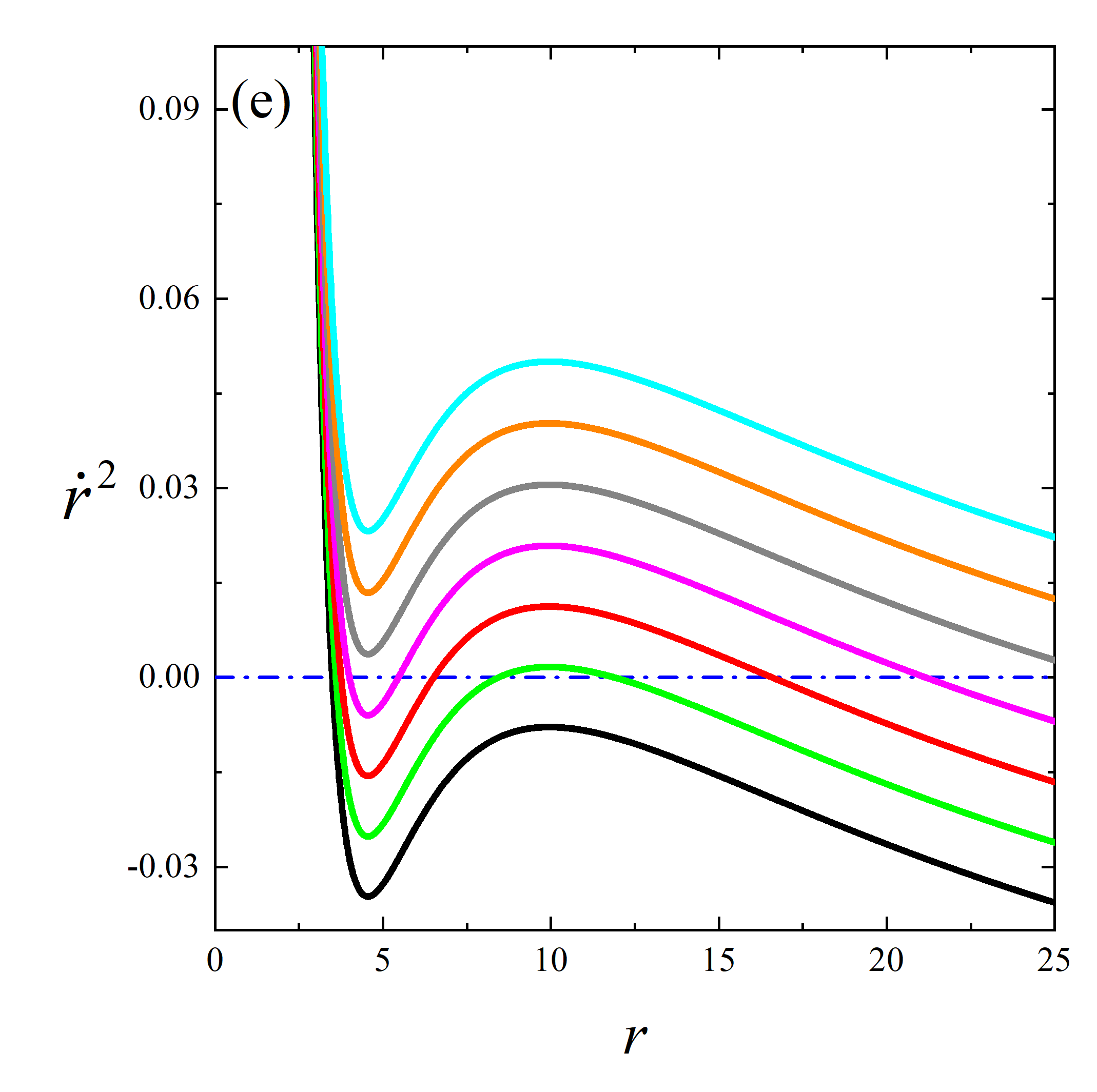}
\includegraphics[width=5cm]{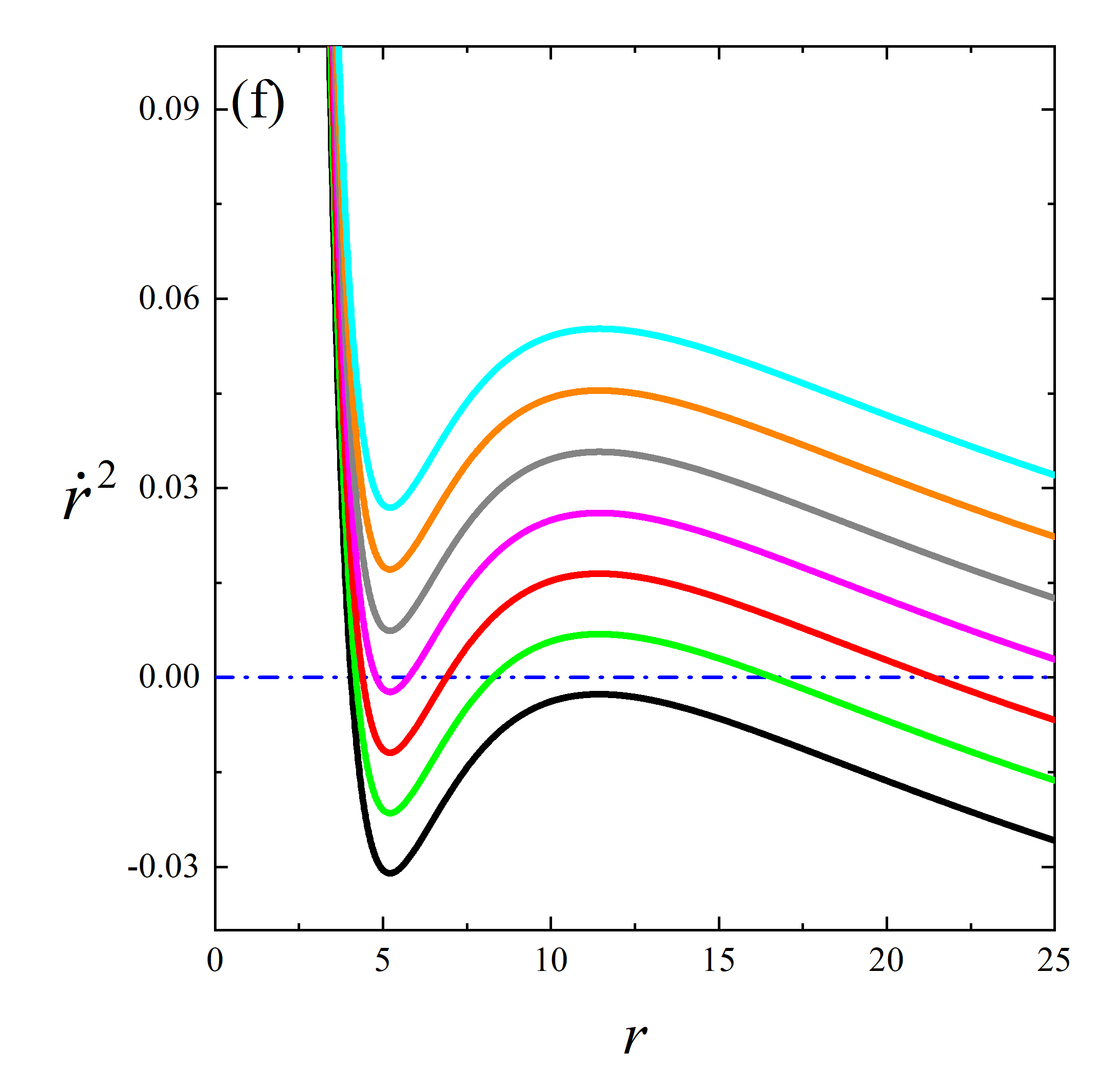}
\caption{Evolution of $\dot{r}^{2}$ with respect to $r$ in different parameter spaces. The first row displays results for $r_{\textrm{s}}=0.2$ with $\rho_{\textrm{s}}$ set to $0.1$, $0.4$, and $0.7$ (from left to right), while the second row shows results for $\rho_{\textrm{s}} = 0.2$ with $r_{\textrm{s}}$ varying as $0.1$, $0.4$, and $0.7$. In each panel, the curves from bottom to top correspond to specific energy $E$ increasing from $0.95$ to $0.98$ with an interval of $0.005$. Each curve is calculated with a fixed angular momentum $L=(L_{\textrm{isco}}+L_{\textrm{mbo}})/2$. The blue horizontal dashed line represents $\dot{r}^{2}=0$. When a curve possesses three intersections with this line, a bound orbit exists, where the two rightmost intersections represent the periastron and apastron, respectively.}}\label{fig3}
\end{figure*}

The results in Fig. 3 demonstrate that the dark matter halo significantly modifies the stable orbital region. For a fixed set of other parameters, increasing either $r_{\textrm{s}}$ or $\rho_{\textrm{s}}$ shifts the curves upward, thereby expanding the bound region for lower-energy orbits. In particular, it can be inferred that stable orbits, which might not exist in a vacuum Schwarzschild spacetime under certain initial conditions, can be generated by increasing either the specific energy of the particle or the dark matter halo parameters. Notably, the influence of $r_{\textrm{s}}$ is more pronounced; for instance, as $r_{\textrm{s}}$ increases from $0.1$ to $0.7$ for the green curves in the second row, the curves shift upward significantly, the separation between the two rightmost zeros widens, and the orbital range consequently expands.

\begin{figure*}
\center{
\includegraphics[width=6cm]{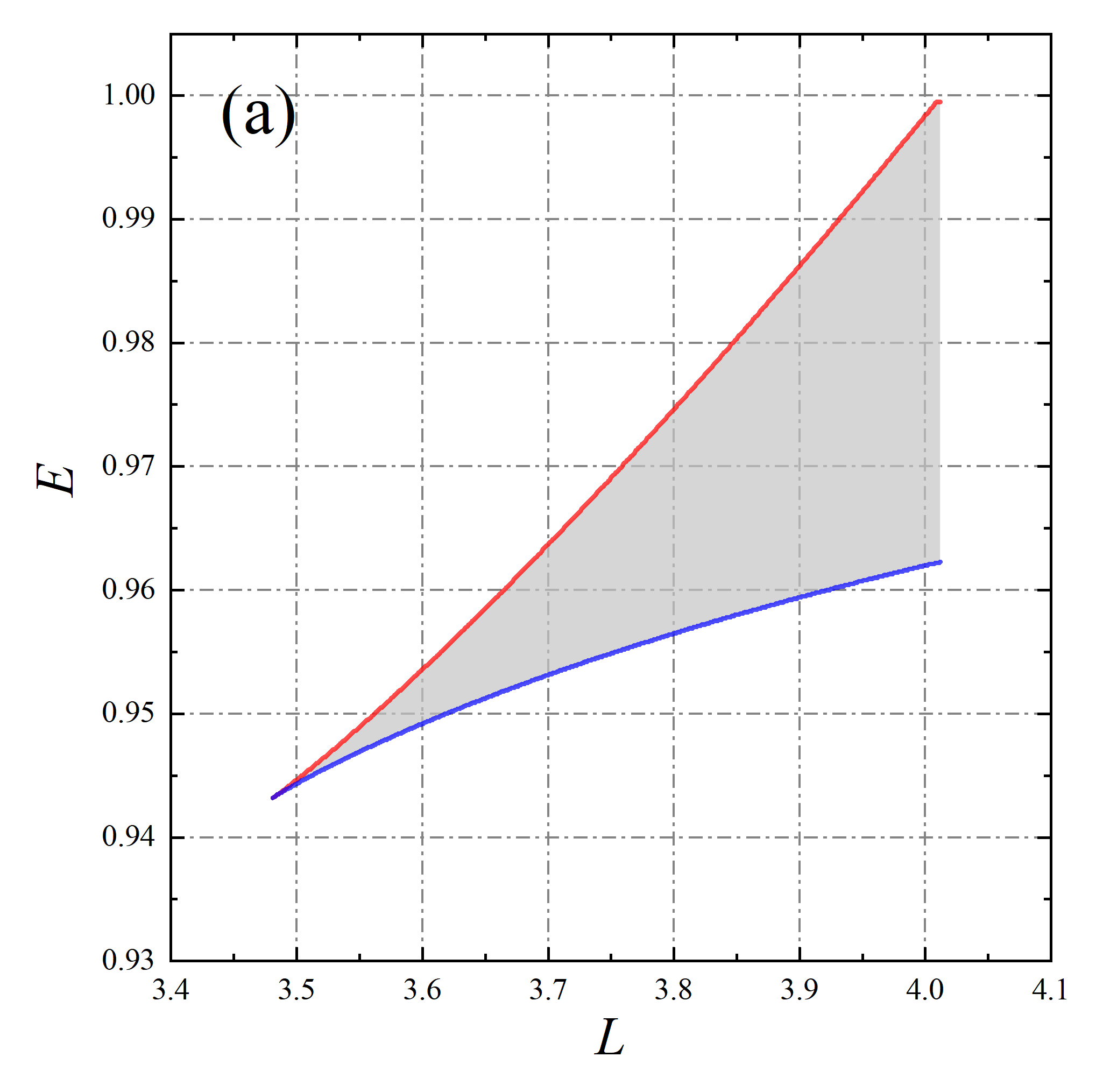}
\includegraphics[width=6cm]{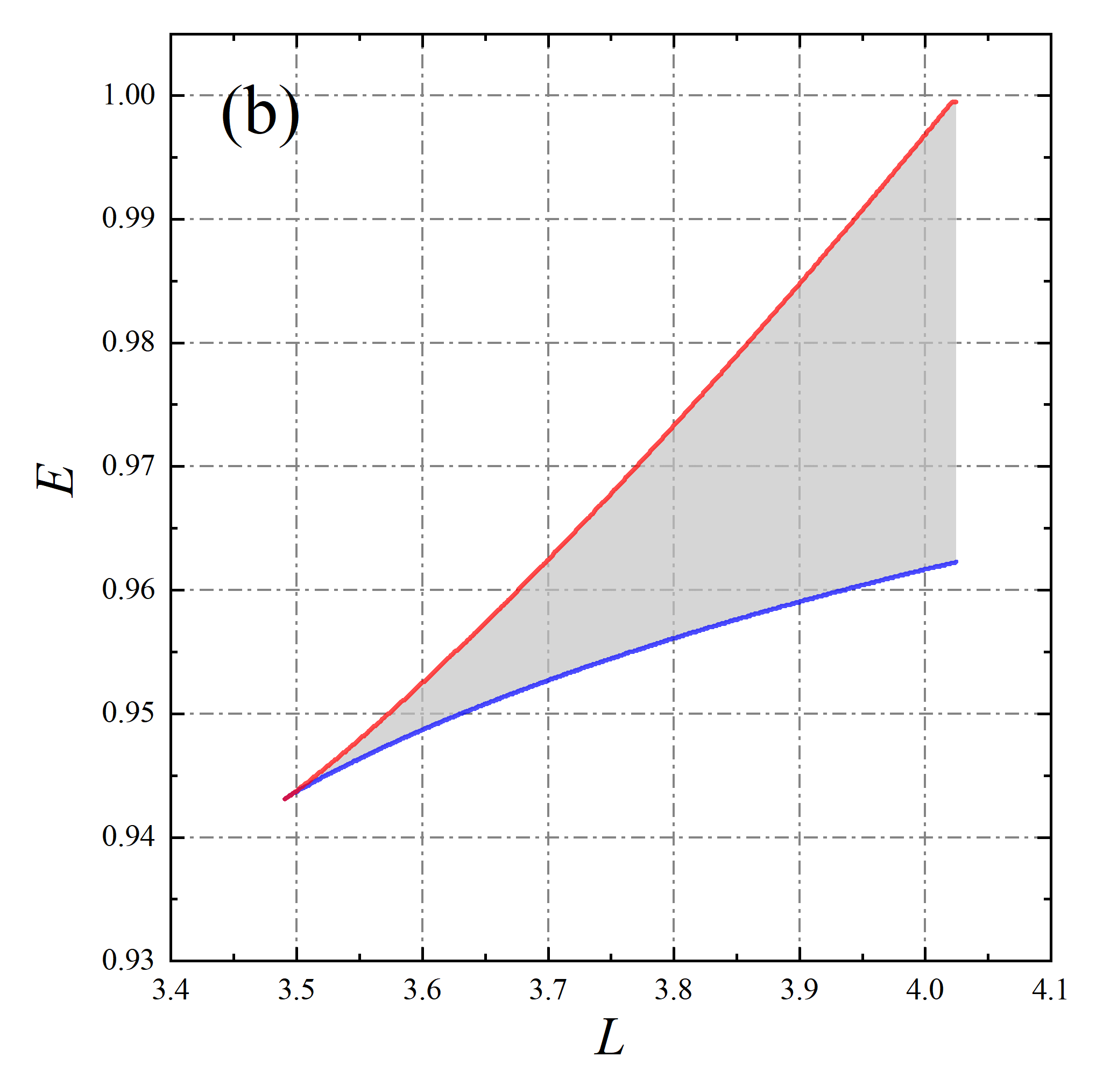}
\includegraphics[width=6cm]{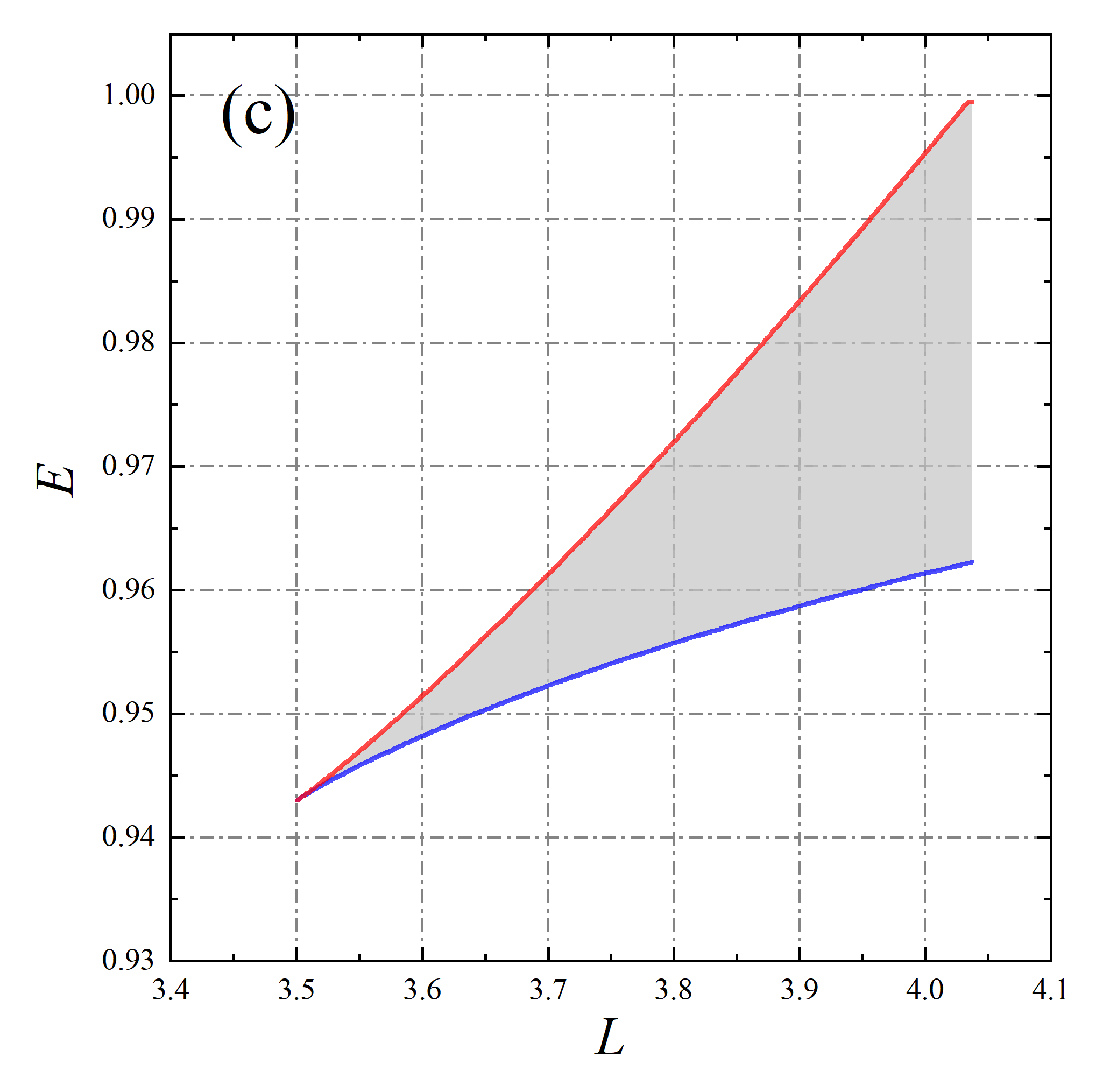}
\includegraphics[width=6cm]{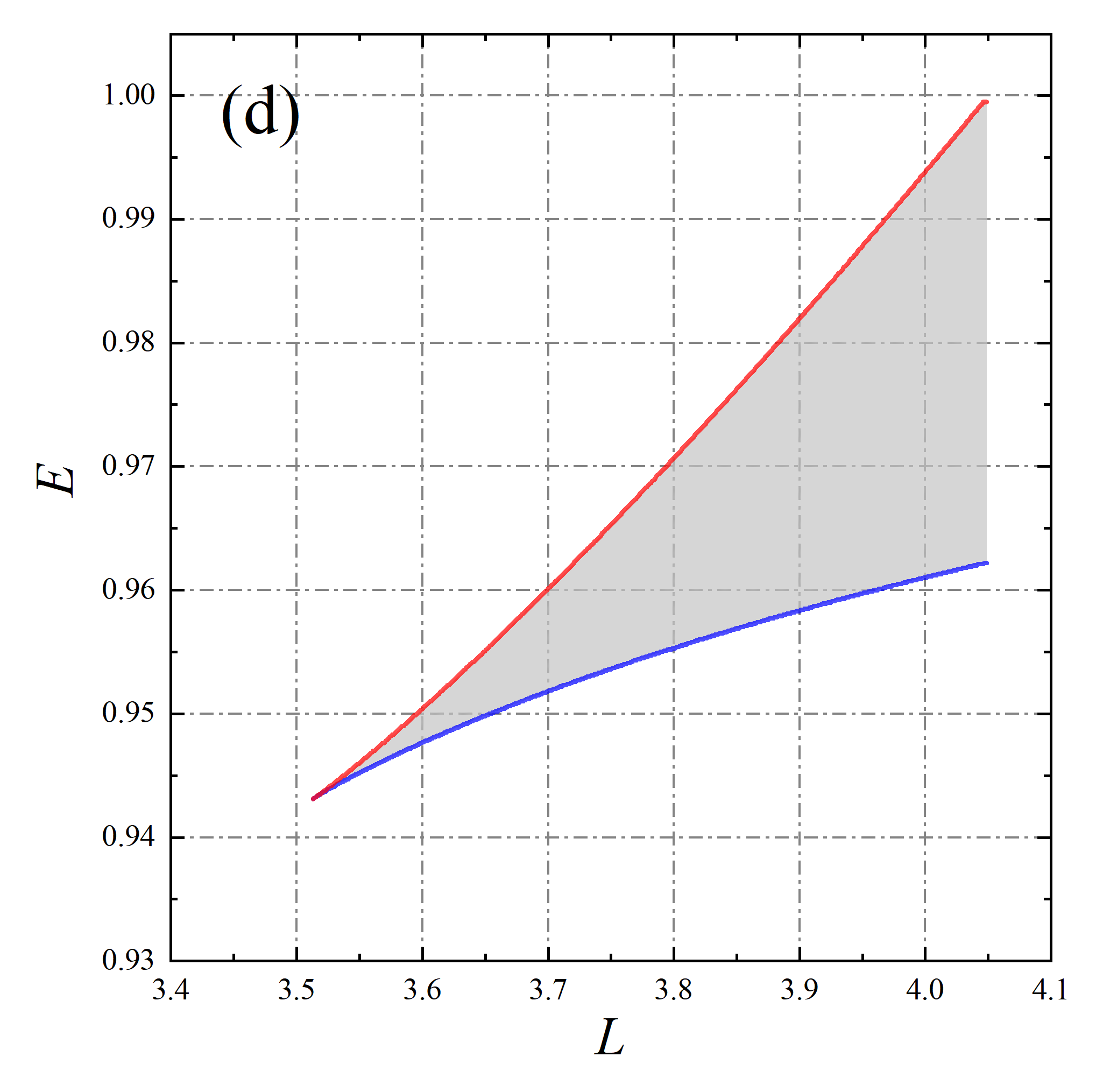}
\caption{Relationship between the allowed specific energy range and specific angular momentum for bound orbits under different dark matter halo density parameters. Panels (a)--(d) correspond to $\rho_{\textrm{s}}=0.1$, $0.2$, $0.3$, and $0.4$, respectively, with a fixed scale radius $r_{\textrm{s}}=0.2$. In each panel, the red and blue curves represent the maximum and minimum specific energy allowed for bound orbits.}}\label{fig4}
\end{figure*}
From another perspective, Fig. 3 reveals a systematic method for identifying bound orbits: given the specific energy $E$, specific angular momentum $L$, and dark matter halo parameters, one must verify whether $\dot{r}^{2}=0$ yields three distinct roots. Following this approach, we investigate how the stability of orbits depends on $E$ and $L$ under different dark matter halo density parameters for a fixed $r_{\textrm{s}}$, as illustrated in Fig. 4. We find that for various angular momenta, the allowable range of $E$ varies; specifically, this range expands as $L$ increases. Consequently, the set of $(E,L)$ pairs that admit bound orbits forms a triangular-like region in the parameter plane. It is also observed that as $\rho_{\textrm{s}}$ increases, this triangular region shifts slightly toward higher angular momenta. In other words, for a given orbital energy, a higher dark matter density necessitates a larger angular momentum to maintain a bound orbit. Furthermore, for a fixed specific angular momentum, increasing the dark matter density results in a contraction of the allowable specific energy range for bound orbits. Similarly, we examine the range of $E$ and $L$ for bound orbits under various halo scale parameters while keeping the density fixed, as shown in Fig. 5. The results are analogous to those in Fig. 4: an increase in $r_{\textrm{s}}$ causes the entire region to shift to the right, and this displacement becomes significantly more pronounced as $r_{\textrm{s}}$ grows larger.
\begin{figure*}
\center{
\includegraphics[width=9cm]{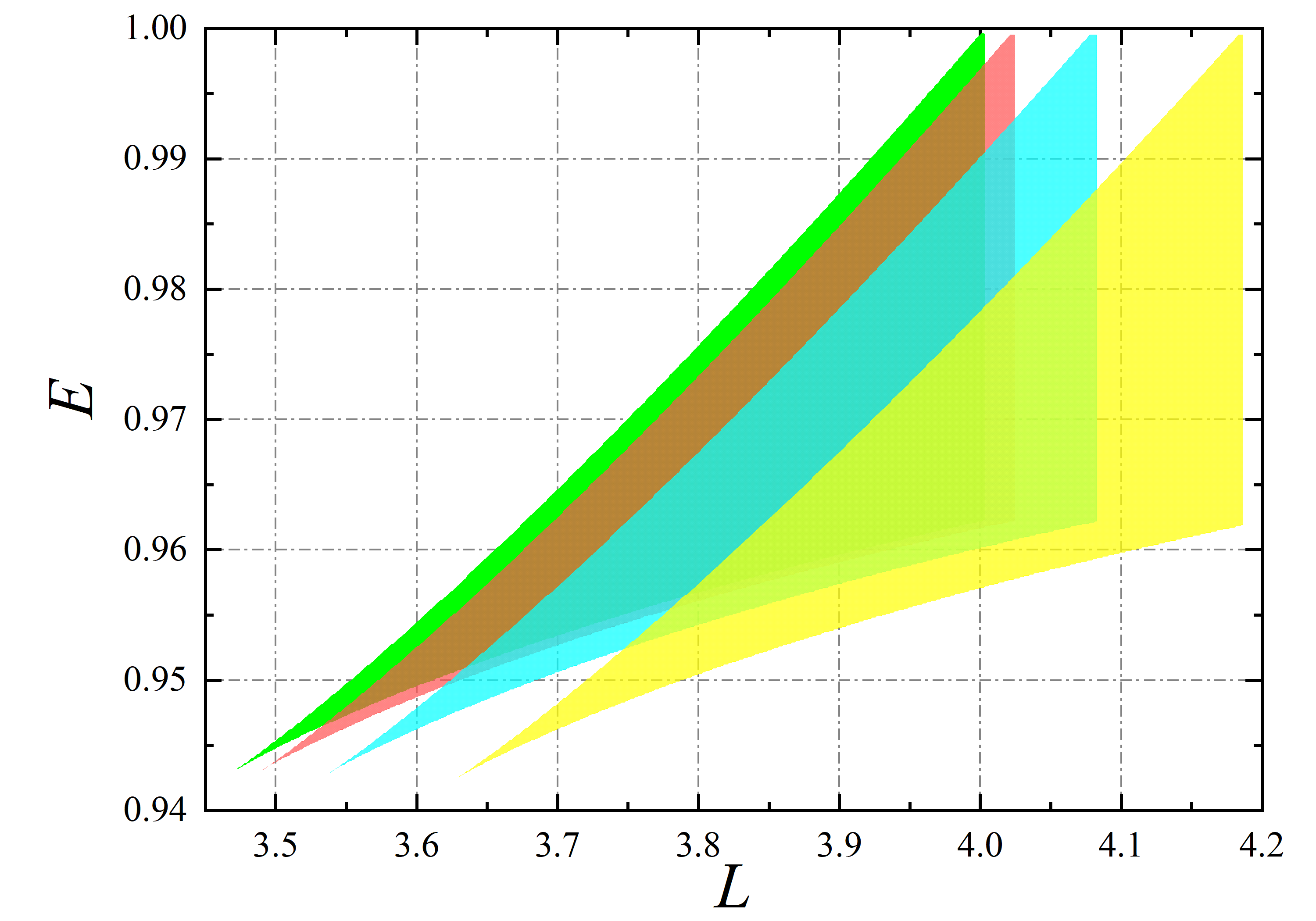}
\caption{Relationship between the specific energy range and specific angular momentum for bound orbits under various dark matter halo scale parameters. The green, red, sky-blue, and yellow curves correspond to $r_{\textrm{s}}=0.1$, $0.2$, $0.3$, and $0.4$, respectively, with the halo density fixed at $\rho_{\textrm{s}}=0.2$.}}\label{fig5}
\end{figure*}
\section{Rational number for closed orbits}
In the framework of general relativity, precise numerical calculations facilitate the identification of orbits that both precess and close. In contrast to generic precessing orbits that never exactly repeat their paths, these quasi-periodic orbits, also referred to as closed or rational orbits, satisfy a commensurability condition between their radial and angular frequencies. Consequently, after a finite number of radial oscillations, the particle returns to its initial angular position, forming a closed loop in configuration space. The topology of such closed orbits is characterized by a rational number $q$. Following the definition provided in \cite{2008PhRvD..77j3005L}, $q$ is expressed as:
\begin{equation}
q = w+\frac{v}{z}. \label{number:q1}
\end{equation}
Here, $w$, $z$, and $v$ are positive integers representing the number of whirls, the number of leaves, and the first vertex hit of the orbit, respectively. Intuitively, the number of whirls $w$ counts the near-circular loops the orbit completes around the black hole near the periastron; the number of leaves $z$ denotes the number of distinct petals or lobes the orbit traces before repeating; and the first vertex hit $v$ specifies which apastron is first revisited after one full radial period. Thus, a set $(z,w,v)$ uniquely characterizes the orbital configuration.

The rational number $q$ is intrinsically related to the accumulated azimuthal angle of the orbit. Specifically, it can be defined as:
\begin{equation}
q = \frac{\Delta\phi}{2\pi} - 1, \label{number:q2}
\end{equation}
where $\Delta\phi$ denotes the accumulated azimuthal angle as the orbit evolves from apastron to periastron and back to the next apastron. This angle is calculated as:
\begin{eqnarray}
\Delta\phi &=& 2\int^{\phi_{2}}_{\phi_{1}}\textrm{d}\phi \nonumber \\
 &=& 2\int^{r_{2}}_{r_{1}}\frac{\dot{\phi}}{\dot{r}}\textrm{d}r. \label{deltaphi}
\end{eqnarray}
where $\phi_{1}$ and $r_{1}$ denote the azimuthal angle and radial coordinate at the periastron, while $\phi_{2}$ and $r_{2}$ are the corresponding values at the apastron. By combining Eqs. \eqref{eq:pphi}, \eqref{veff1}, and \eqref{veff2}, it becomes evident that calculating $\Delta\phi$ between the two turning points, and subsequently determining the rational number $q$, requires only the specification of the specific energy $E$, specific angular momentum $L$, and the dark matter halo parameters. On the other hand, we can also compute the proper time required for a timelike particle to travel from apastron to periastron (or vice versa), which can be expressed as
\begin{eqnarray}\label{periodictime}
\Delta\tau=\int^{r_{2}}_{r_{1}}\frac{1}{\dot{r}}\textrm{d}r.
\end{eqnarray}
Then, the proper time needed for one complete rational orbit is $\tau=2z\Delta\tau$.

Next, we introduce a parameter $\varepsilon \in [0,1]$ to select the specific angular momentum $L$ between $L_{\textrm{isco}}$ and $L_{\textrm{mbo}}$:
\begin{equation}
L = L_{\textrm{isco}} + \varepsilon\left(L_{\textrm{mbo}}-L_{\textrm{isco}}\right).
\end{equation}
By examining whether the radial kinetic term $\dot{r}^{2}=0$ has three distinct roots, the allowable range of specific energy $[E_{\textrm{min}},E_{\textrm{max}}]$ for bound orbits can be determined for any given $L$. Fig. 6 illustrates the evolution of the rational number $q$ within the interval $[E_{\textrm{min}},E_{\textrm{max}}]$ for different values of $\varepsilon$ and dark matter halo parameters. It is evident that all curves share a consistent trend: $q$ starts from a relatively small value, increases gradually with $E$ across most of the interval, and terminates with a sharp rise as $E$ approaches $E_{\textrm{max}}$. This behavior indicates that a higher specific energy leads to an increased number of leaves and whirls before the orbit closes. We also observe that increasing $\varepsilon$ shifts the entire curve to the right, implying that a larger specific angular momentum requires higher energy to maintain orbital stability. More importantly, the range of the rational number $q$ expands during this process. For instance, in Fig. 6(a), the condition $q > 1.25$ always holds; however, when $\varepsilon$ increases to $0.5$, as shown in Fig. 6(c), the range of $q$ is significantly broadened. This extension implies that orbital configurations with $w=0$ can emerge at higher angular momenta. A comparison between the upper and lower rows reveals that the presence of a dark matter halo shifts the curves to the left. Furthermore, the scale parameter $r_{\textrm{s}}$ exerts a more substantial impact on the orbital structure than the density parameter $\rho_{\textrm{s}}$. Additionally, an increase in $L$ appears to suppress the influence of the dark matter halo parameters on the rational number distribution.
\begin{figure*}
\center{
\includegraphics[width=4.5cm]{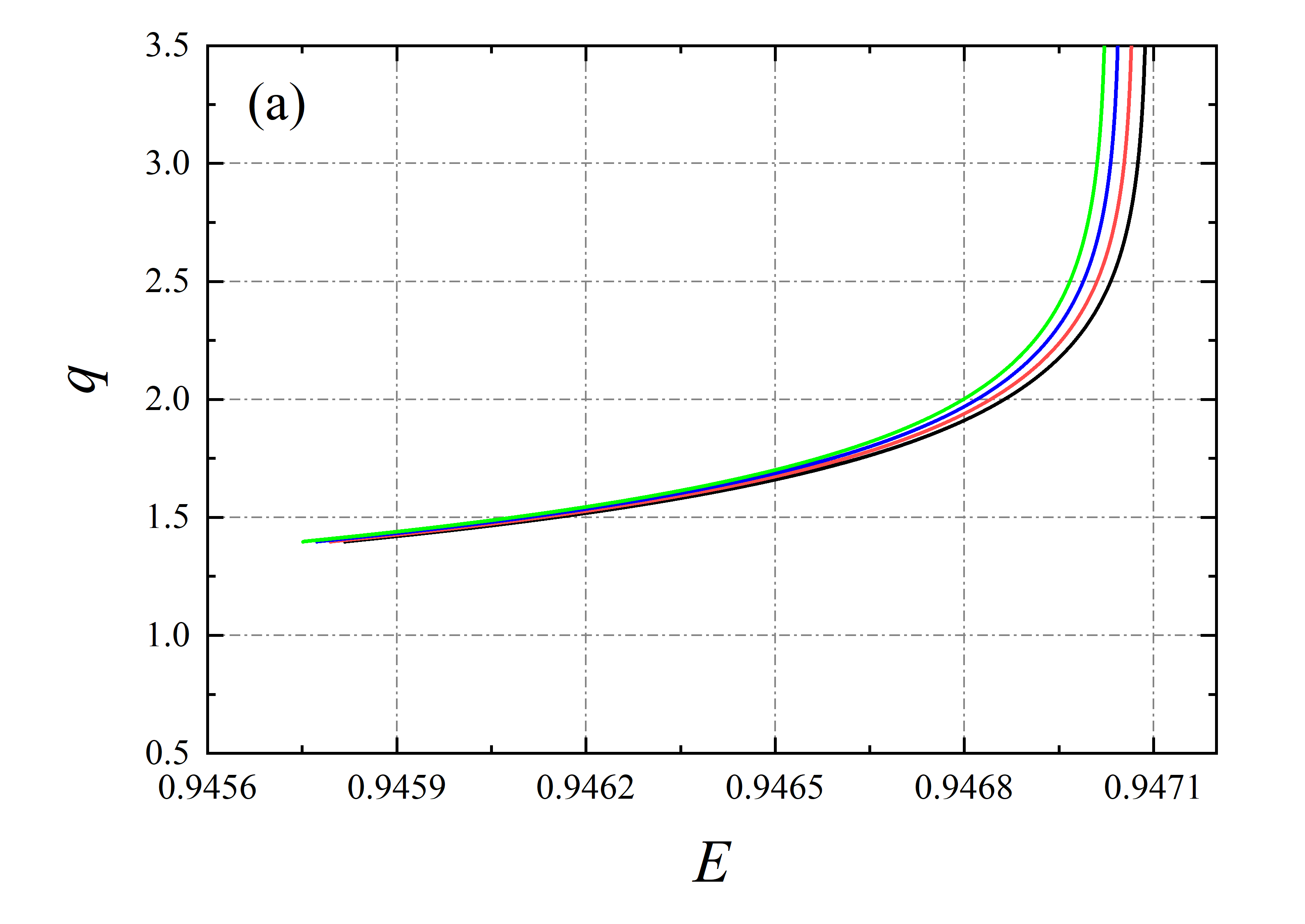}
\includegraphics[width=4.5cm]{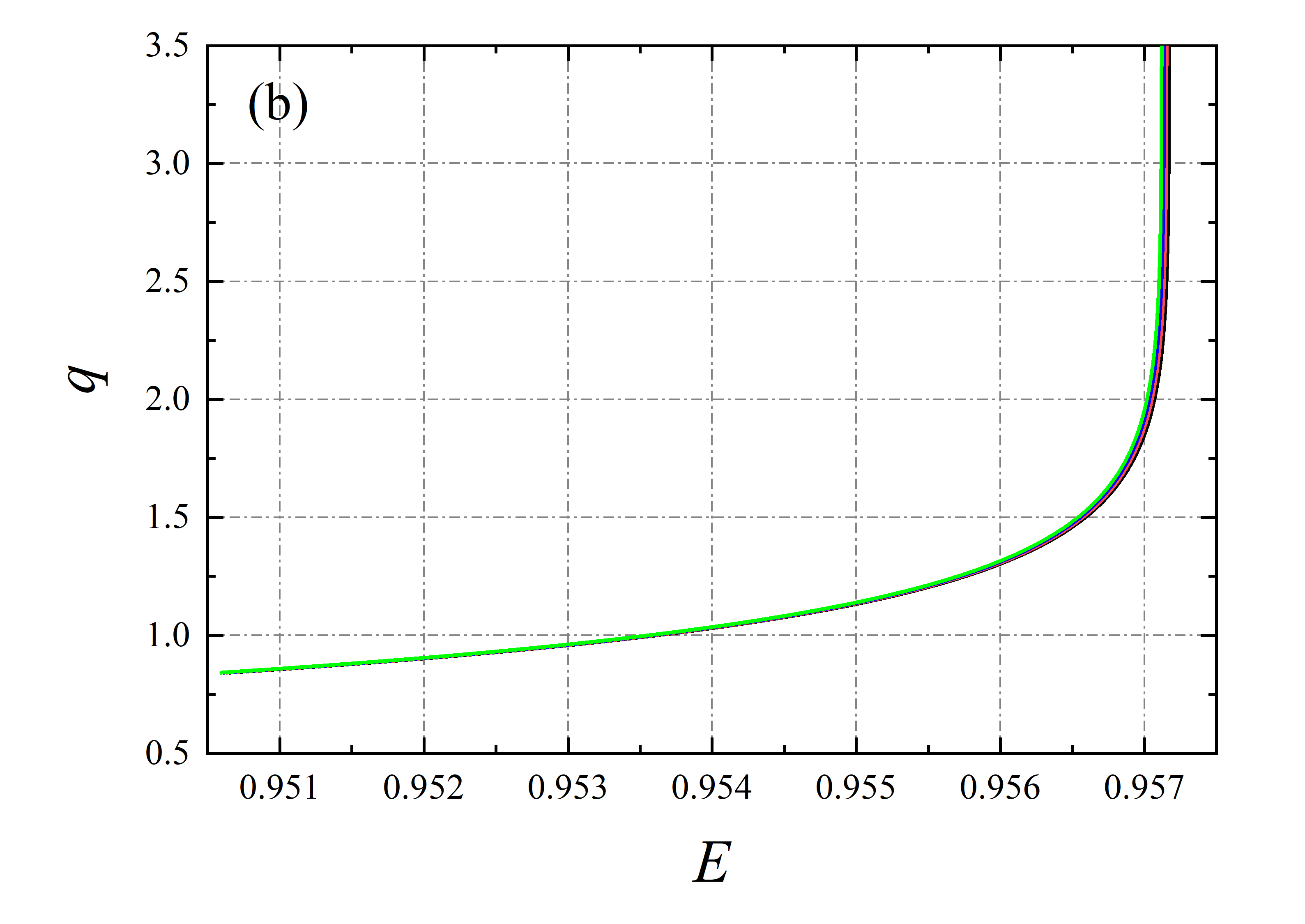}
\includegraphics[width=4.5cm]{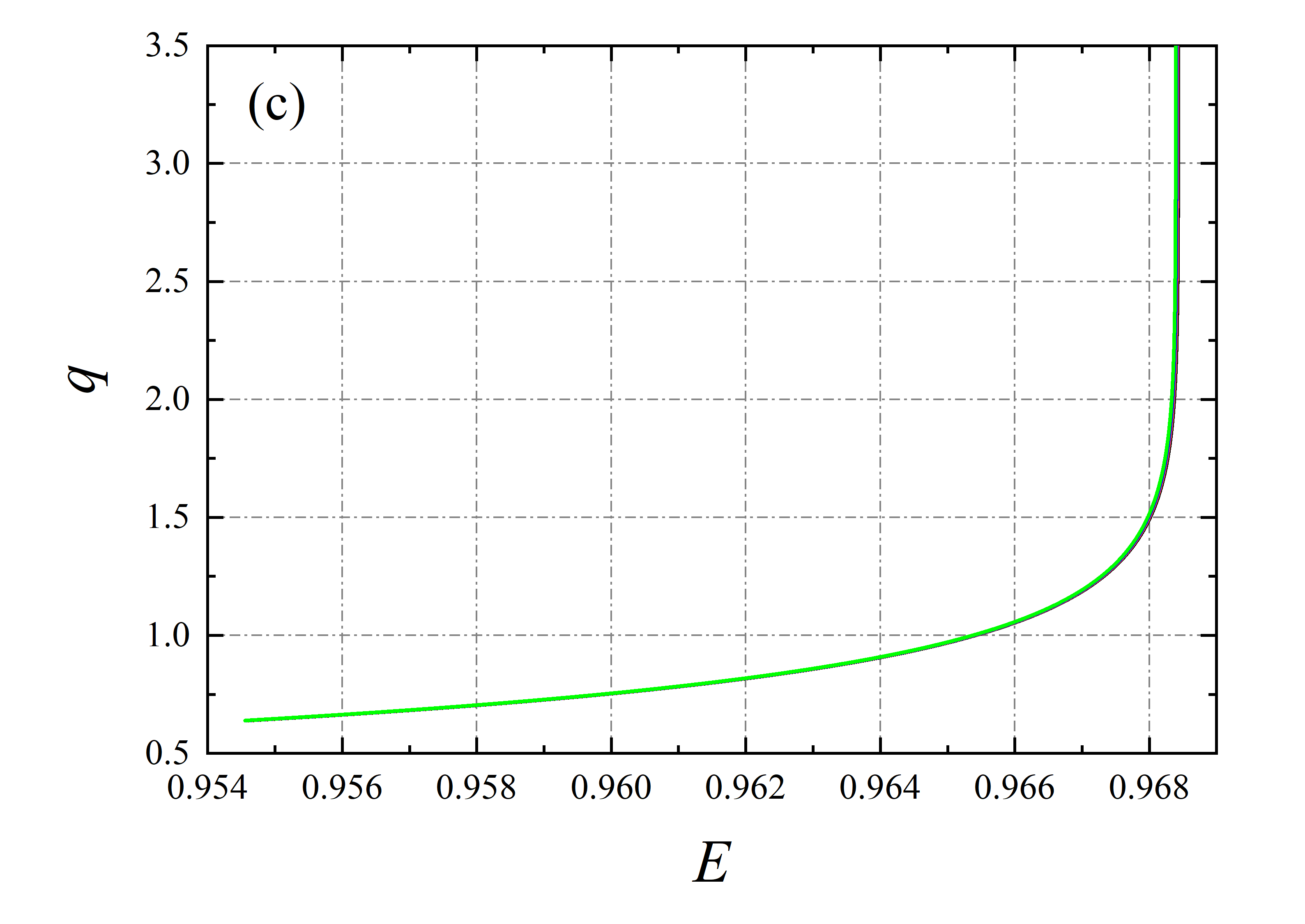}
\includegraphics[width=4.5cm]{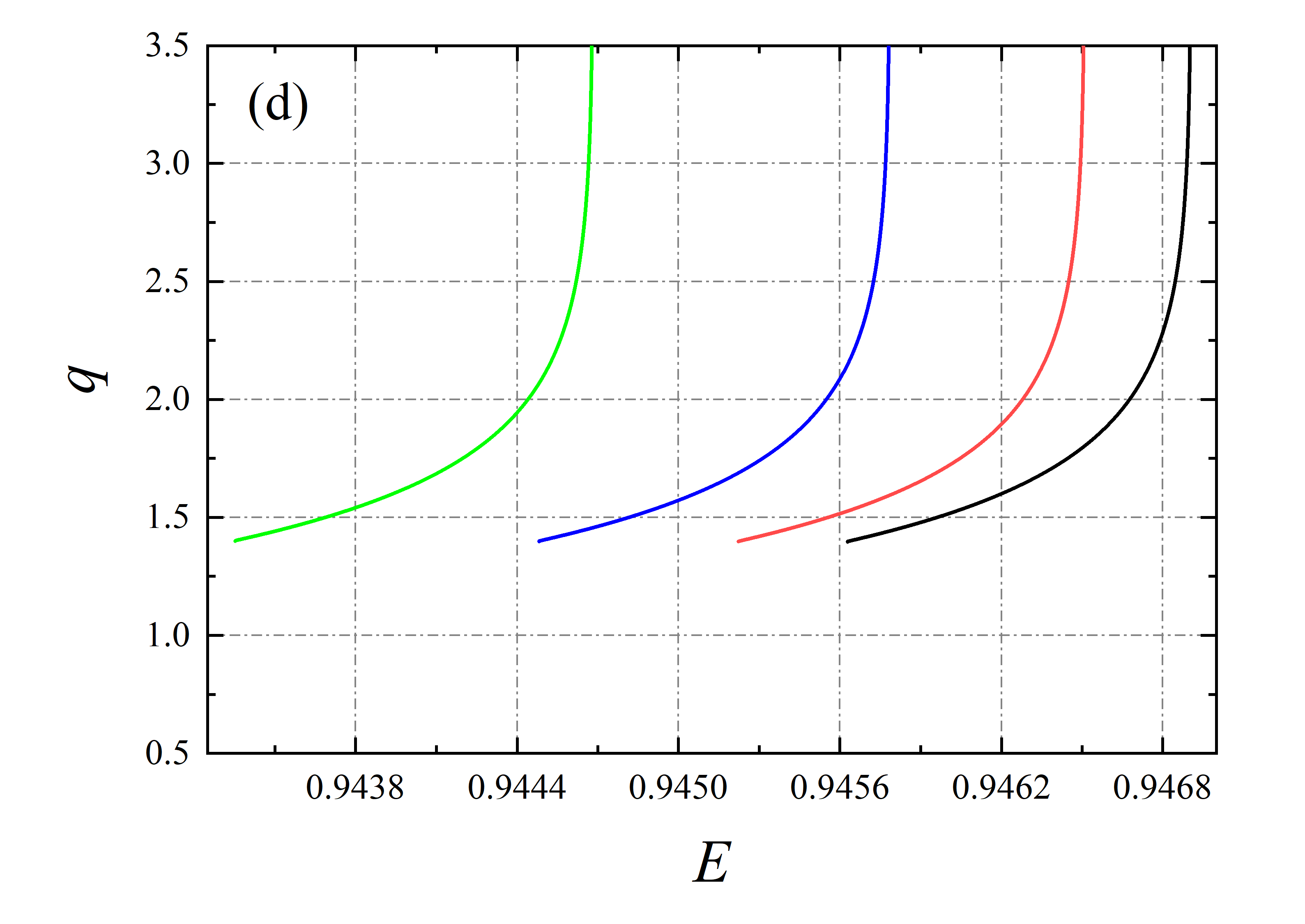}
\includegraphics[width=4.5cm]{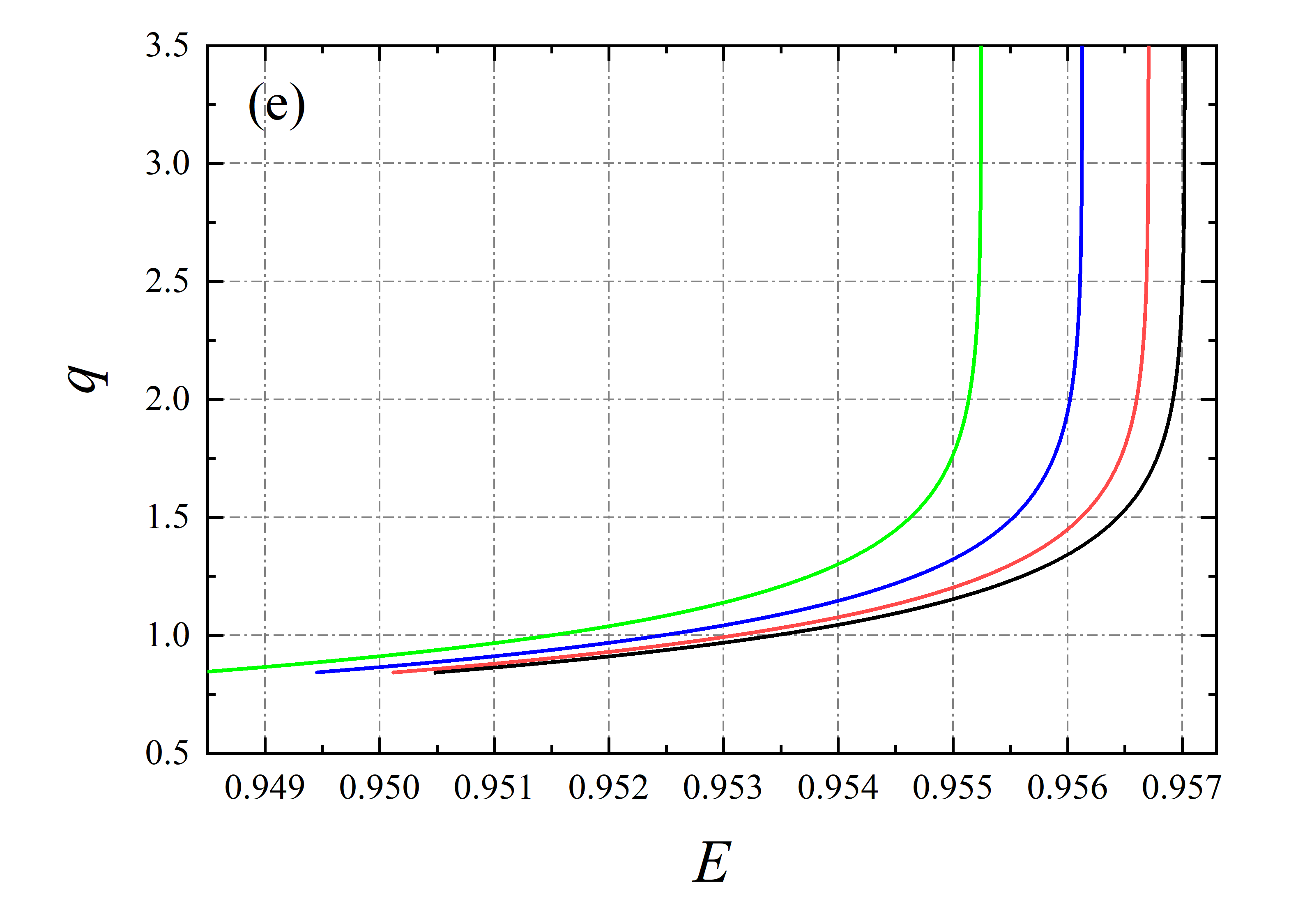}
\includegraphics[width=4.5cm]{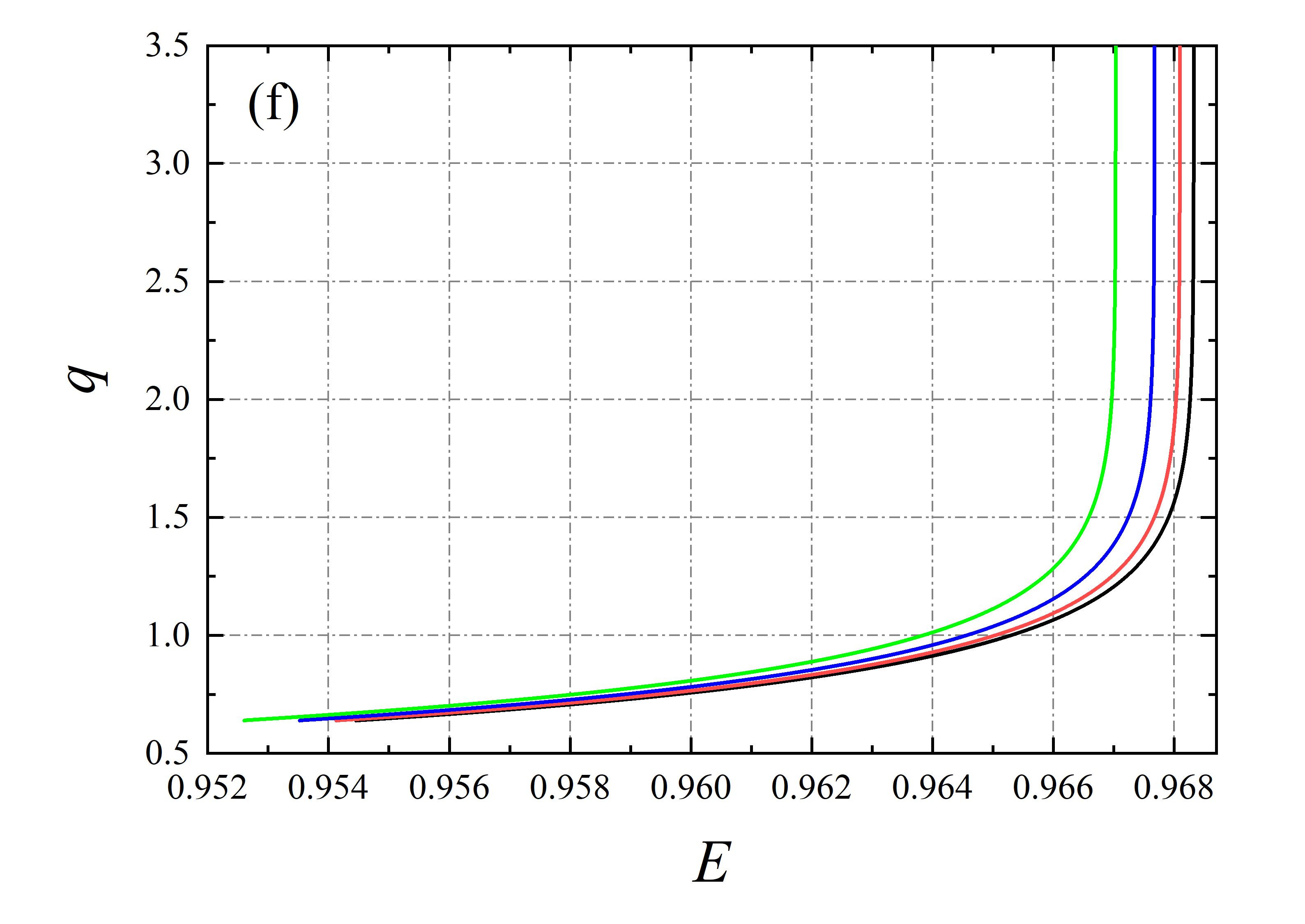}
\caption{Distribution of the rational number $q$ as a function of specific energy $E$ across different parameter spaces. From left to right, the panels correspond to $\varepsilon =$ $0.1$, $0.3$, and $0.5$. The upper row shows results for fixed $r_{\textrm{s}}=0.2$ with $\rho_{\textrm{s}}$ varying from $0.1$ to $0.4$ (black, red, blue, and green curves, respectively). The lower row displays results for fixed $\rho_{\textrm{s}}=0.2$ with $r_{\textrm{s}}$ varying from $0.3$ to $0.6$. All curves exhibit a slow initial growth followed by a rapid surge as $E$ approaches $E_{\textrm{max}}$.}}\label{fig6}
\end{figure*}
\section{Orbits, gravitational waves, and light curves}
The general procedure for identifying closed rational orbits in curved spacetimes is as follows: given a specific orbital configuration $(z,w,v)$, we first calculate the corresponding rational number $q$ and the accumulated azimuthal angle $\Delta\phi$. Subsequently, for a specified set of dark matter halo parameters and specific angular momentum, we determine the required specific energy $E$ by performing an interpolation based on the integral in Eq. \eqref{deltaphi}. Once $E$ is obtained, the initial generalized coordinates and conjugate momenta associated with the chosen orbital configuration are uniquely determined. To simulate the trajectories, we employ OCTOPUS, a numerical toolkit developed in \cite{Hu:2025lyp}. The underlying integration is performed using a sixth-order fixed-step Runge-Kutta method (RK6).

\begin{figure*}
\center{
\includegraphics[width=3.1cm]{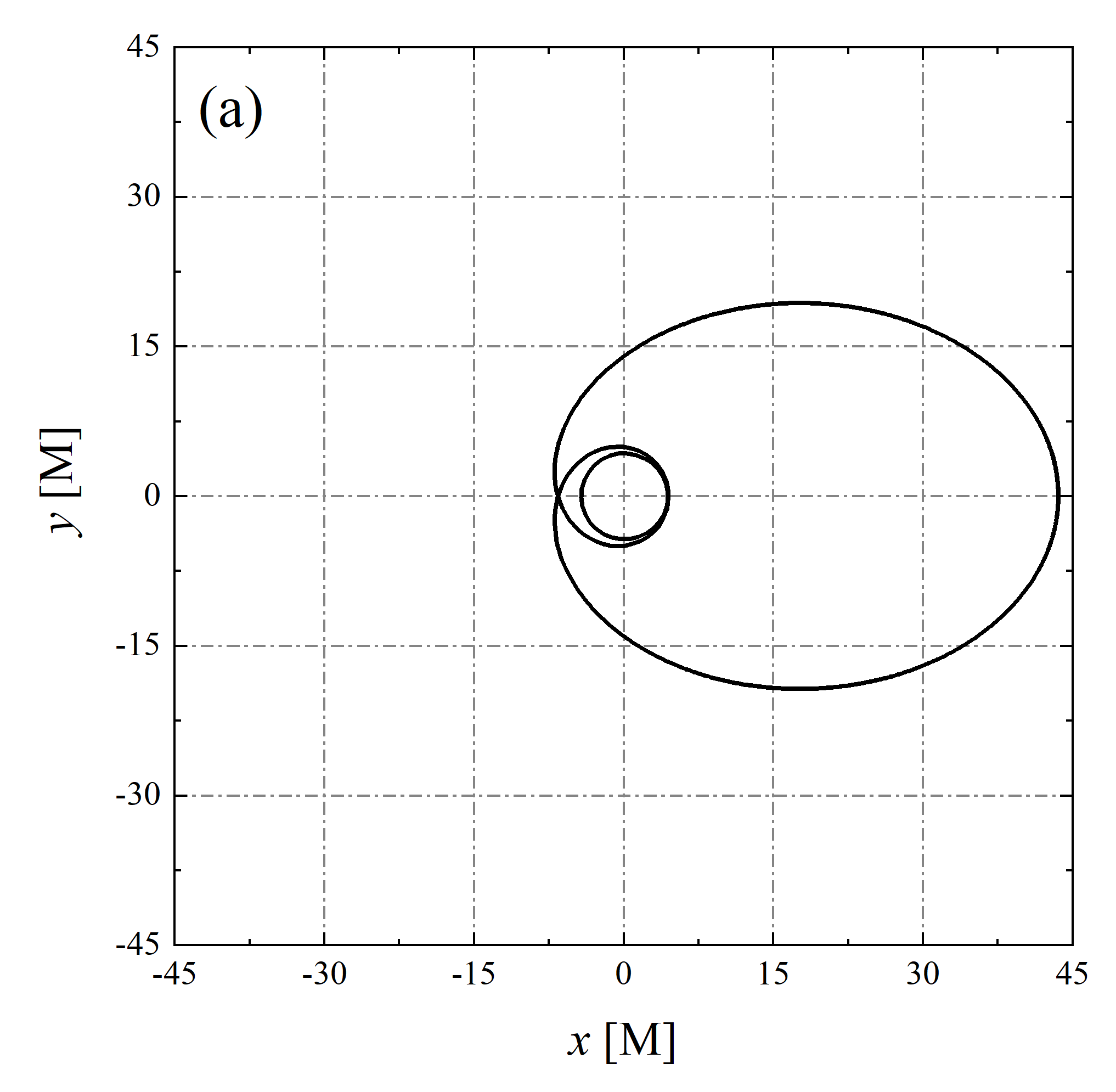}
\includegraphics[width=3.1cm]{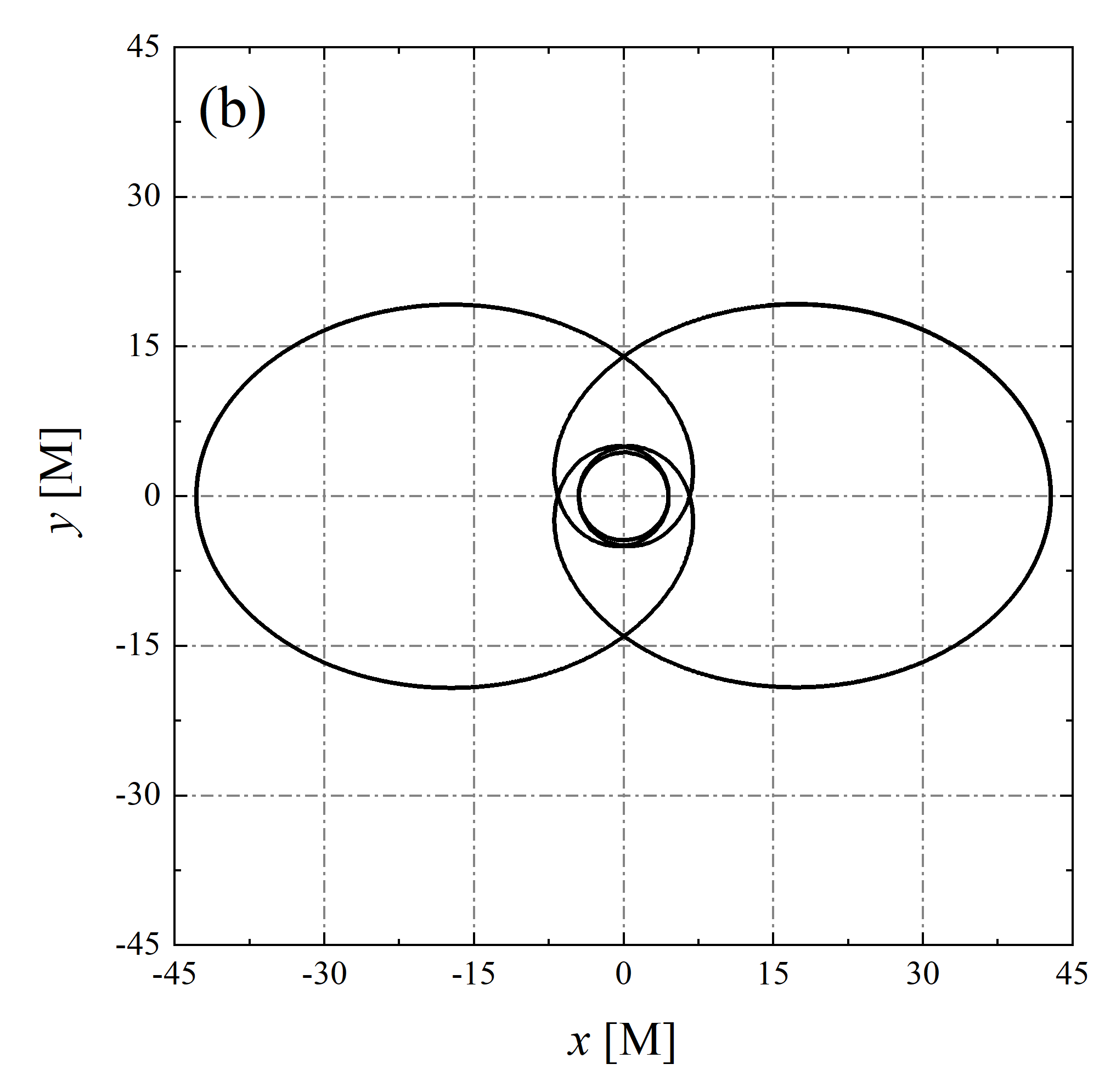}
\includegraphics[width=3.1cm]{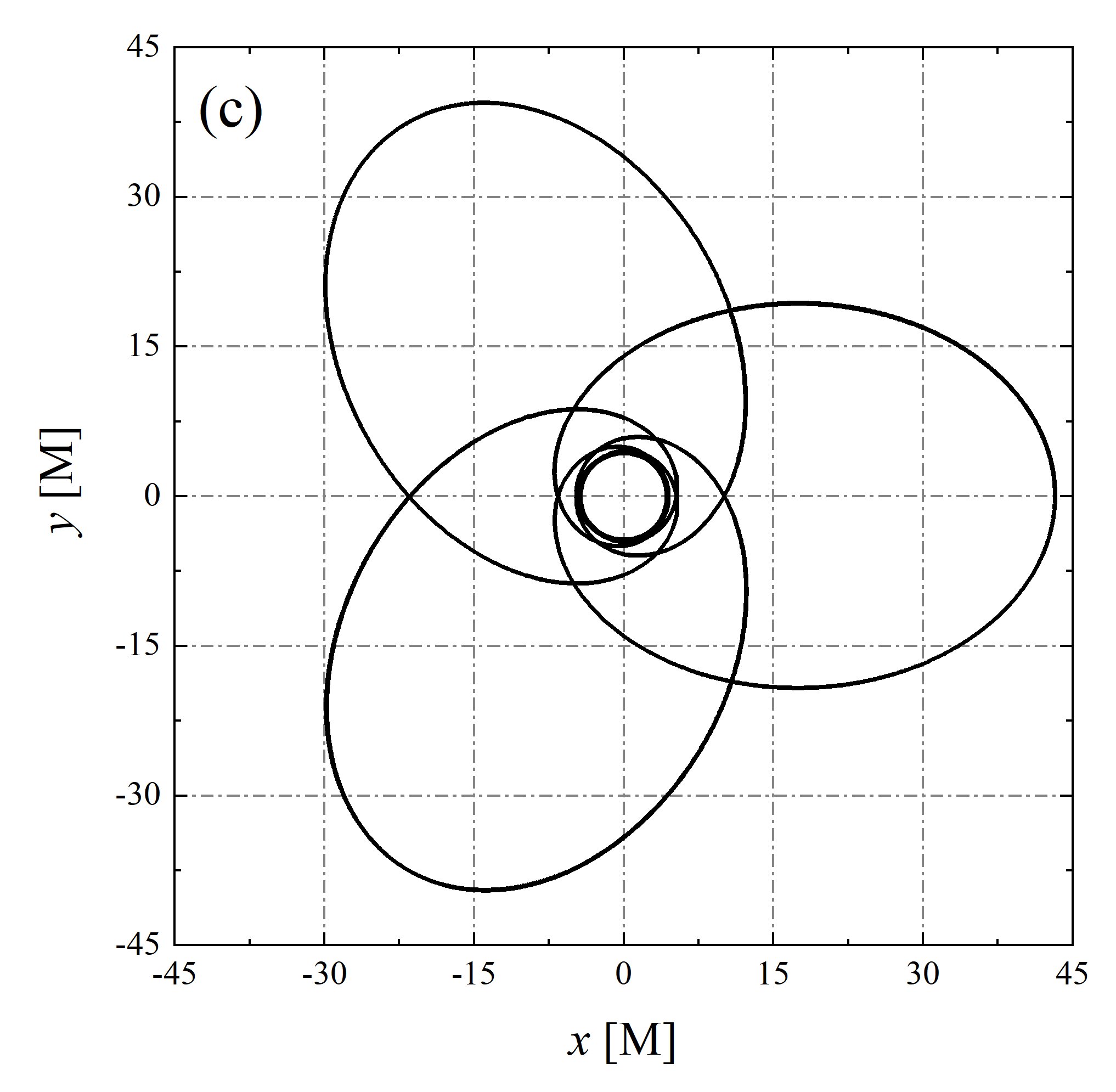}
\includegraphics[width=3.1cm]{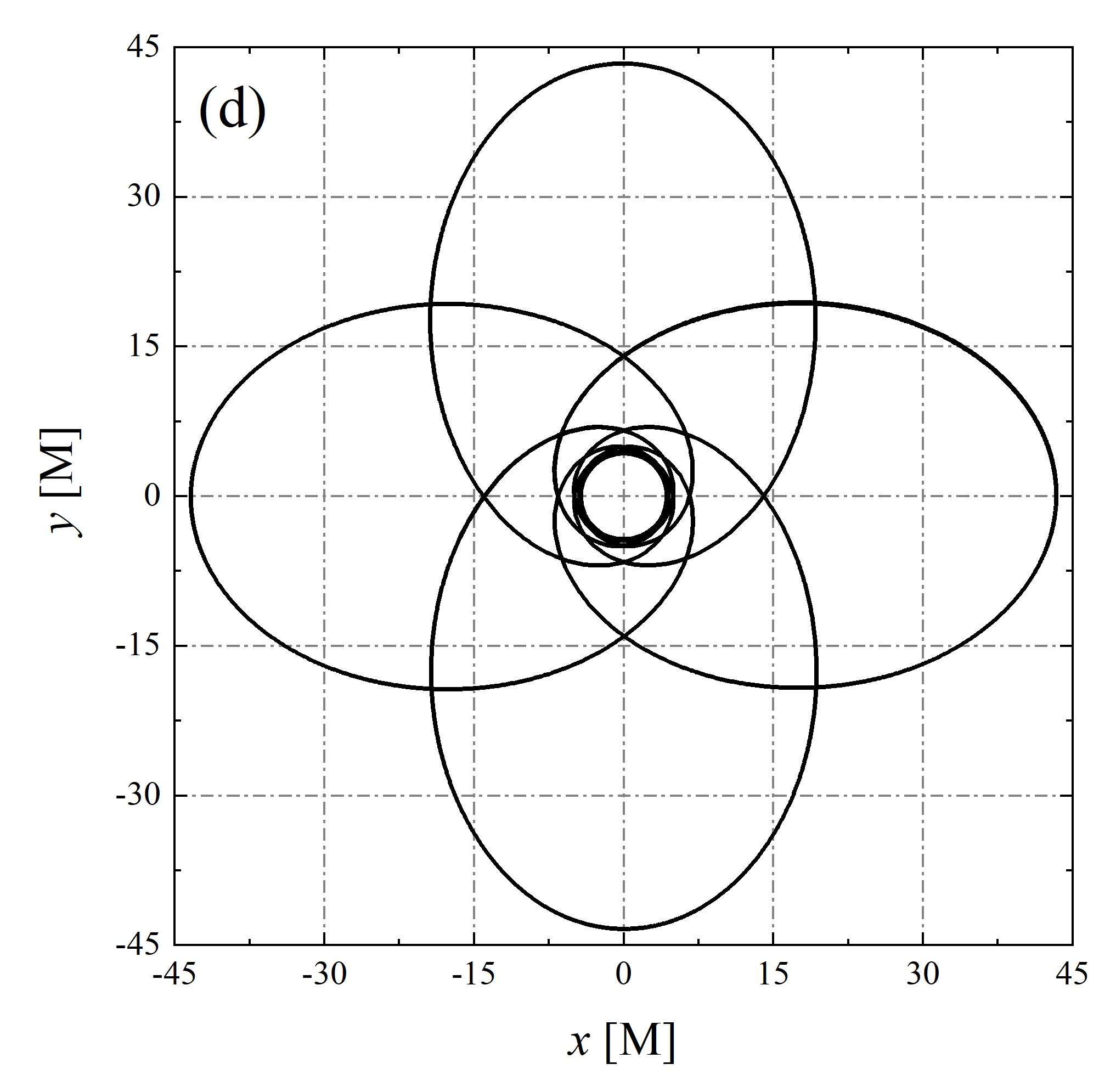}
\includegraphics[width=3.1cm]{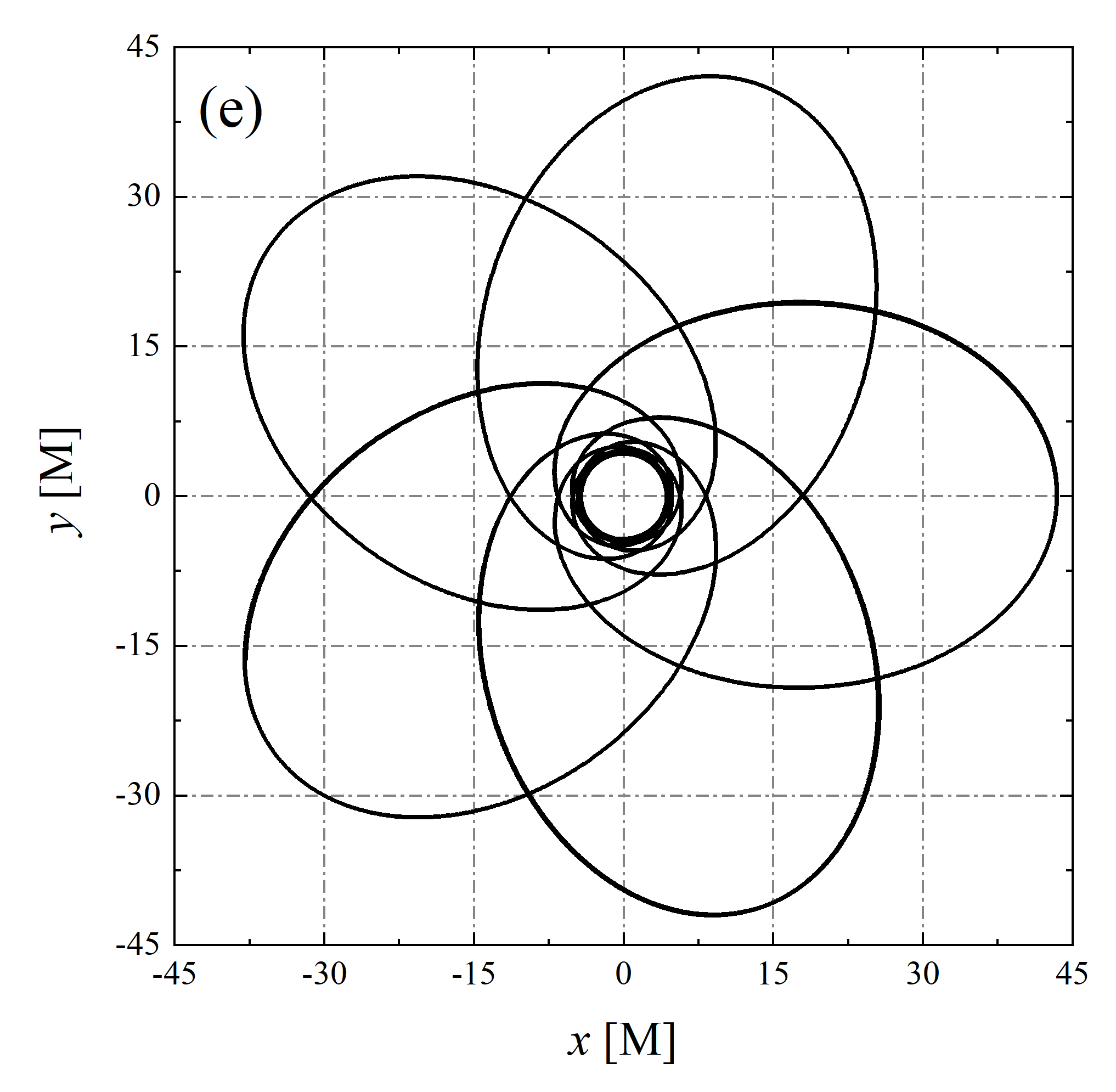}
\includegraphics[width=3.1cm]{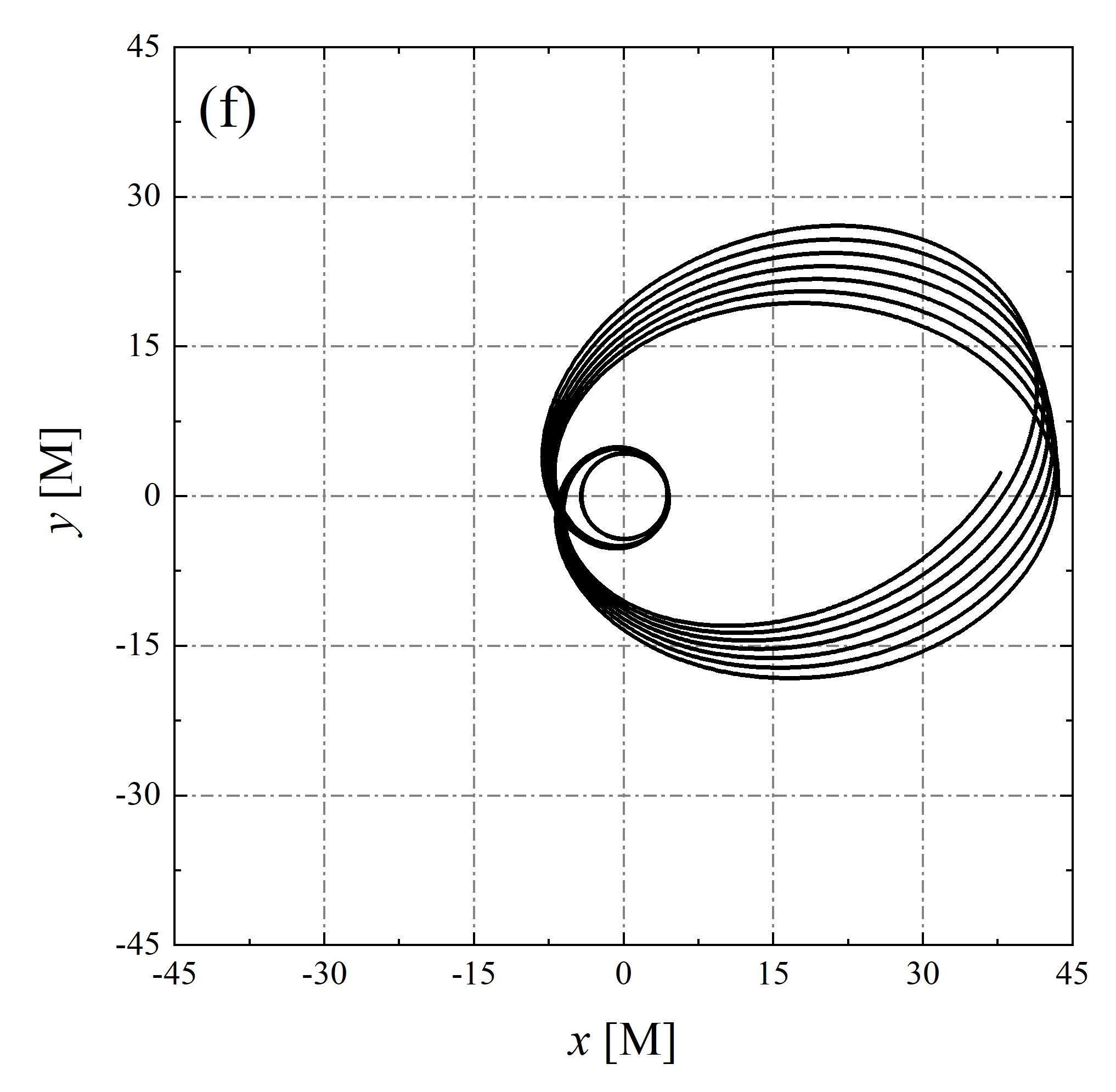}
\includegraphics[width=3.1cm]{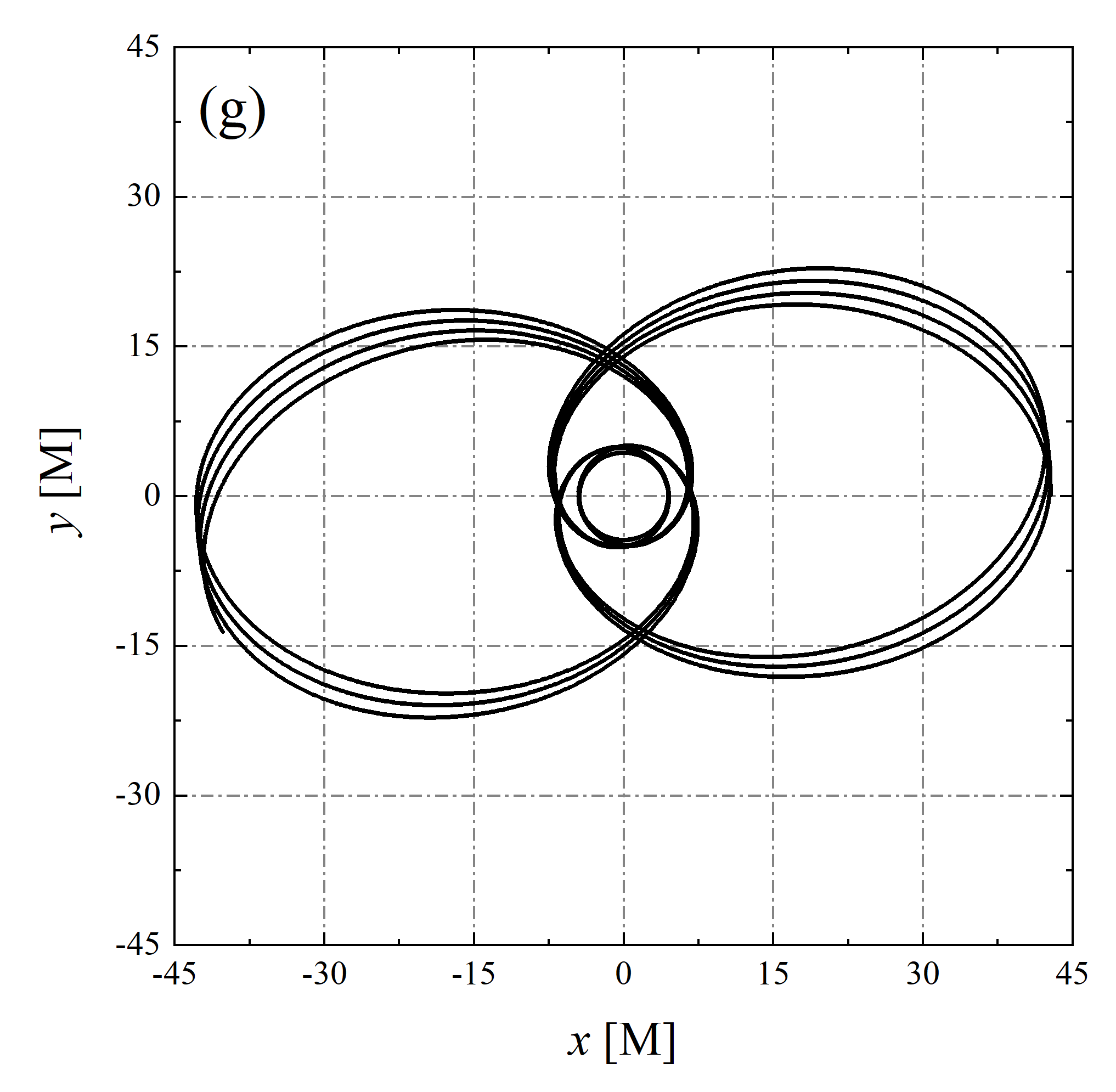}
\includegraphics[width=3.1cm]{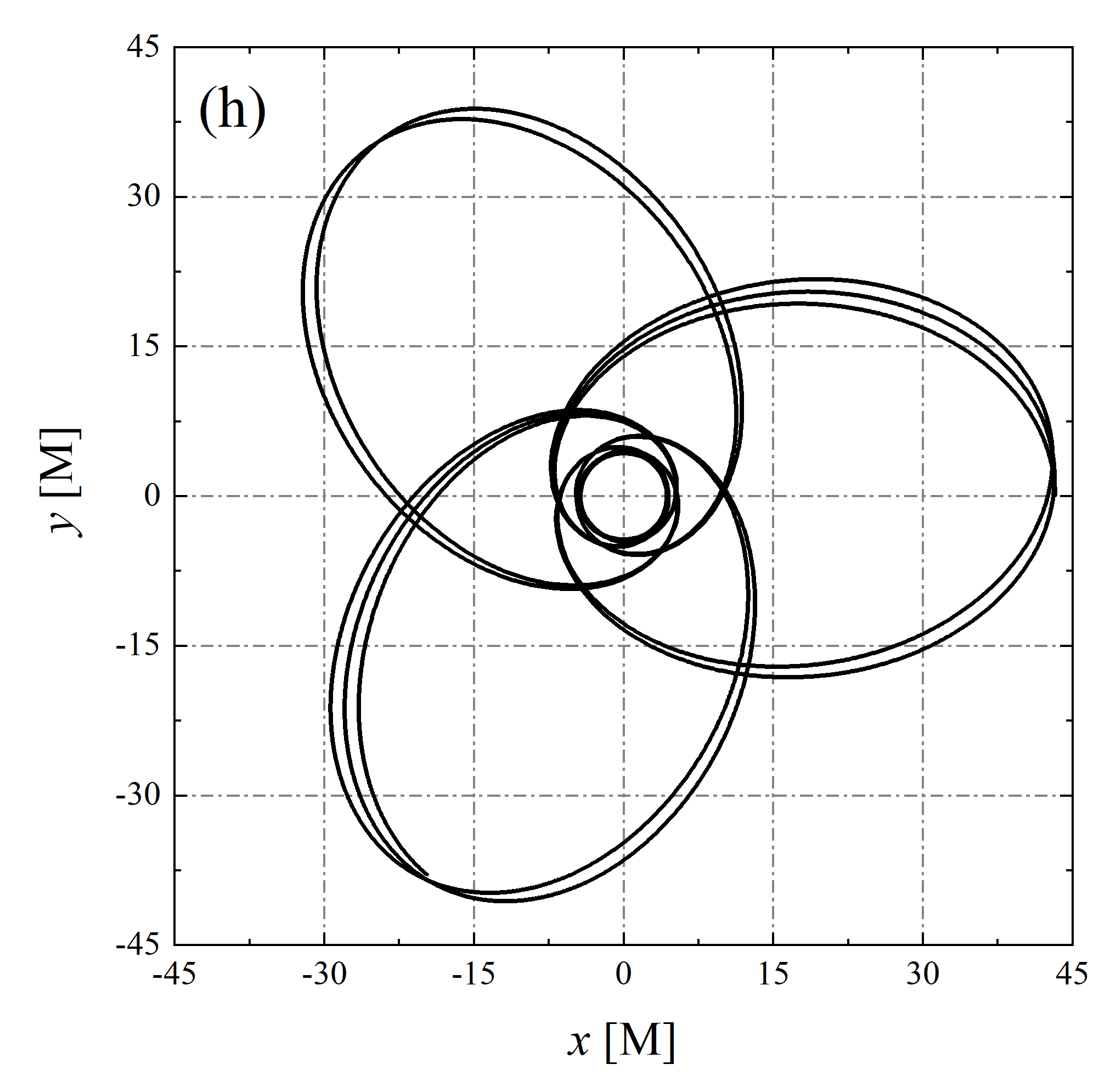}
\includegraphics[width=3.1cm]{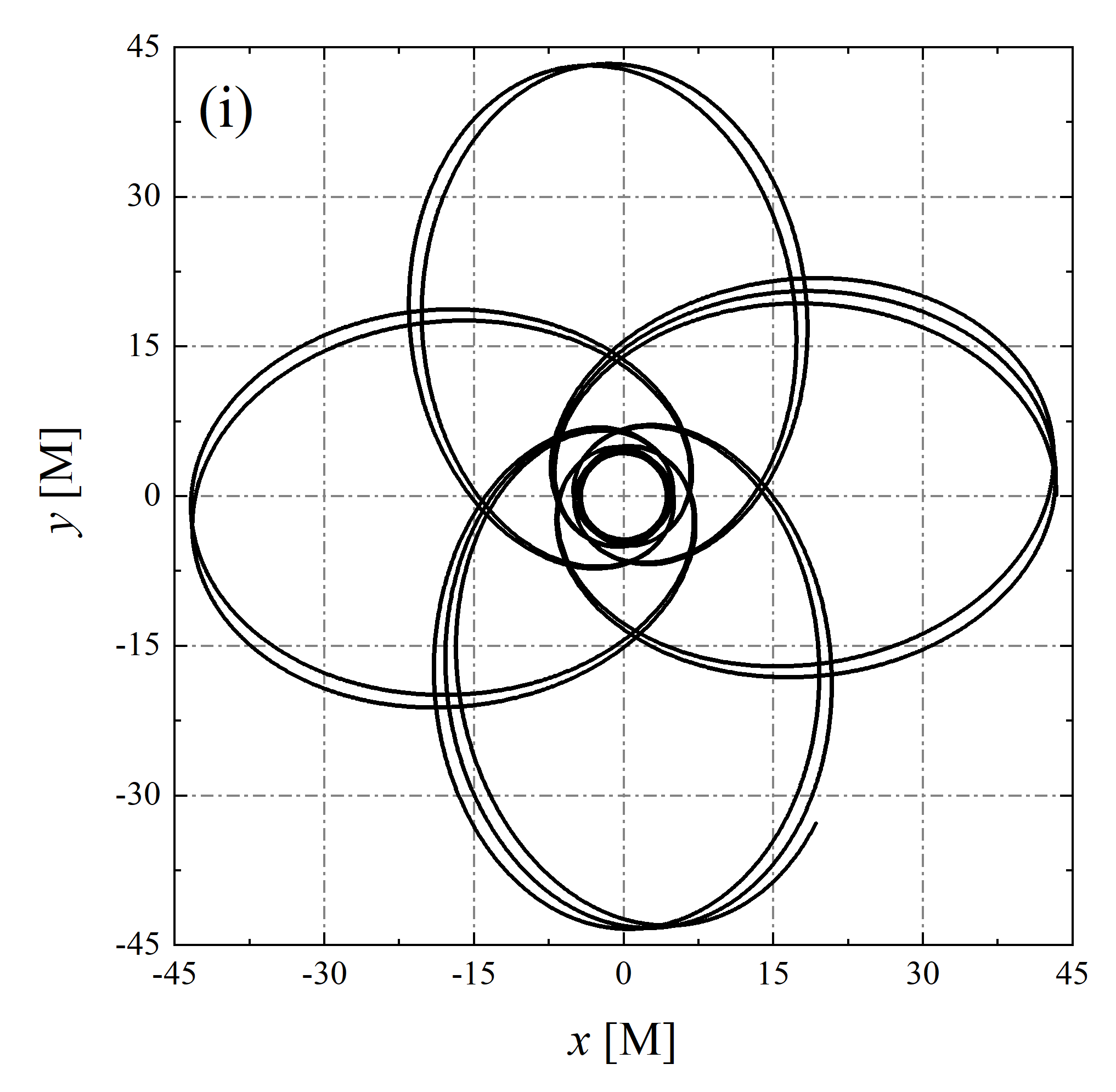}
\includegraphics[width=3.1cm]{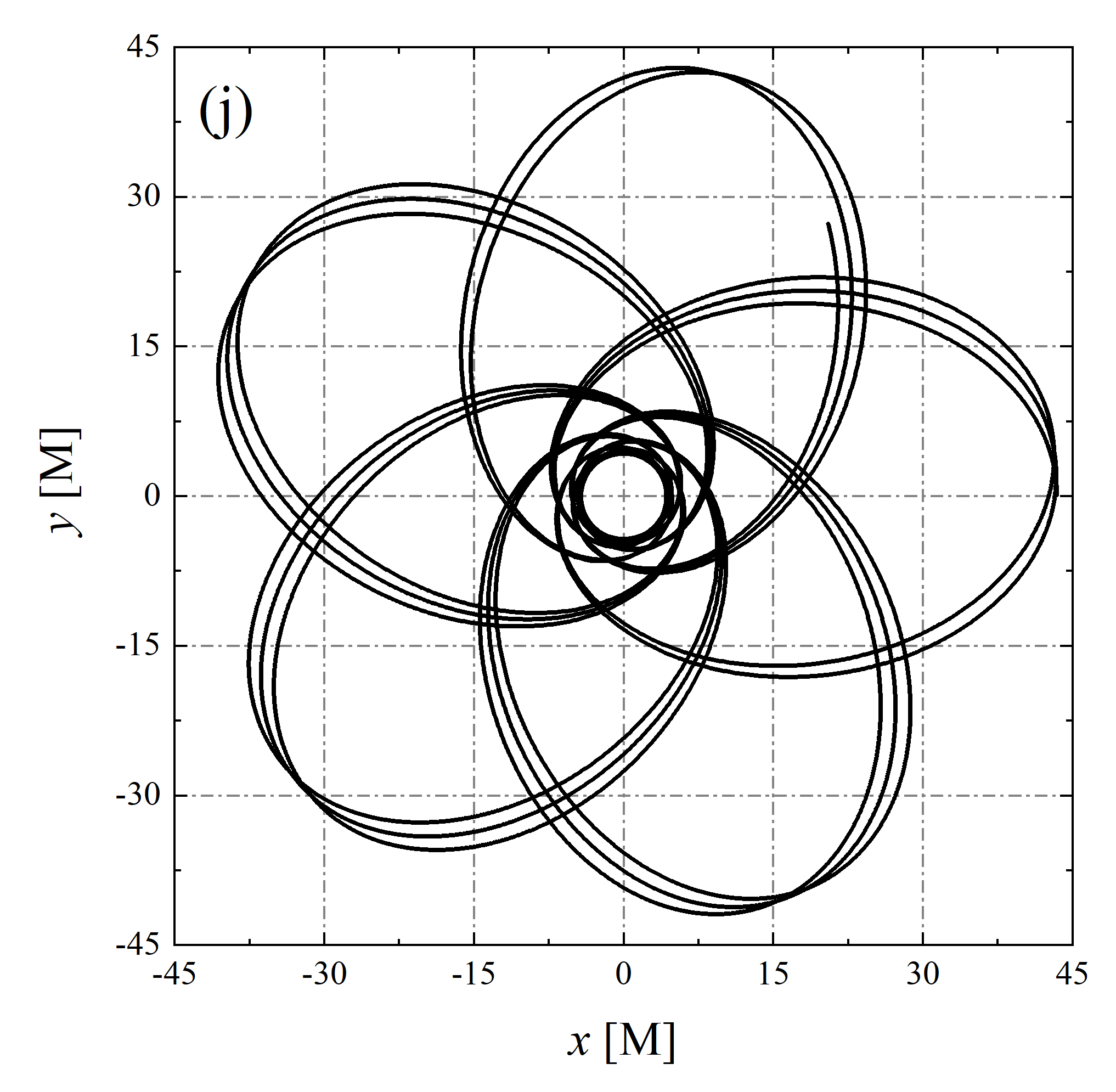}
\includegraphics[width=3.1cm]{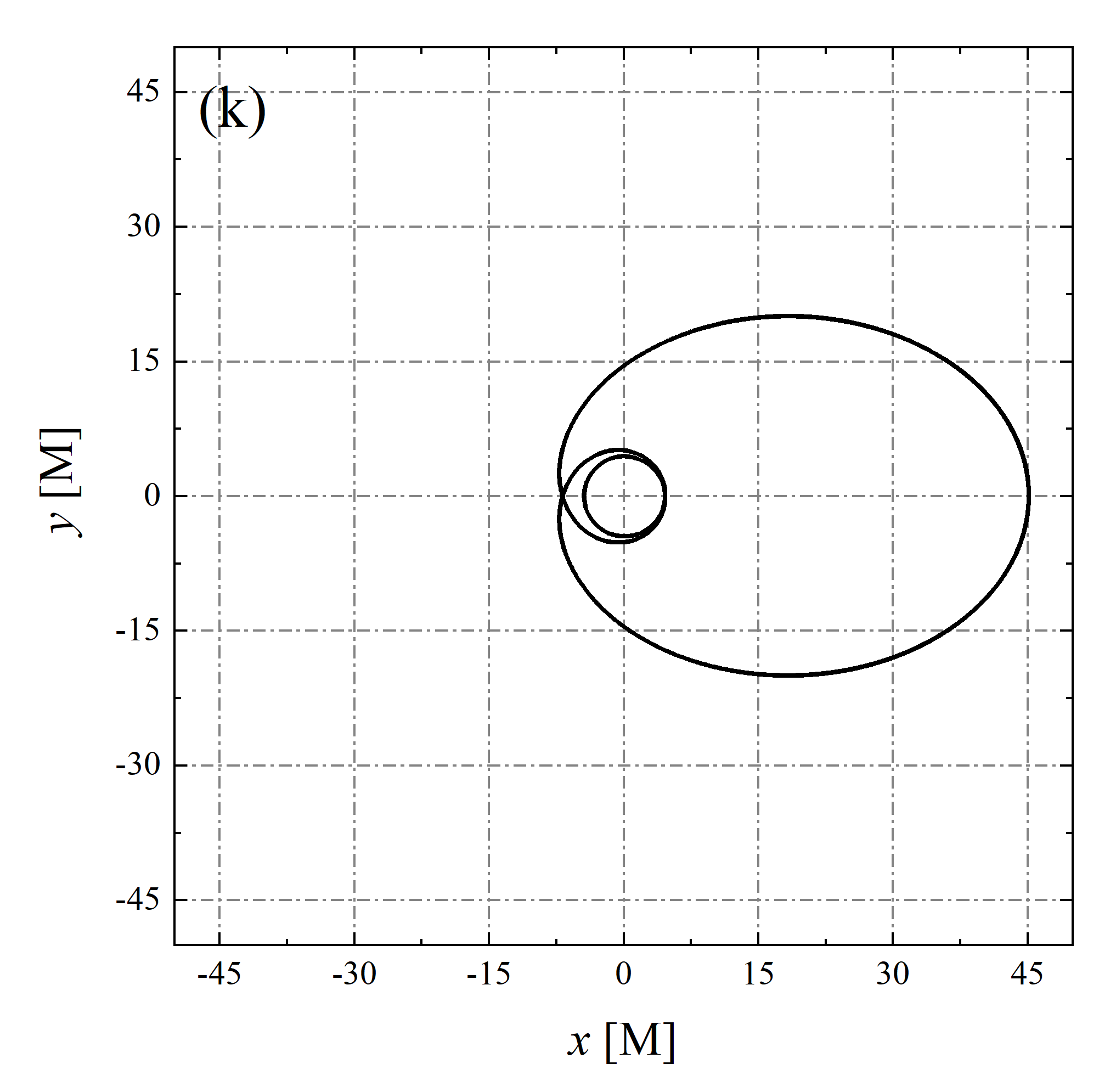}
\includegraphics[width=3.1cm]{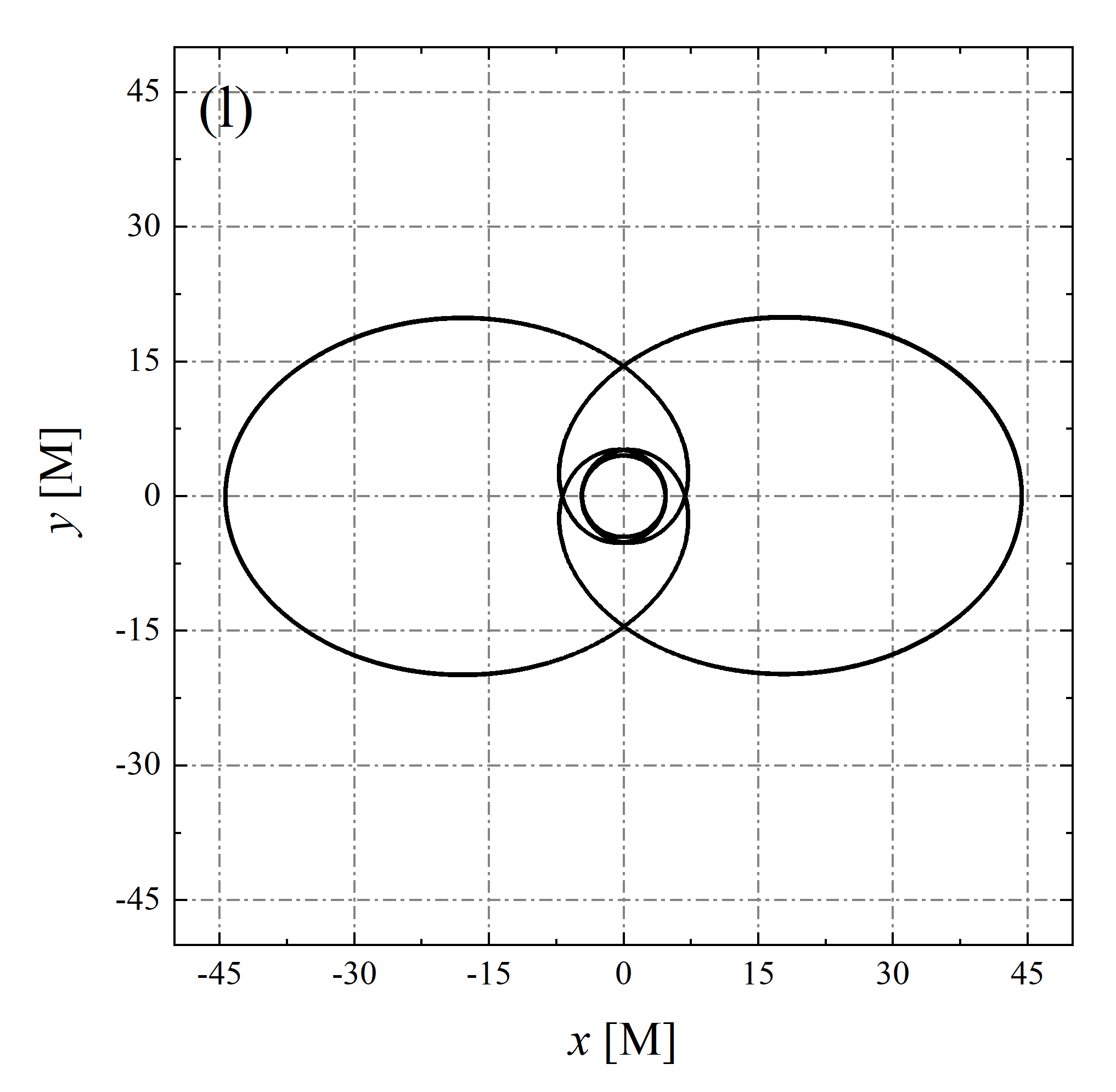}
\includegraphics[width=3.1cm]{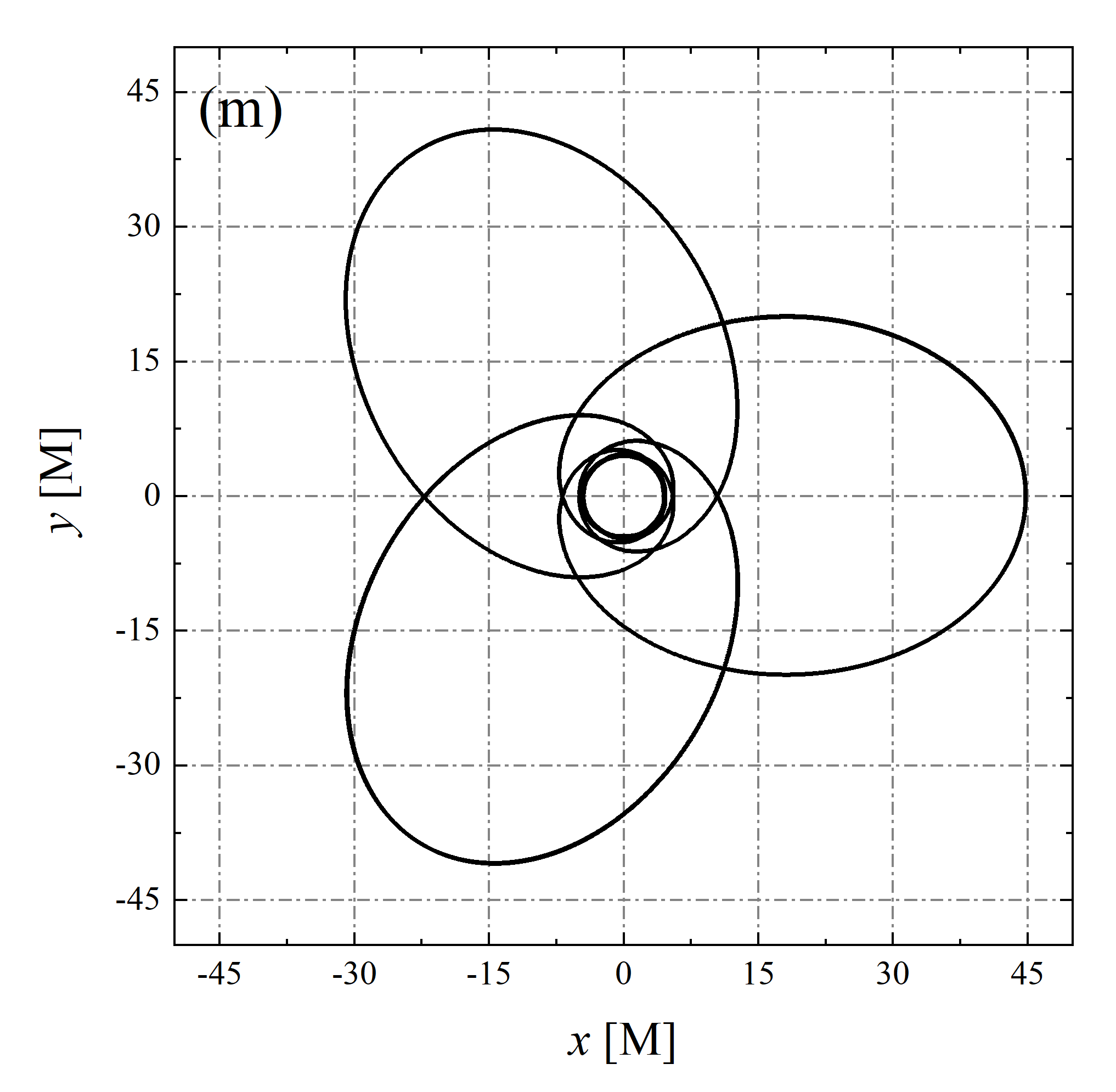}
\includegraphics[width=3.1cm]{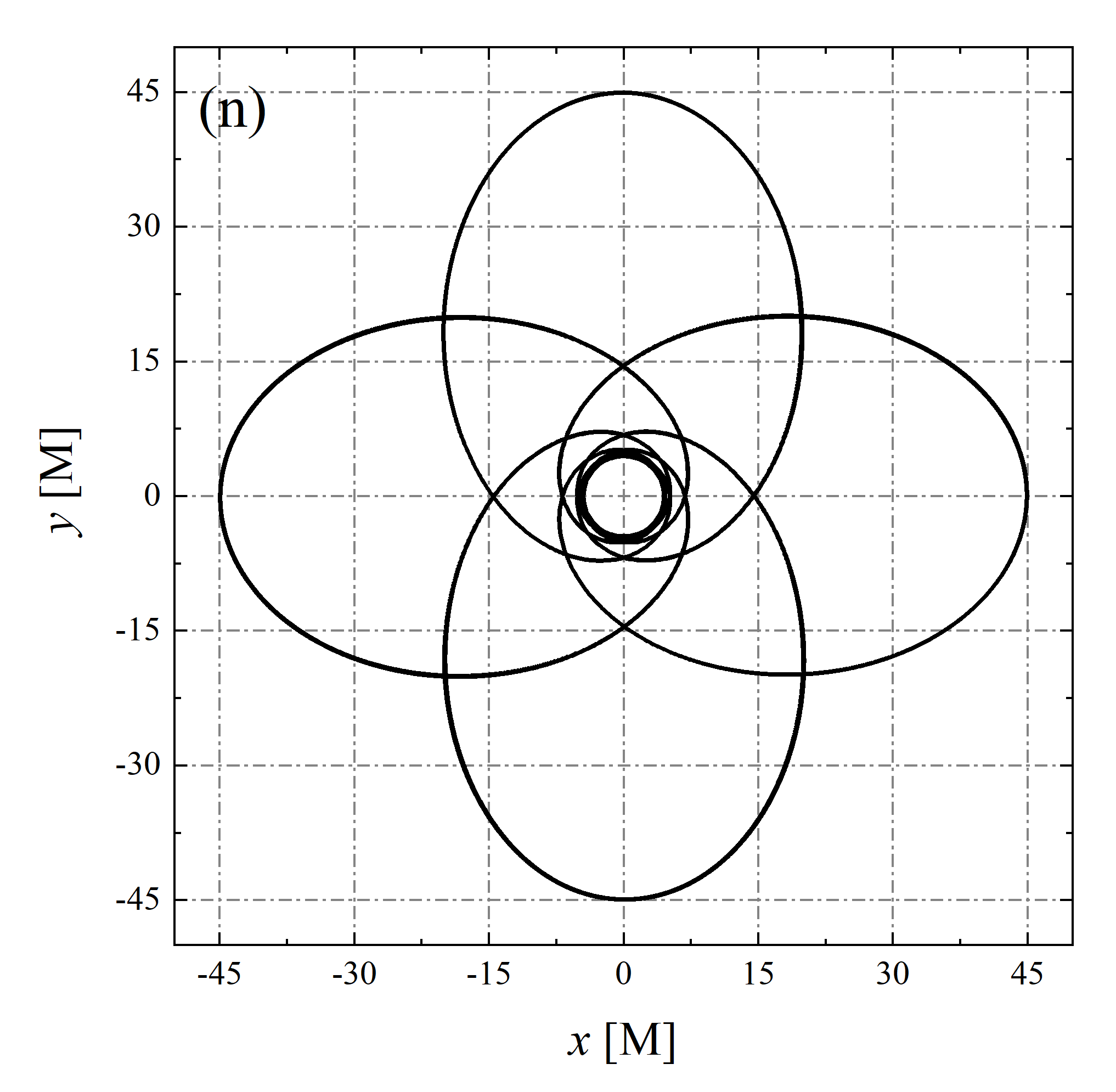}
\includegraphics[width=3.1cm]{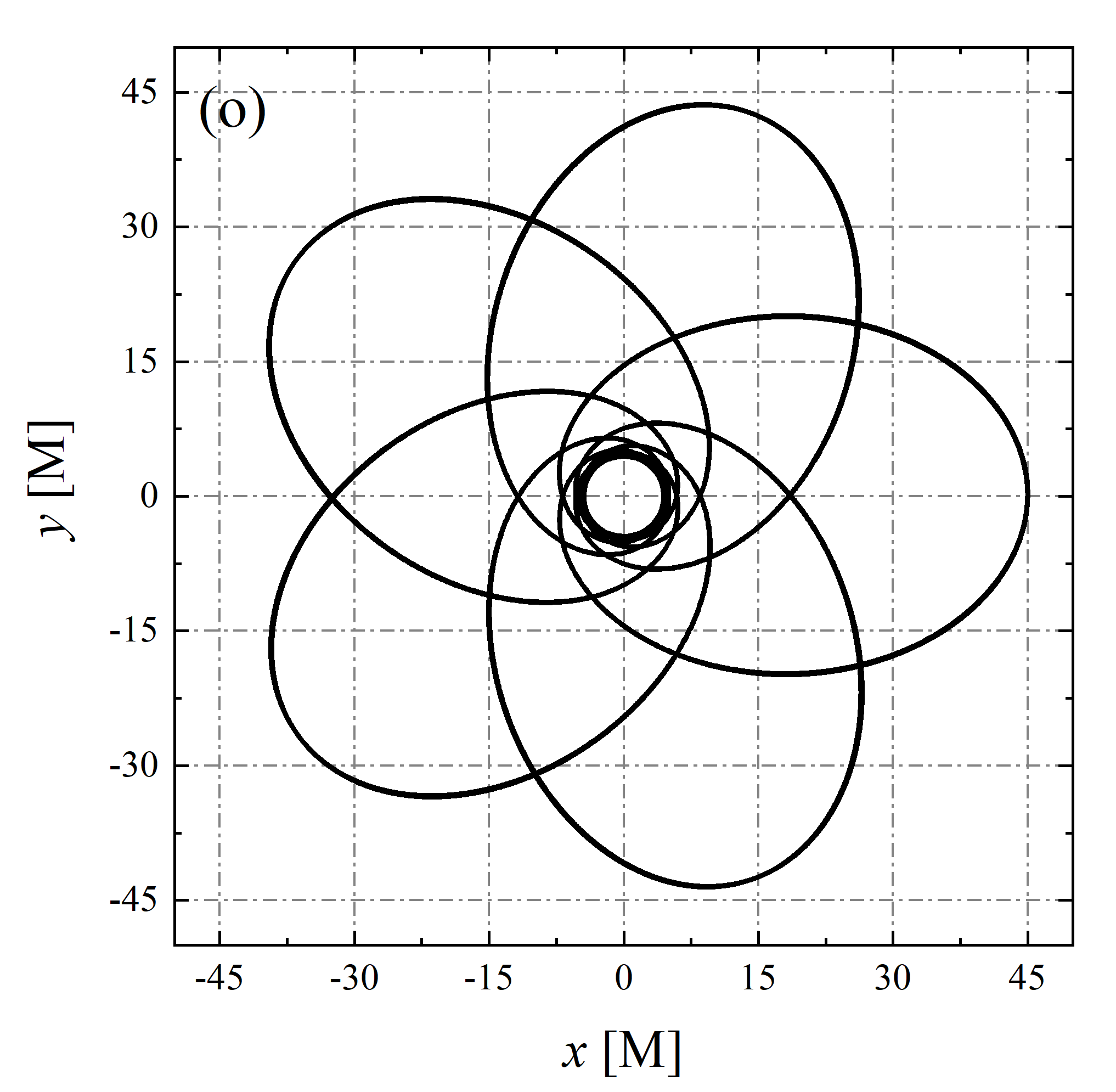}
\includegraphics[width=3.1cm]{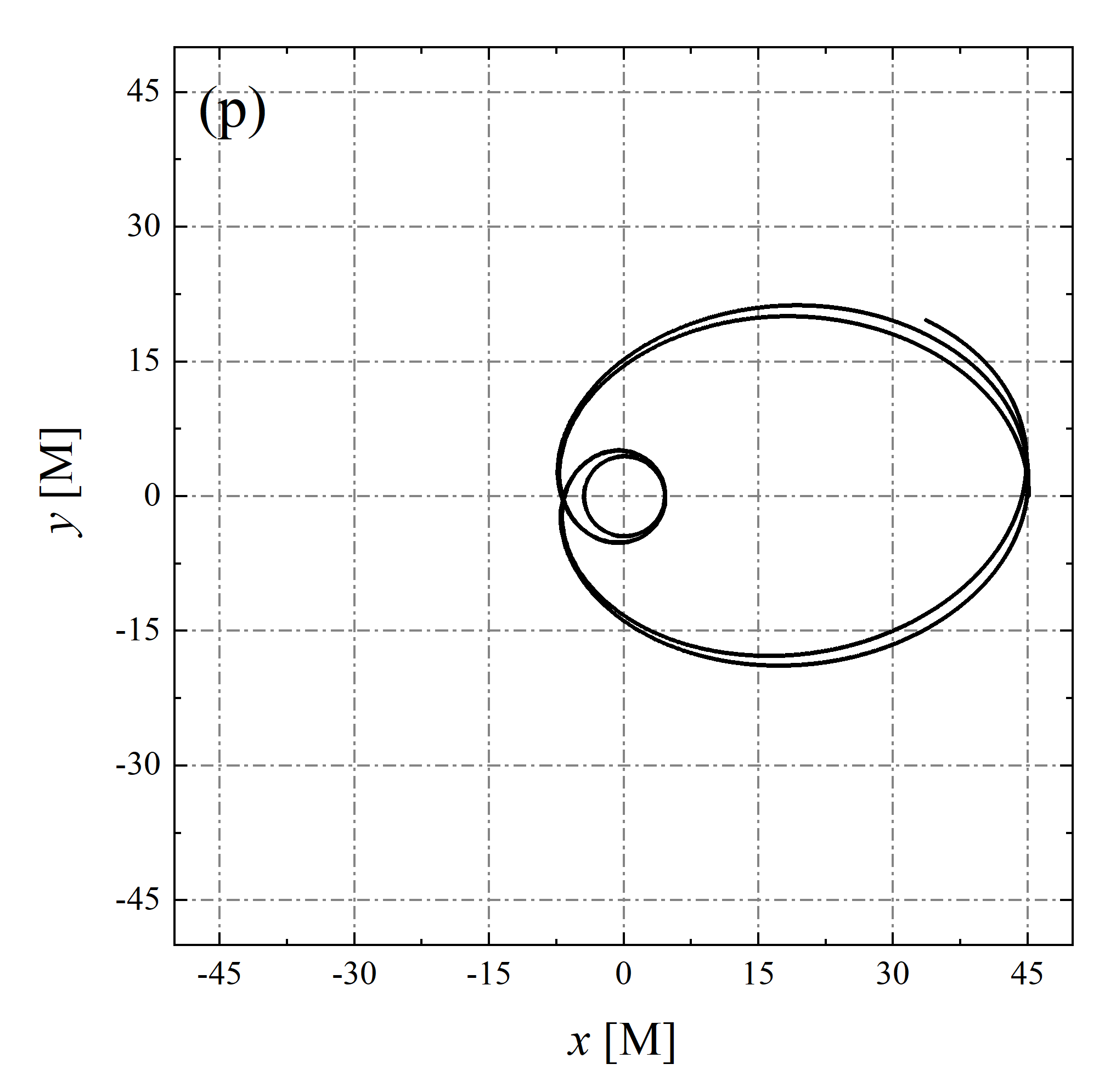}
\includegraphics[width=3.1cm]{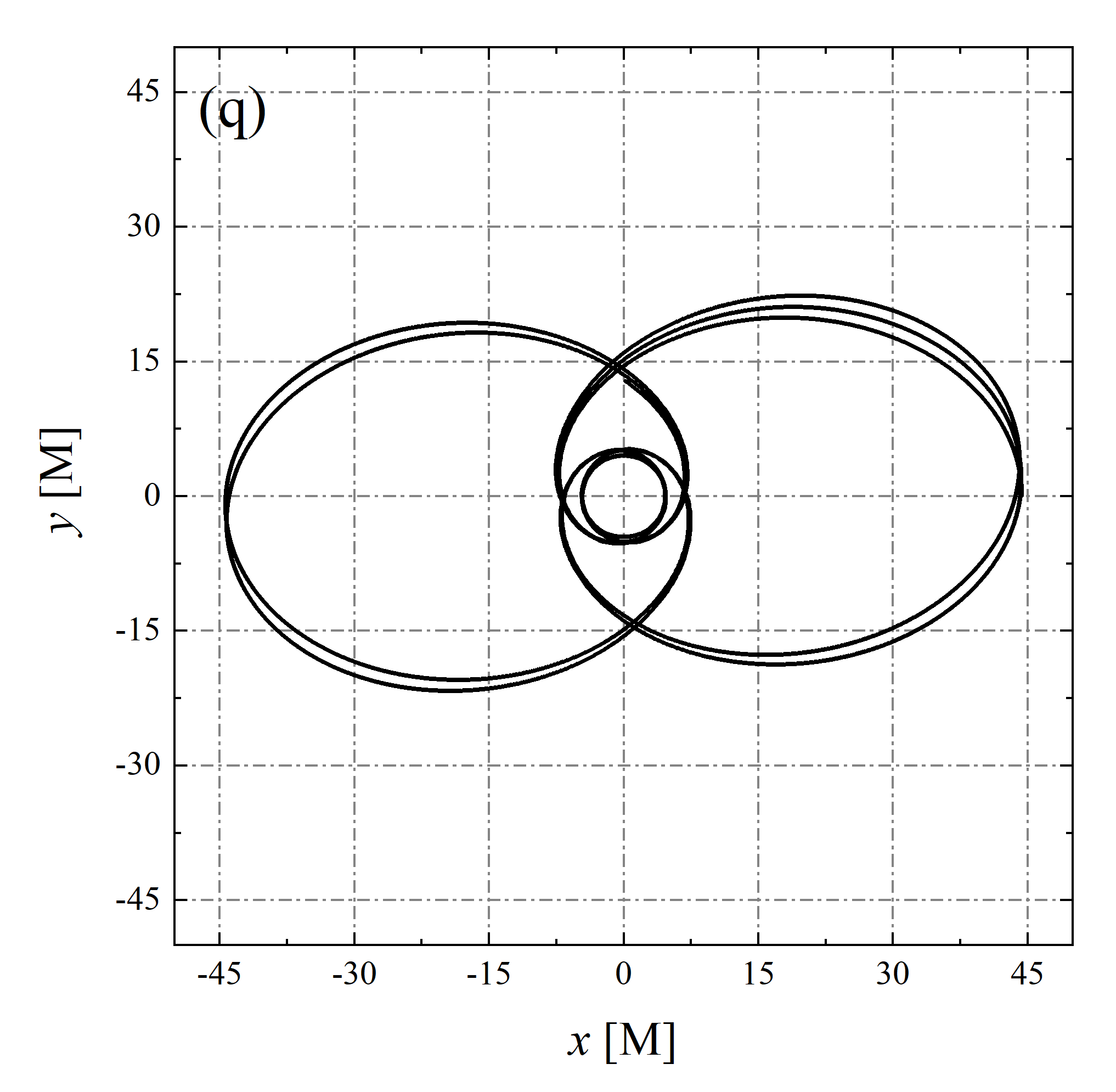}
\includegraphics[width=3.1cm]{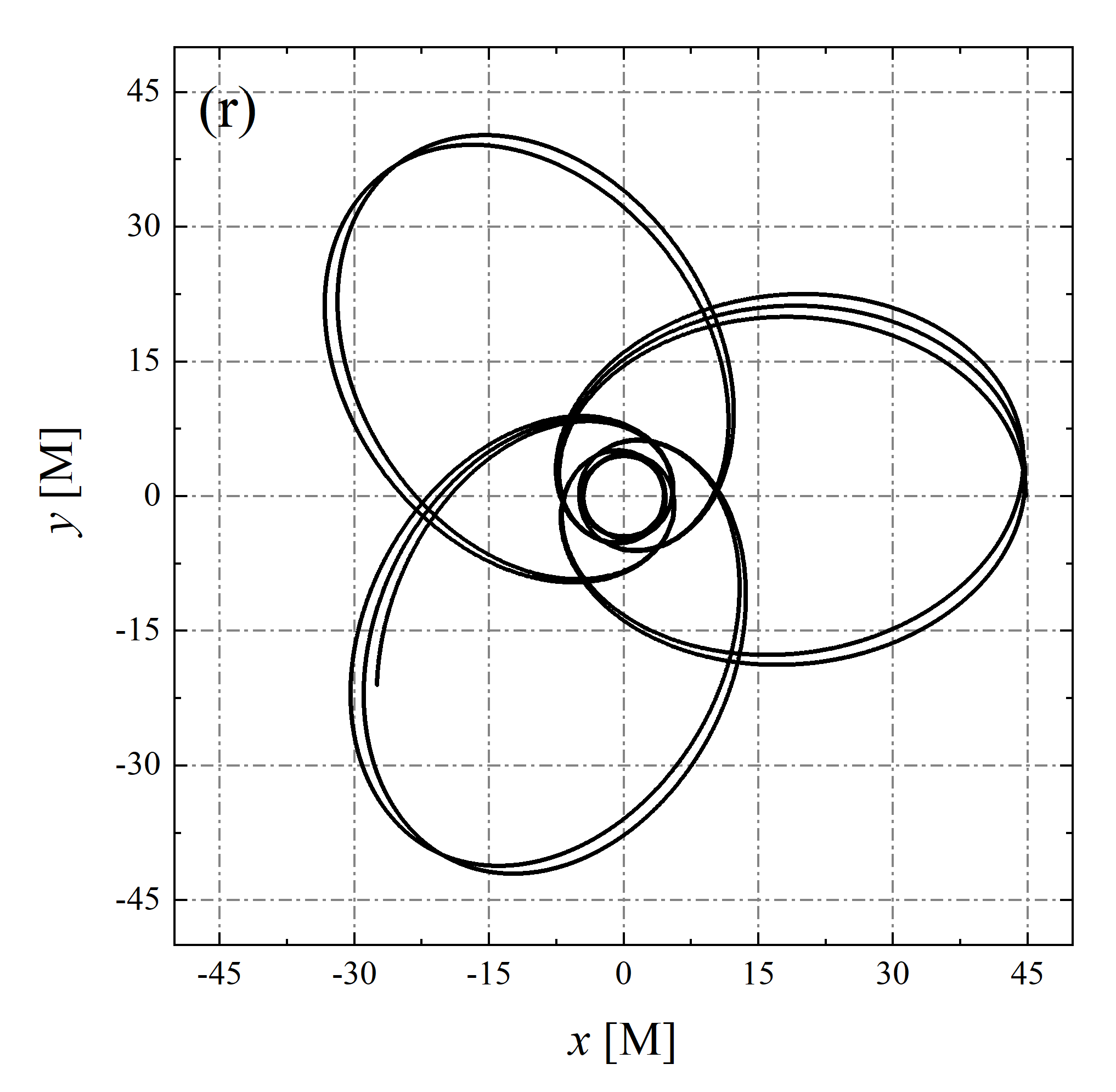}
\includegraphics[width=3.1cm]{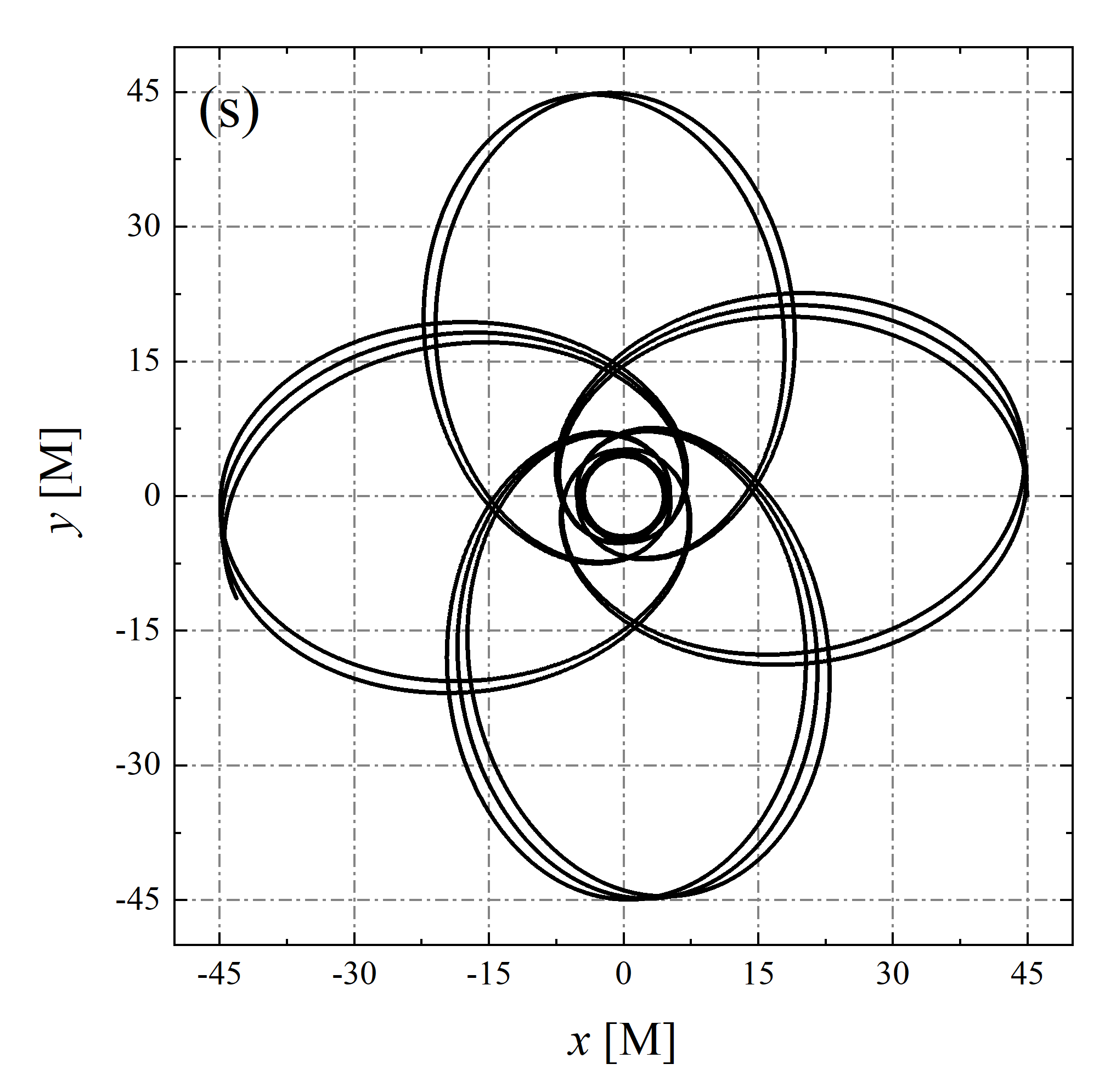}
\includegraphics[width=3.1cm]{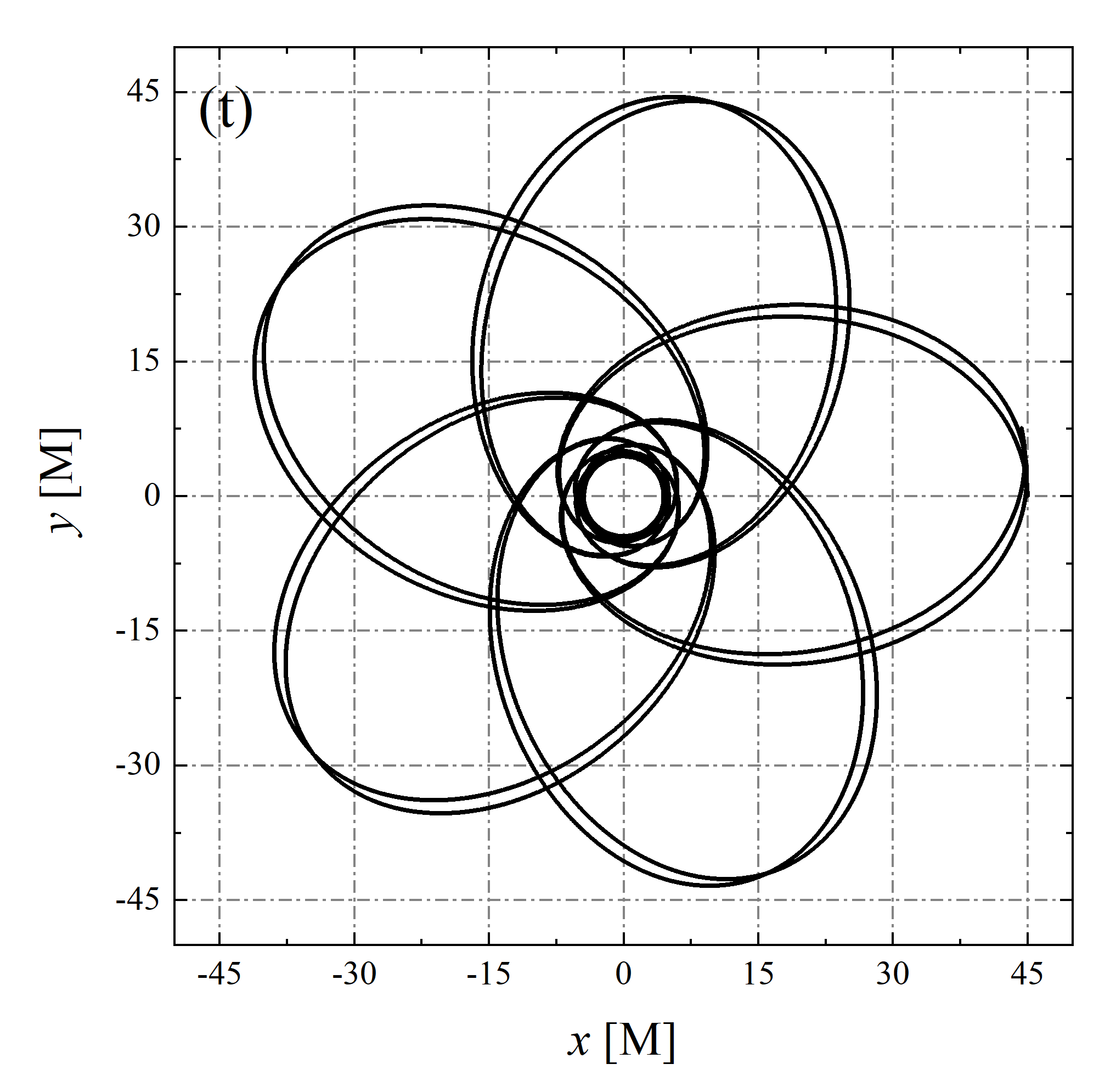}
\caption{Strictly closed orbits and their non-closed counterparts under various dark matter halo parameters and rational numbers $q$. In the first two rows, the halo parameters are $(r_{\textrm{s}},\rho_{\textrm{s}})=(0.2,0.1)$; in the last two rows, the scale parameter is increased to $r_{\textrm{s}}=0.5$. The second and fourth rows represent non-closed counterparts obtained by adding a small perturbation $\delta=1/(100z)$ to the rational numbers of the first and third rows, respectively. the leaf parameter $z$ varies from $1$ to $5$. The angular momentum parameter is fixed at $\varepsilon=0.7$.}}\label{fig7}
\end{figure*}
We fixed $\varepsilon=0.7$ and simulated quasi-periodic orbits corresponding to various dark matter halo parameters and rational numbers $q$, as illustrated in Fig. 7. Within this figure, the second and fourth rows display non-closed orbits obtained by applying a small perturbation to the rational numbers of the first and third rows, respectively. Specifically, the configuration parameters for panels (a) and (k) are $(z,w,v)=(1,1,1)$, corresponding to $q=2$. For panels (b)--(e) and (l)--(o), the configuration parameters are $(z,1,z-1)$, where $z$ ranges from $2$ to $5$ (from left to right), corresponding to $q= (2z-1)/z$. It is evident that the number of orbital leaves (petals) perfectly coincides with the value of $z$. If we follow a fixed sequence and denote the segment of an orbit from one apastron to the subsequent periastron as $\Psi$, then two such $\Psi$ segments constitute a single leaf. By introducing a small increment $\delta=1/(100z)$ to the rational number---for instance, changing $q=1+1/1$ in panel (a) to $q=1+101/100$ in panel (f)---the proportionality between the azimuthal and radial oscillations is broken. Consequently, the orbits fail to close and exhibit counter-clockwise precession. As expected, after sufficient time evolution, the envelope of such non-closed precessing orbits forms a circular boundary. We also observe that a larger value of $z$ generally corresponds to a smaller degree of precession.

Furthermore, by increasing the dark matter halo scale parameter from $r_{\textrm{s}}=0.2$ (first row) to $r_{\textrm{s}}=0.5$, the resulting orbital evolution is presented in the third row. Although the orbits remain closed and their configurations are consistent with those in the first row, a significant orbital amplification is captured. For example, the apastron radius of the single-leaf orbit in panel (a) is clearly less than $r=45$, whereas the corresponding value in panel (k) is nearly equal to $r=45$. This is because increasing the dark matter halo size enhances the gravitational field intensity, requiring stable orbits to shift further from the black hole to persist. At the same time, because the modification of the spacetime by the dark matter halo exhibits scale symmetry, the commensurability ratio between the azimuthal and radial oscillations is strictly conserved. As a result, the overall orbital configuration is independent of the increase in the dark matter halo parameters. Importantly, these results suggest that the orbital structure is primarily determined by the rational number $q$ rather than the dark matter halo parameters.

\begin{figure*}
\center{
\includegraphics[width=5cm]{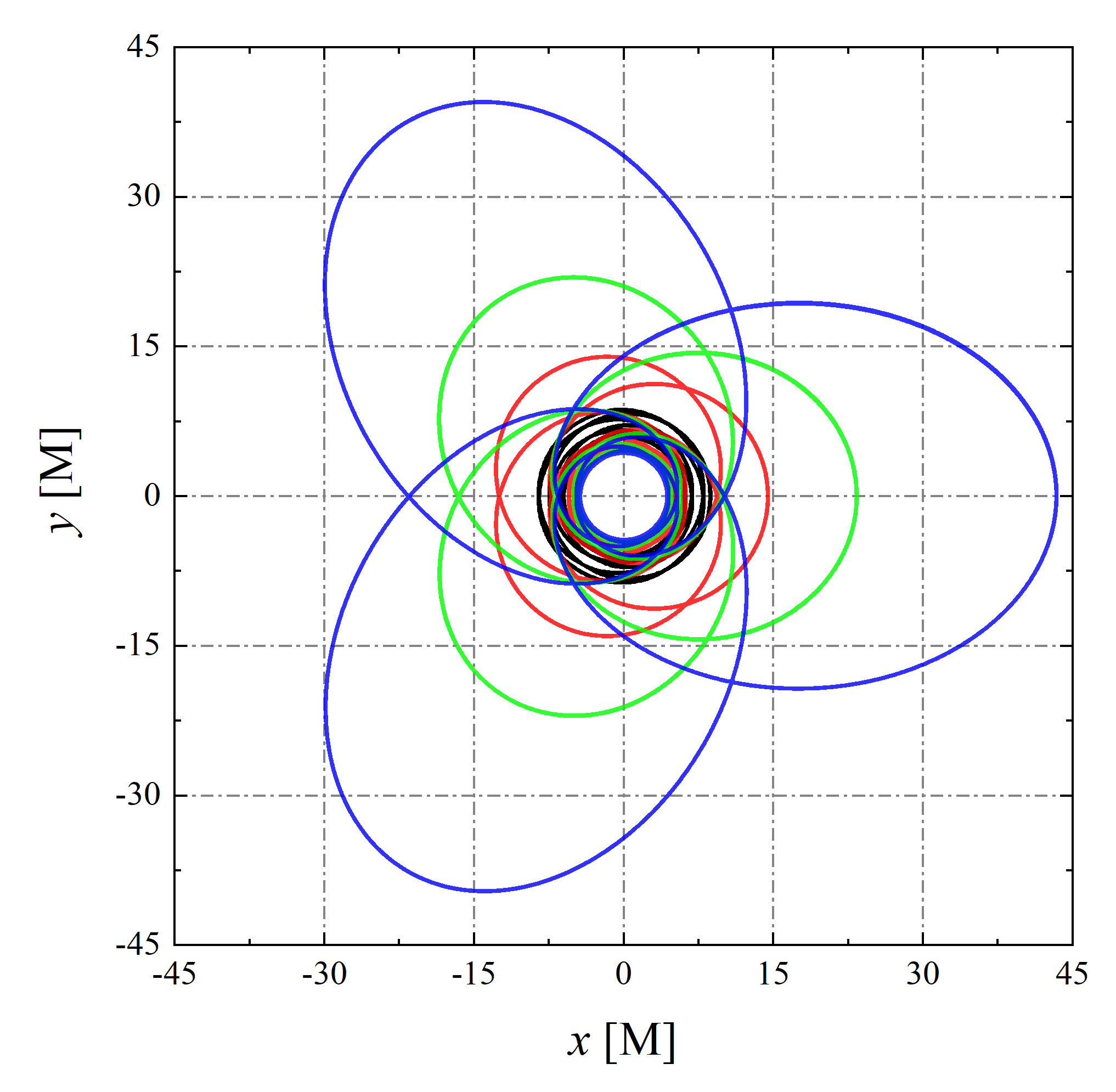}
\caption{Influence of the parameter $\varepsilon$ on the orbital configuration. From the innermost to the outermost curves, $\varepsilon$ takes the values $0.1$, $0.3$, $0.5$, and $0.7$, respectively. Here, the dark matter halo parameters are fixed at $(r_{\textrm{s}},\rho_{\textrm{s}})=(0.2,0.2)$, and the orbital configuration is set to $(z,w,v)=(3,1,2)$. It is observed that increasing $\varepsilon$ expands the orbital scale while preserving the original configuration.}}\label{fig8}
\end{figure*}
Similar to the effects of dark matter halo parameters, the orbital angular momentum also plays a critical role in expanding or contracting the orbital scale. As illustrated in Fig. 8, under the conditions of a fixed configuration $(z,w,v)=(3,1,2)$ and constant dark matter halo parameters, an increase in $\varepsilon$ causes the orbit to evolve from black to red, then to green, and finally to blue. Throughout this process, the amplification of the orbit is clearly visible while maintaining the three-leaf configuration, and the leaf-like characteristics become increasingly distinct. This stems from the fact that an increase in angular momentum expands the bounded region of the orbit, leading to a more distant apastron and larger leaves. Notably, when the angular momentum is close to $L_{\textrm{isco}}$ ($\varepsilon=0.1$), the three-leaf features of the black orbit are difficult to identify. This difficulty arises because a smaller angular momentum results in a narrower gap between the periastron and apastron, leading to a more circularized trajectory where the leaves are less pronounced.

Particle orbits encode crucial spacetime parameters and are closely associated with various high-energy astrophysical phenomena. Therefore, it is highly desirable to explore their potential observational signatures in depth. Interestingly, a timelike particle can be treated as a luminous source that emits continuous electromagnetic radiation as it orbits the central black hole. Simultaneously, the entire dynamical system can be modeled as an extreme mass ratio inspiral (EMRI), which serves as a persistent source of gravitational radiation encoding valuable intrinsic spacetime properties and environmental information \cite{2024JCAP...10..047K,2024JCAP...01..035R,2025arXiv250800516Z,2026arXiv260103374Z}. Consequently, time-domain signals, such as gravitational waves and light curves, offer invaluable opportunities to observationally probe these orbital dynamics. Following this line of reasoning, based on the diverse orbital configurations obtained within the target spacetime, we simulate the corresponding gravitational waves and light curves to uncover their intrinsic observational features.

Following the formalism provided in \cite{2007PhRvD..75b4005B}, the two gravitational wave polarization states, $h_{+}$ and $h_{\times}$, emitted by a timelike particle orbiting in the equatorial plane of a black hole are given by:
\begin{eqnarray}
h_{+} = -\frac{2\eta}{D_{L}}\frac{GM}{c^{2}r}\left(1+\cos^{2}\iota\right)\cos\left(2\phi+2\zeta\right), \label{h1} \\
h_{\times} = -\frac{4\eta}{D_{L}}\frac{GM}{c^{2}r}\cos\iota\sin\left(2\phi+2\zeta\right). \label{h2}
\end{eqnarray}
Here, $\iota$ and $\zeta$ denote the inclination angle and the longitude of the orbit's periastron, respectively. In the subsequent calculations, both of these angular parameters are fixed at $\pi/4$. The parameter $D_{L}$ represents the luminosity distance from the observer to the source. The symmetric mass ratio of the extreme mass ratio inspiral system is denoted by $\eta$, which is defined as $\eta=Mm/(M+m)^{2}$. Since our primary objective is to qualitatively reveal the underlying correlation between the orbital morphology and the corresponding gravitational waveforms, we set both $\eta$ and $D_{L}$ to unity for simplicity. Once the radial coordinate $r$ and the azimuthal angle $\phi$ of the orbit are obtained at any given time, the two gravitational wave polarization states can be directly computed.

\begin{figure*}
\center{
\includegraphics[width=4.2cm]{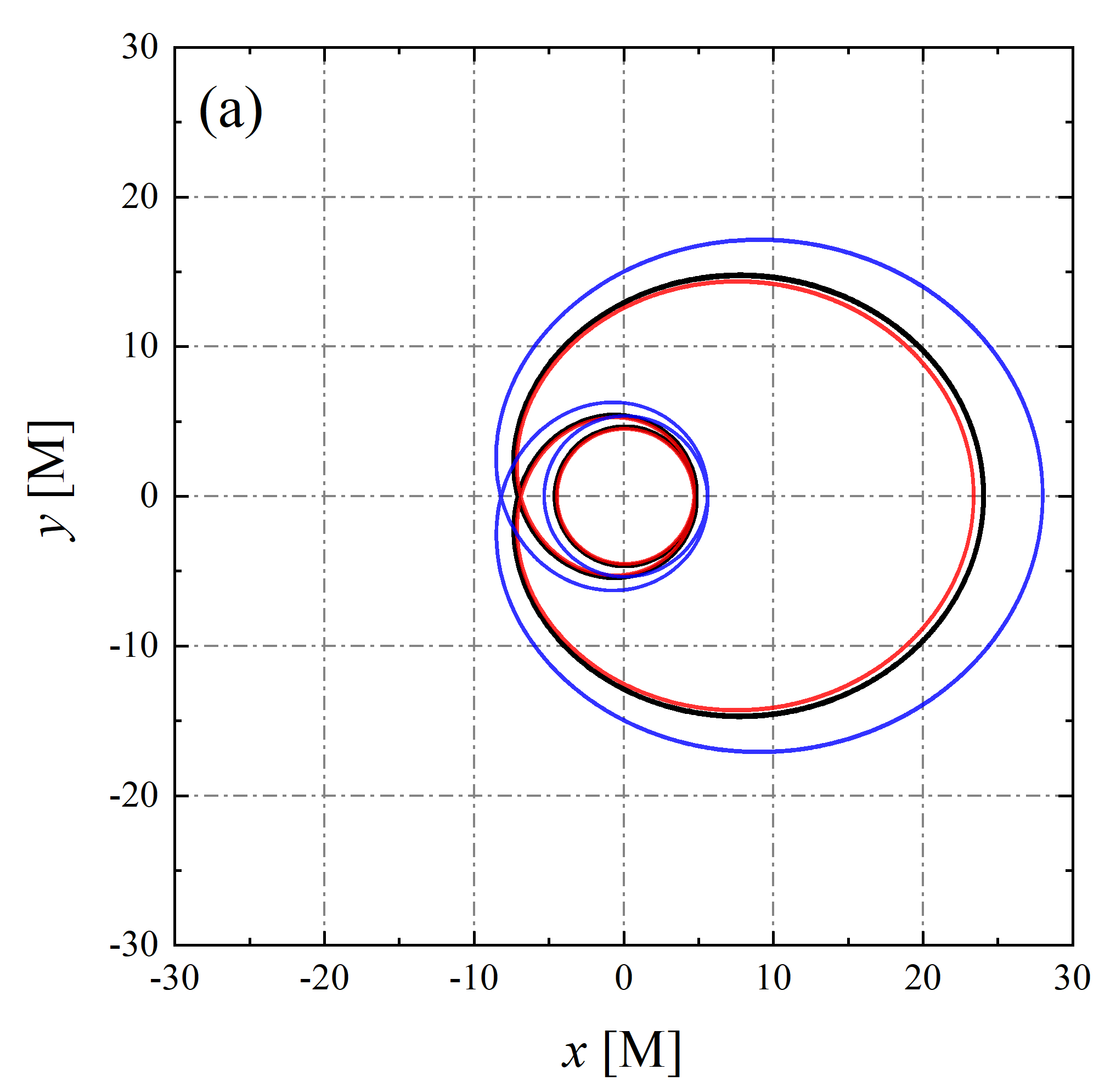}
\includegraphics[width=10.5cm]{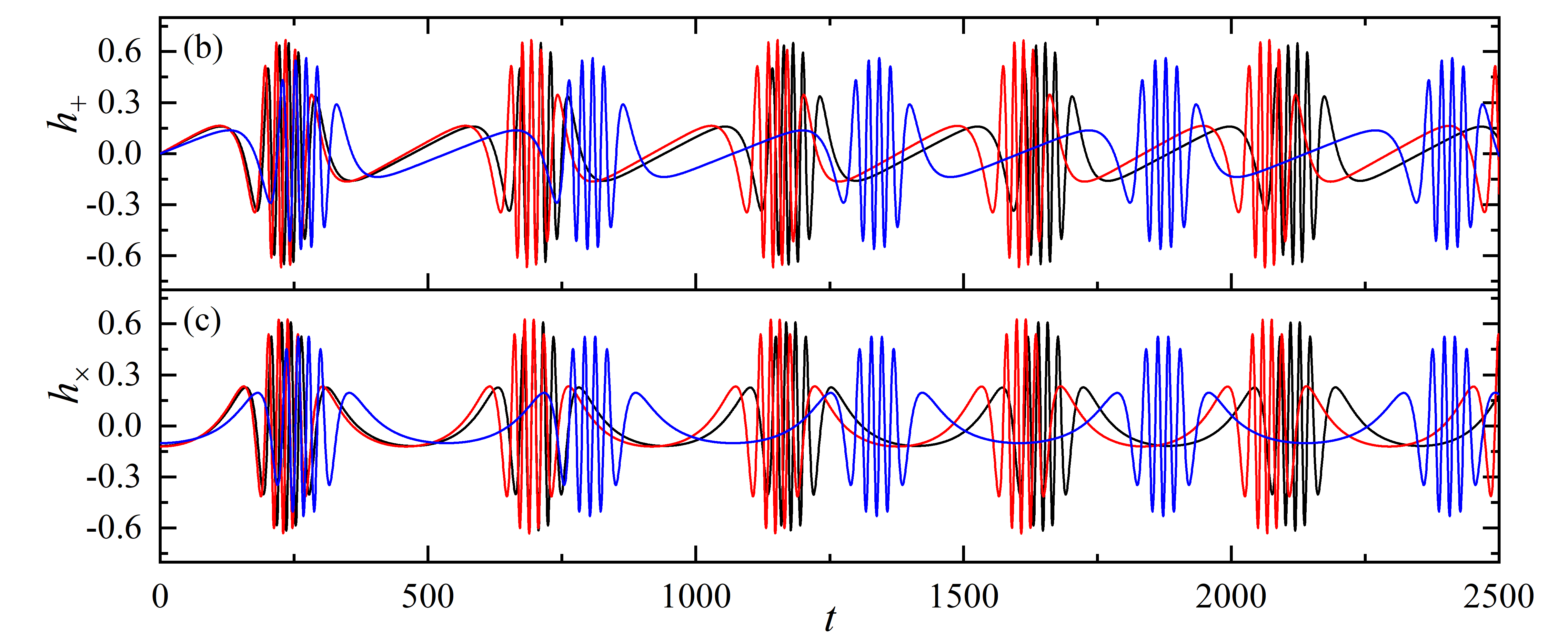}
\includegraphics[width=4.2cm]{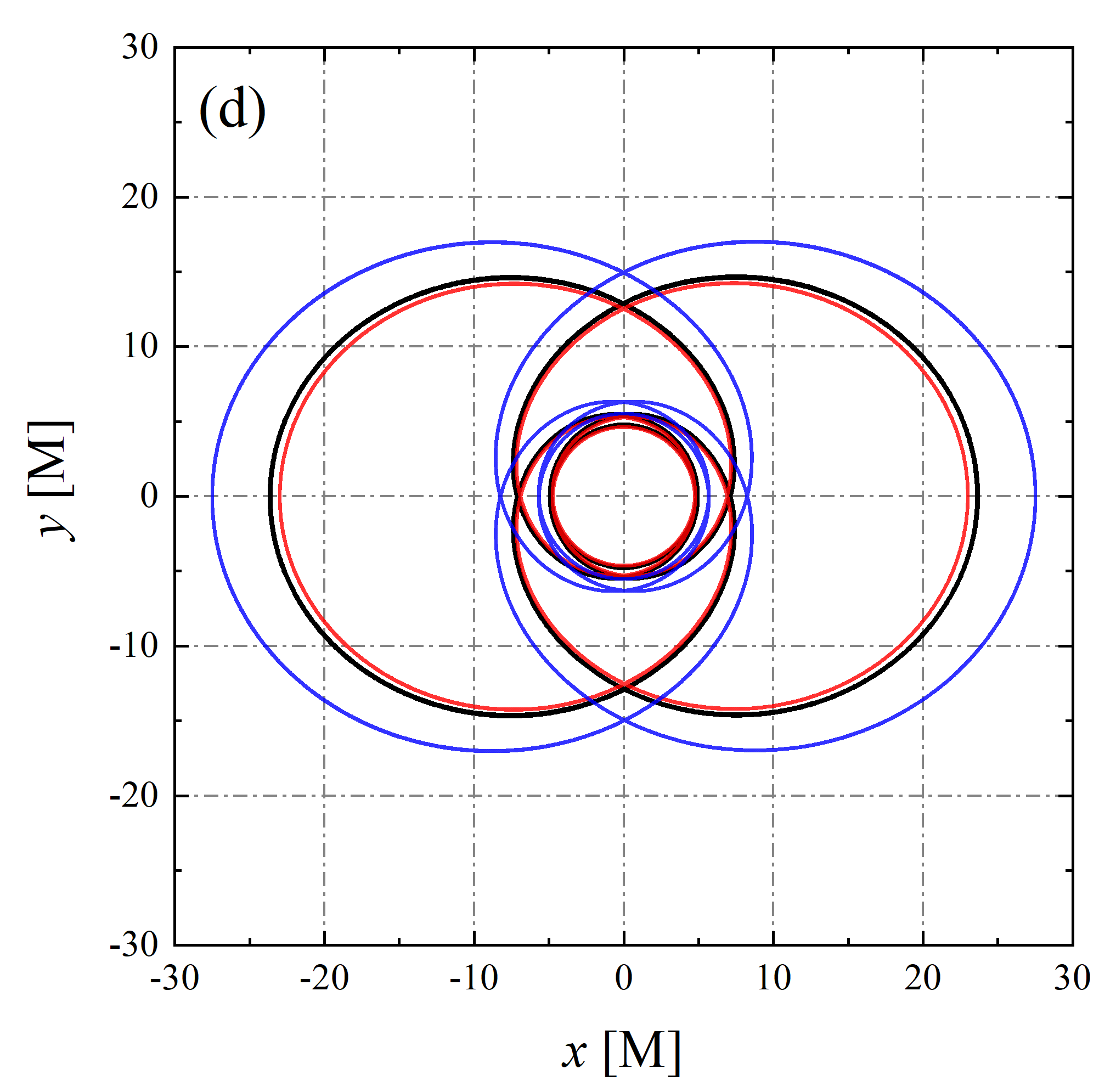}
\includegraphics[width=10.5cm]{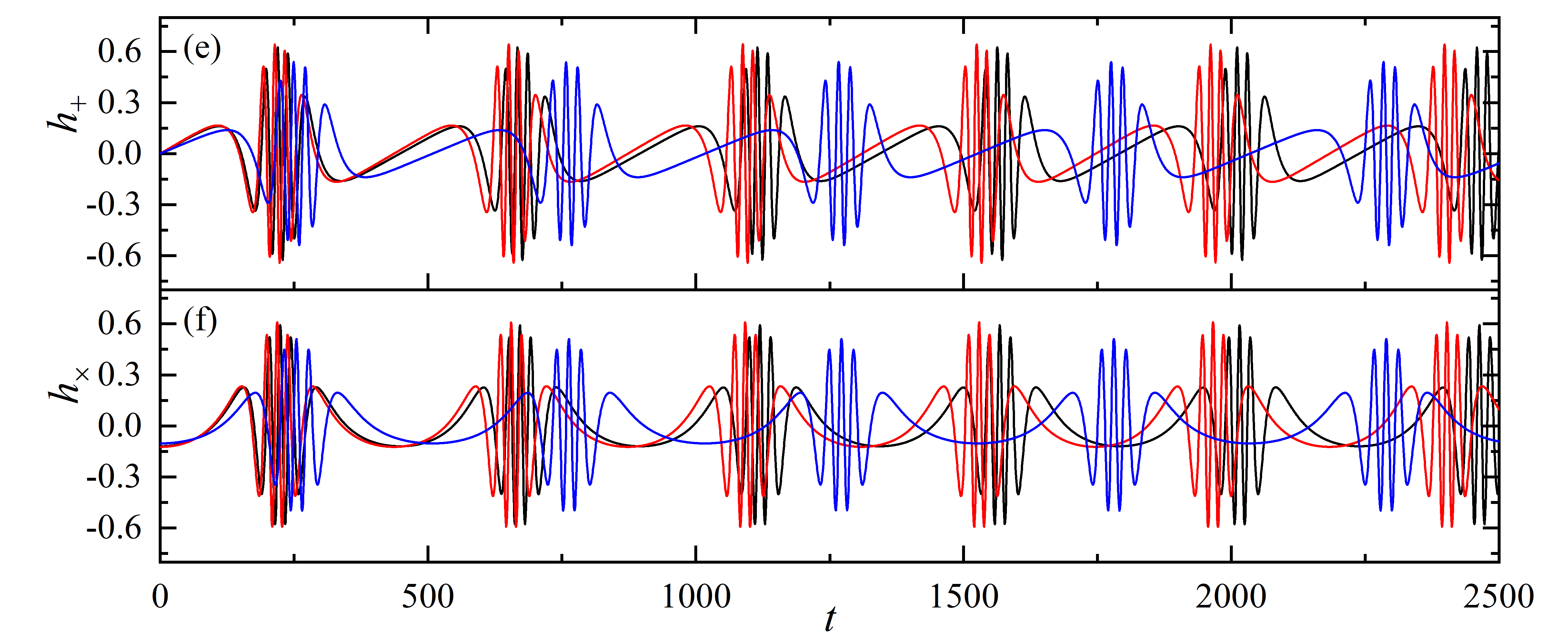}
\includegraphics[width=4.2cm]{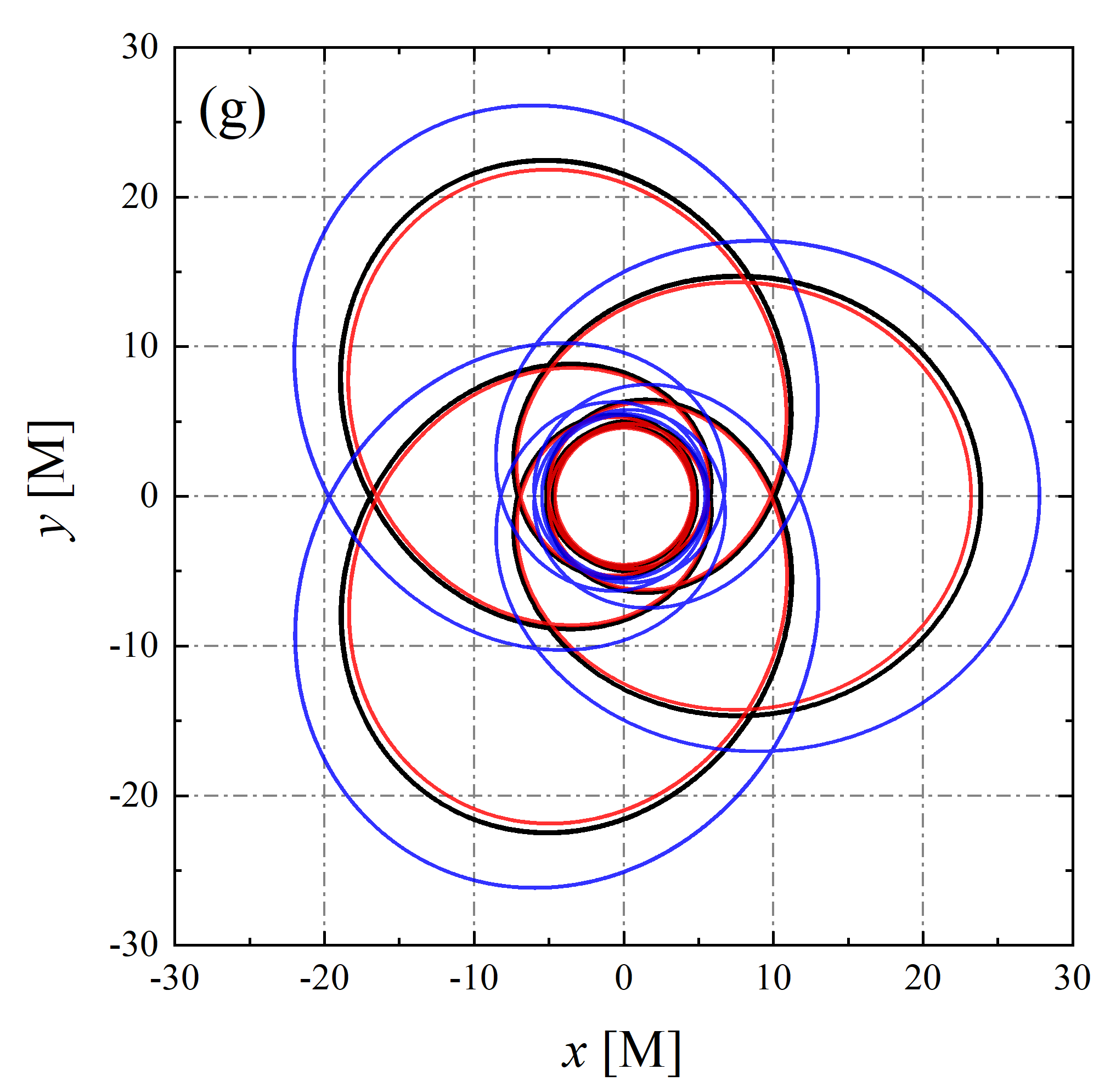}
\includegraphics[width=10.5cm]{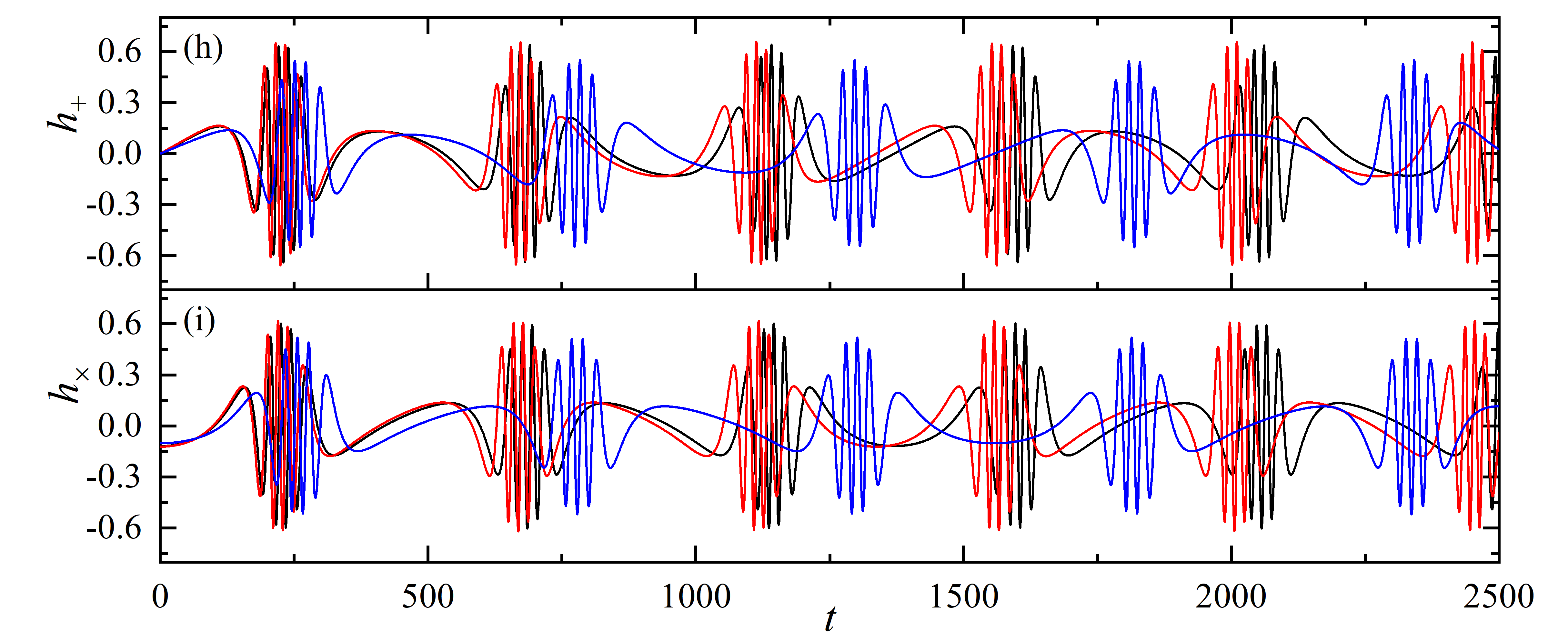}
\includegraphics[width=4.2cm]{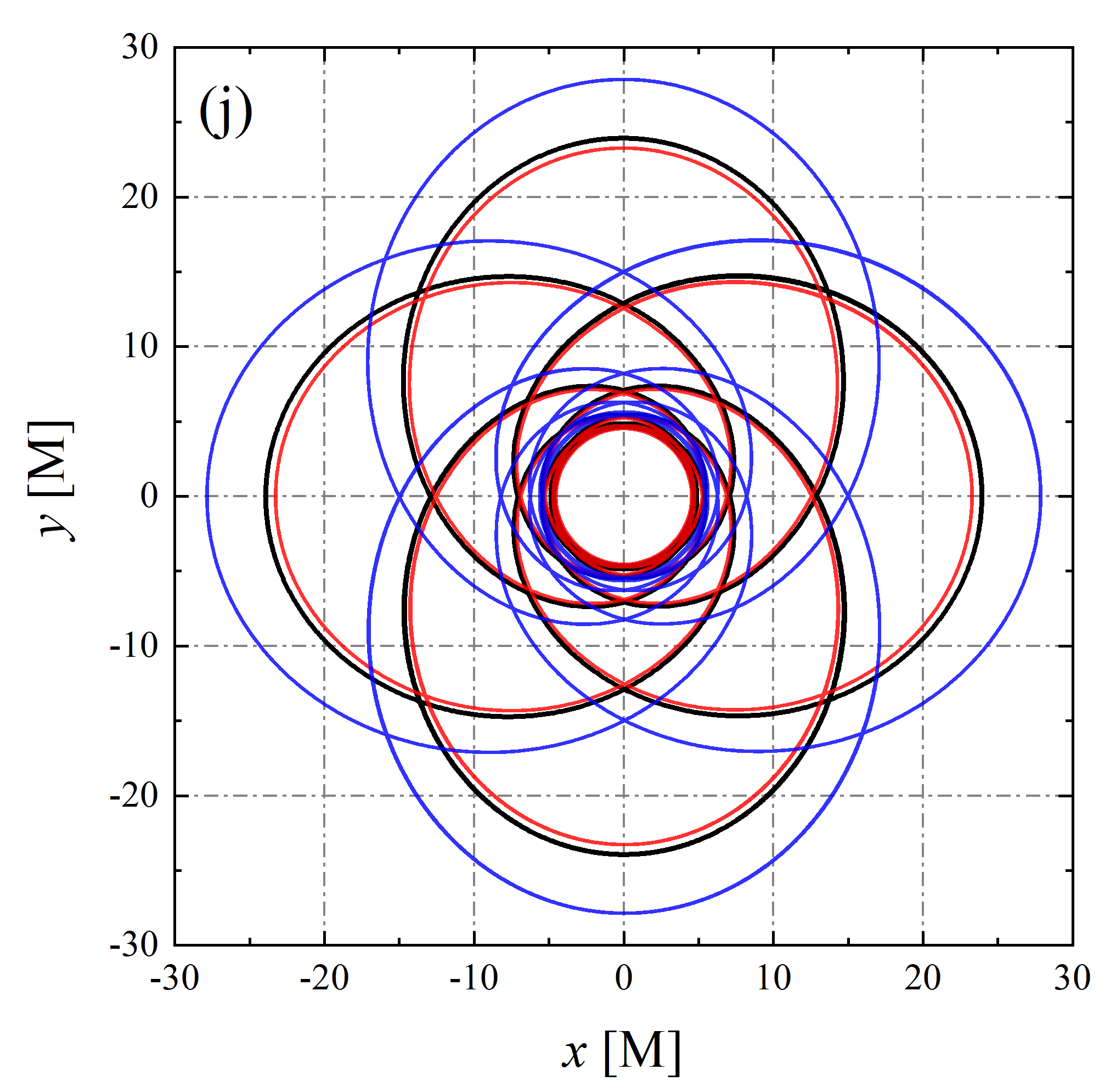}
\includegraphics[width=10.5cm]{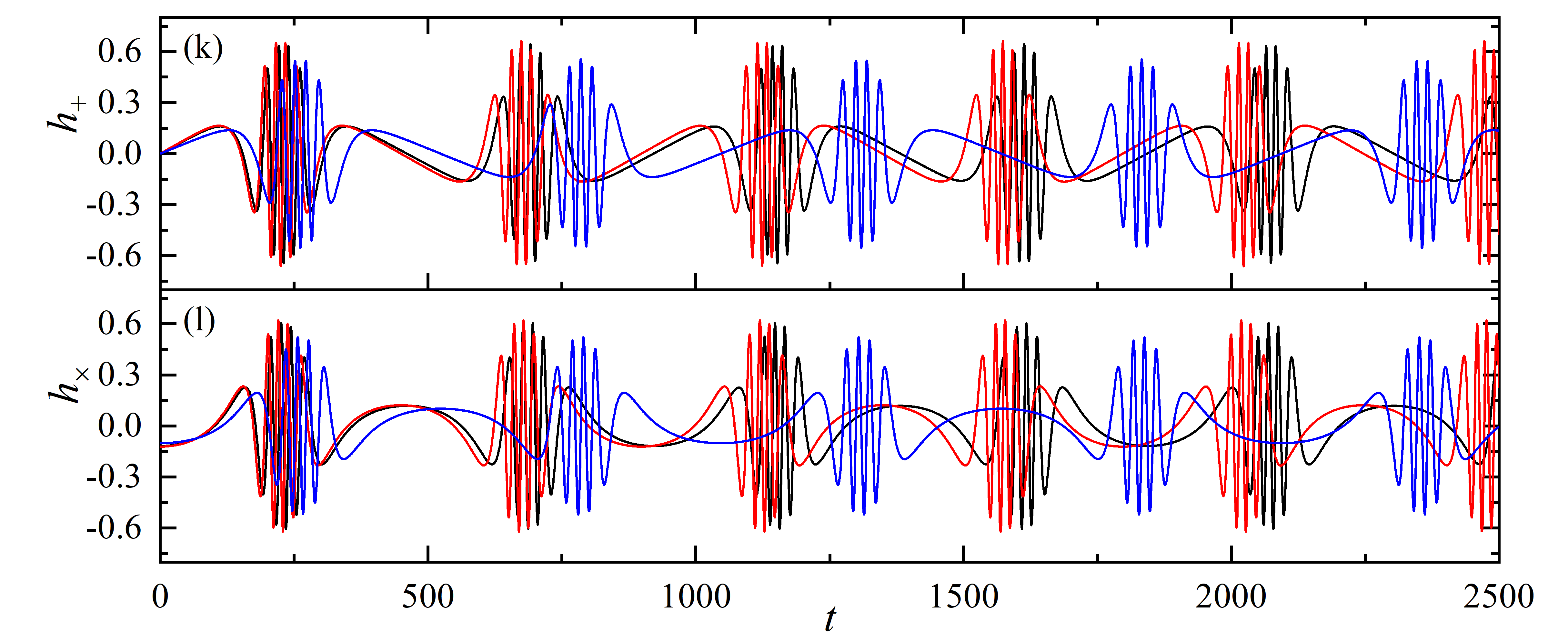}
\includegraphics[width=4.2cm]{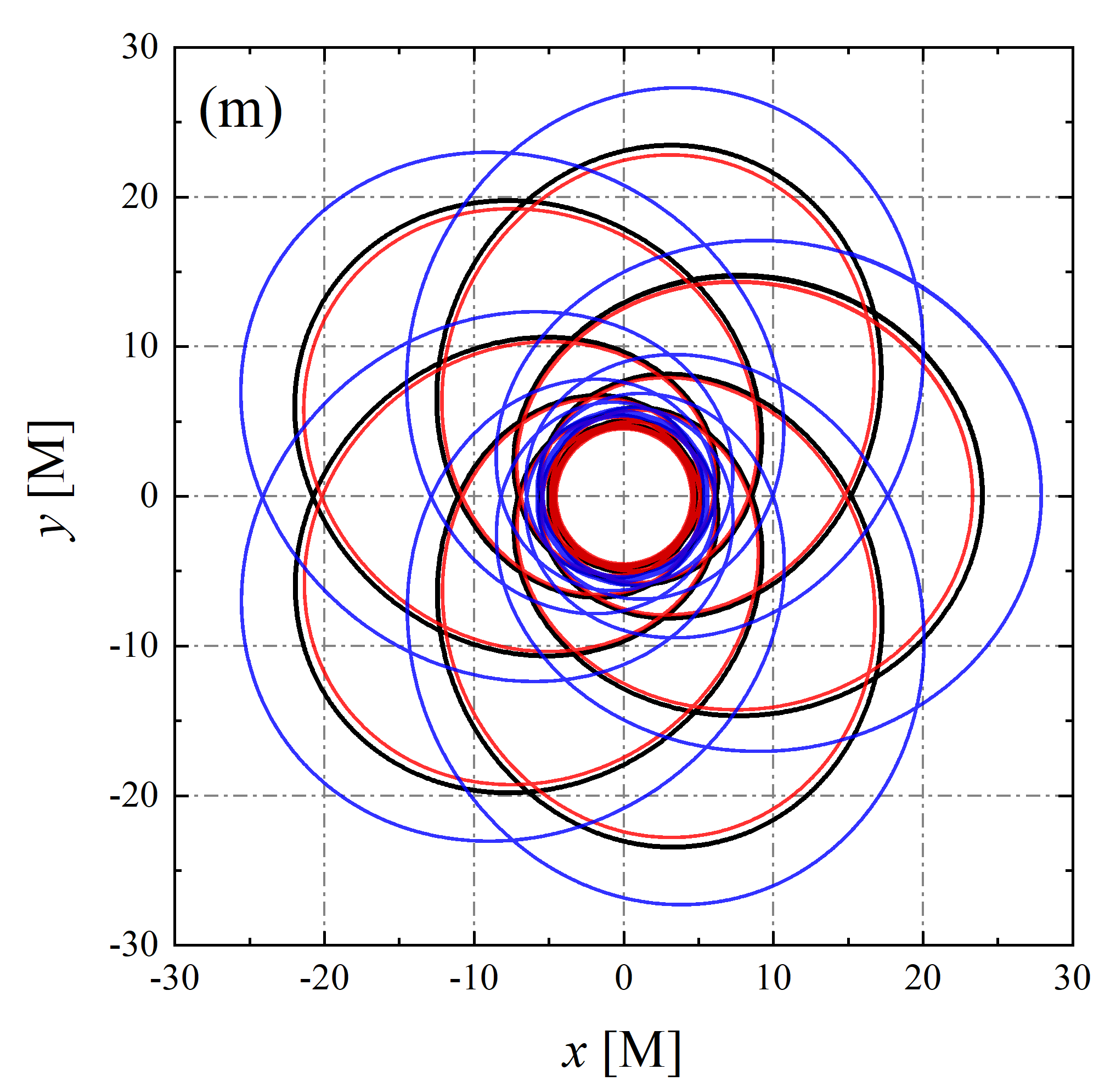}
\includegraphics[width=10.5cm]{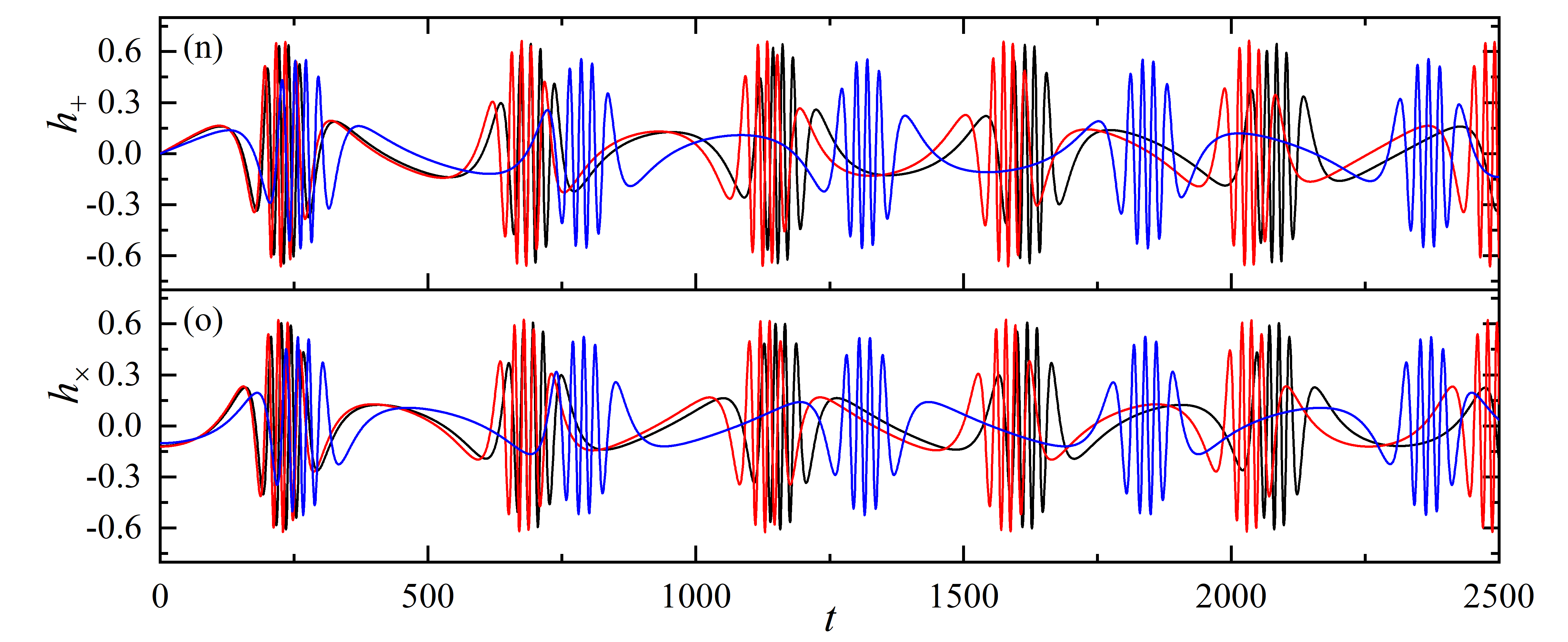}
\caption{Closed orbits (left column) and the corresponding gravitational wave waveforms (right column) across different parameter spaces. In each panel, the red, black, and blue curves correspond to the dark matter halo parameters $(r_{\textrm{s}},\rho_{\textrm{s}})$ $=$ $(0,0)$, $(0.3,0.3)$, and $(0.6,0.3)$, respectively. The angular momentum parameter $\varepsilon$ is fixed at $0.5$.}}\label{fig9}
\end{figure*}
With $\varepsilon$ fixed at $0.5$, we simulated the orbits for $z=1$, $2$, $3$, $4$, and $5$ along with their corresponding gravitational wave polarization states, as illustrated in Fig. 9. In these plots, the blue and black curves represent the dark matter halo parameters $(r_{\textrm{s}},\rho_{\textrm{s}})$ of $(0.6,0.3)$ and $(0.3,0.3)$, respectively, while the red curves denote the pure Schwarzschild case. In the orbital plots of each row, it is evident that the introduction of the dark matter halo preserves the orbital configuration while causing an overall amplification of the orbital scale. This expansion becomes particularly significant when $(r_{\textrm{s}},\rho_{\textrm{s}})=(0.6,0.3)$, an amplification behavior that is highly consistent with the results presented in Fig. 7. Simultaneously, we find that the two gravitational wave polarization signals exhibit clear quasi-periodicity for any given $z$. These signals are primarily composed of two distinct components. The first component consists of large-amplitude, high-frequency oscillations, which often manifest as several closely packed, sharp peaks within a short time scale; these are primarily contributed by the orbital motion near the periastron. The second component comprises smooth, broad bumps that connect the high-frequency oscillations, which are mainly generated during the orbit's passage near the apastron. This implies that by analyzing the high-frequency oscillations and broad bumps within the waveform, one can roughly infer the degree of orbital complexity.

It is worth noting that within our simulated time scales, the locations and morphologies of the high-frequency oscillations and broad bumps in the gravitational wave signals remain remarkably similar across different orbits, making it relatively difficult to distinguish specific orbital features based solely on these waveforms. Nevertheless, the impact of the dark matter halo parameters on the waveforms is unmistakable: the introduction of the dark matter halo induces a noticeable phase delay in the gravitational wave signals. As illustrated by the contrast between the red and blue curves, this delay effect becomes increasingly pronounced with larger dark matter halo parameters and longer orbital evolution times. 
\begin{figure*}
\center{
\includegraphics[height=3.5cm]{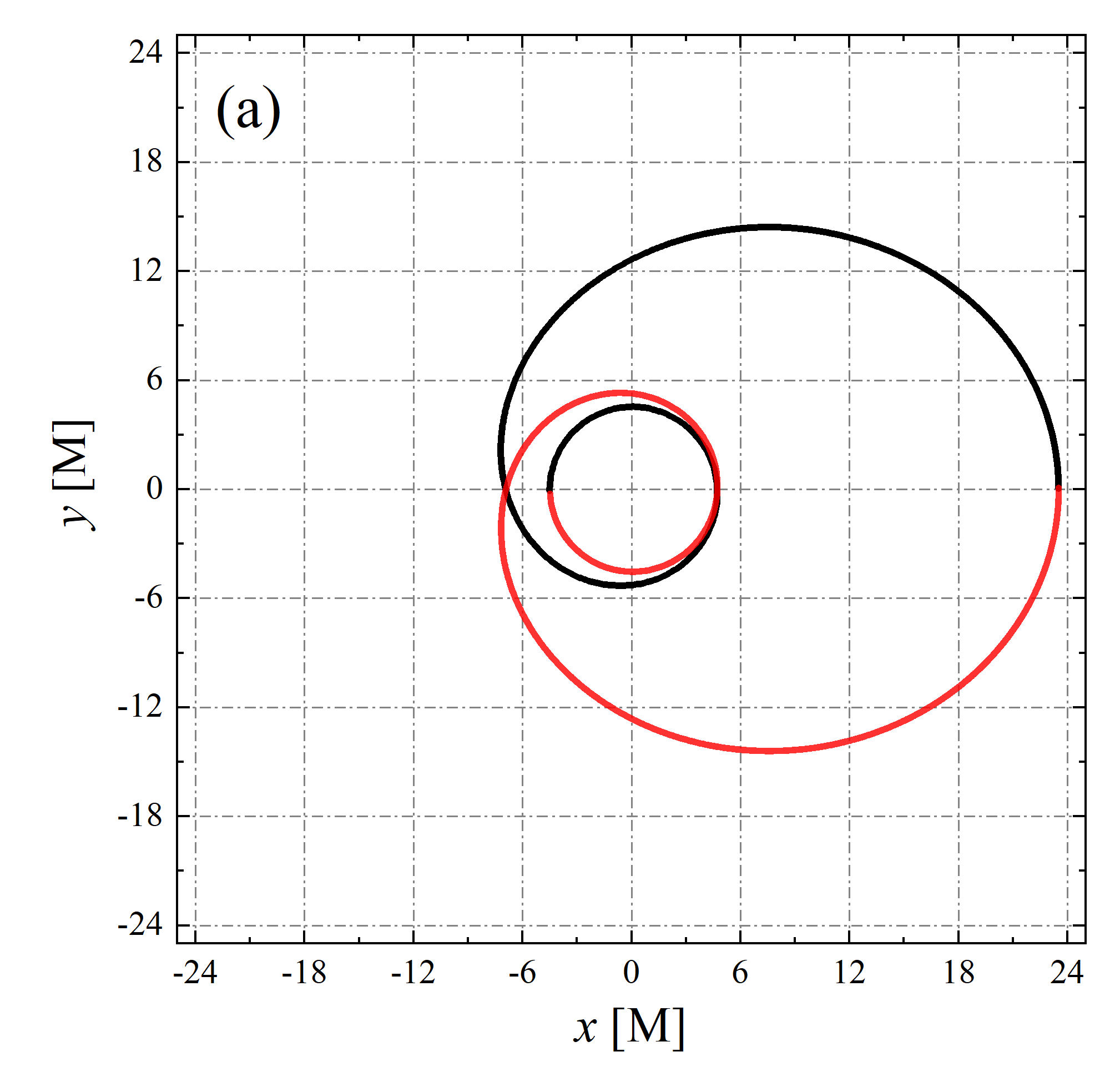}
\includegraphics[height=3.5cm]{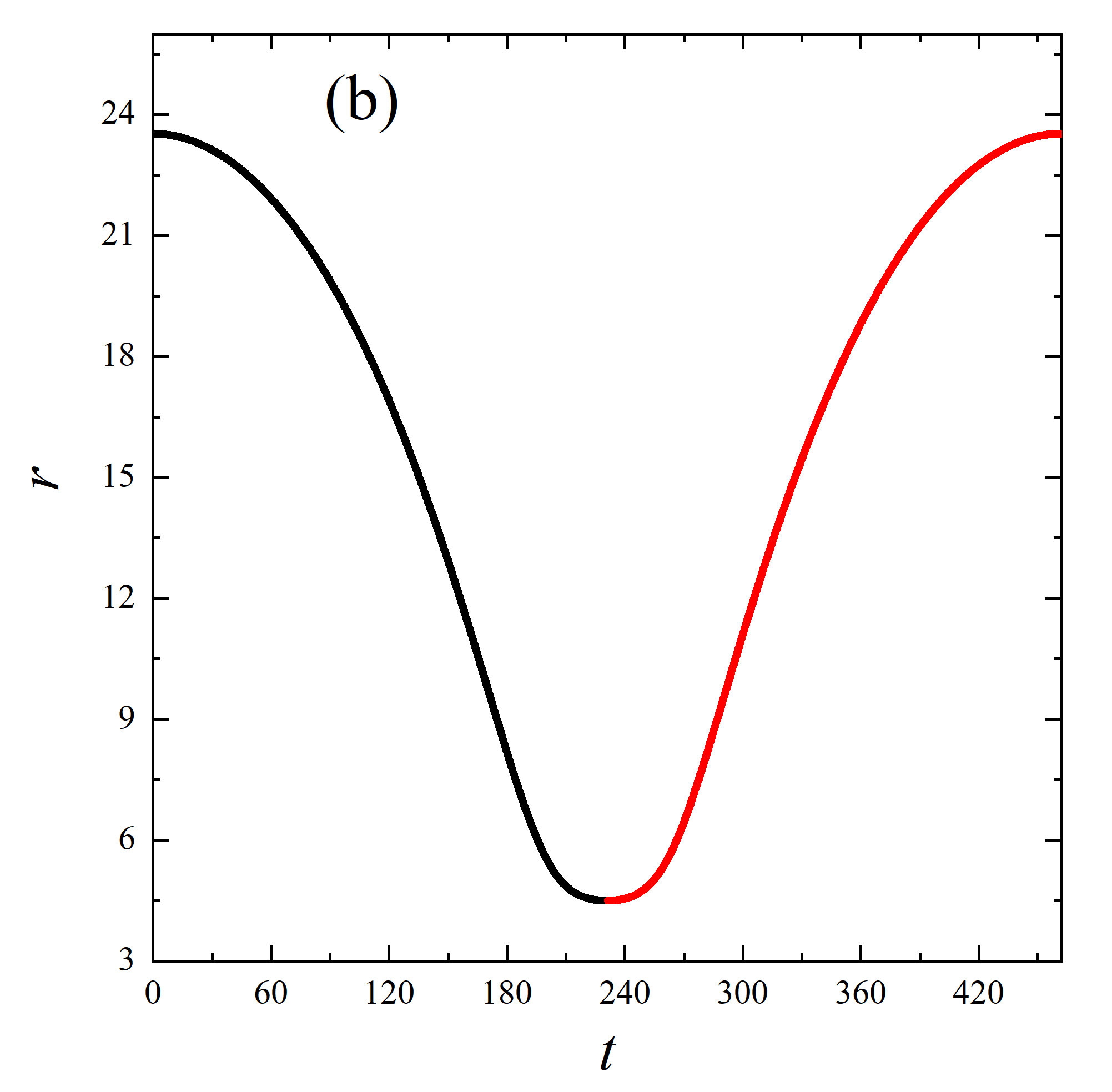}
\includegraphics[height=3.5cm]{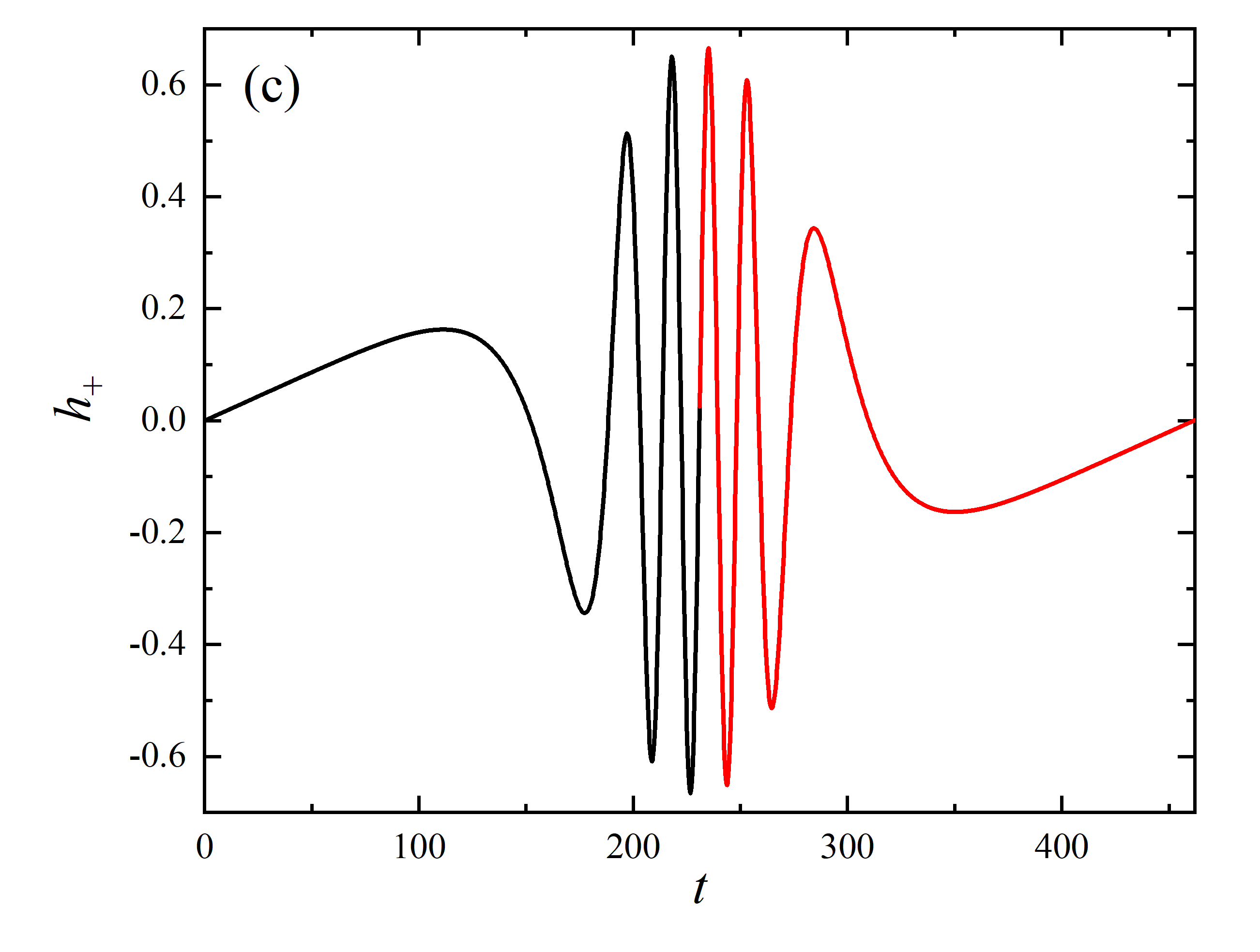}
\includegraphics[height=3.5cm]{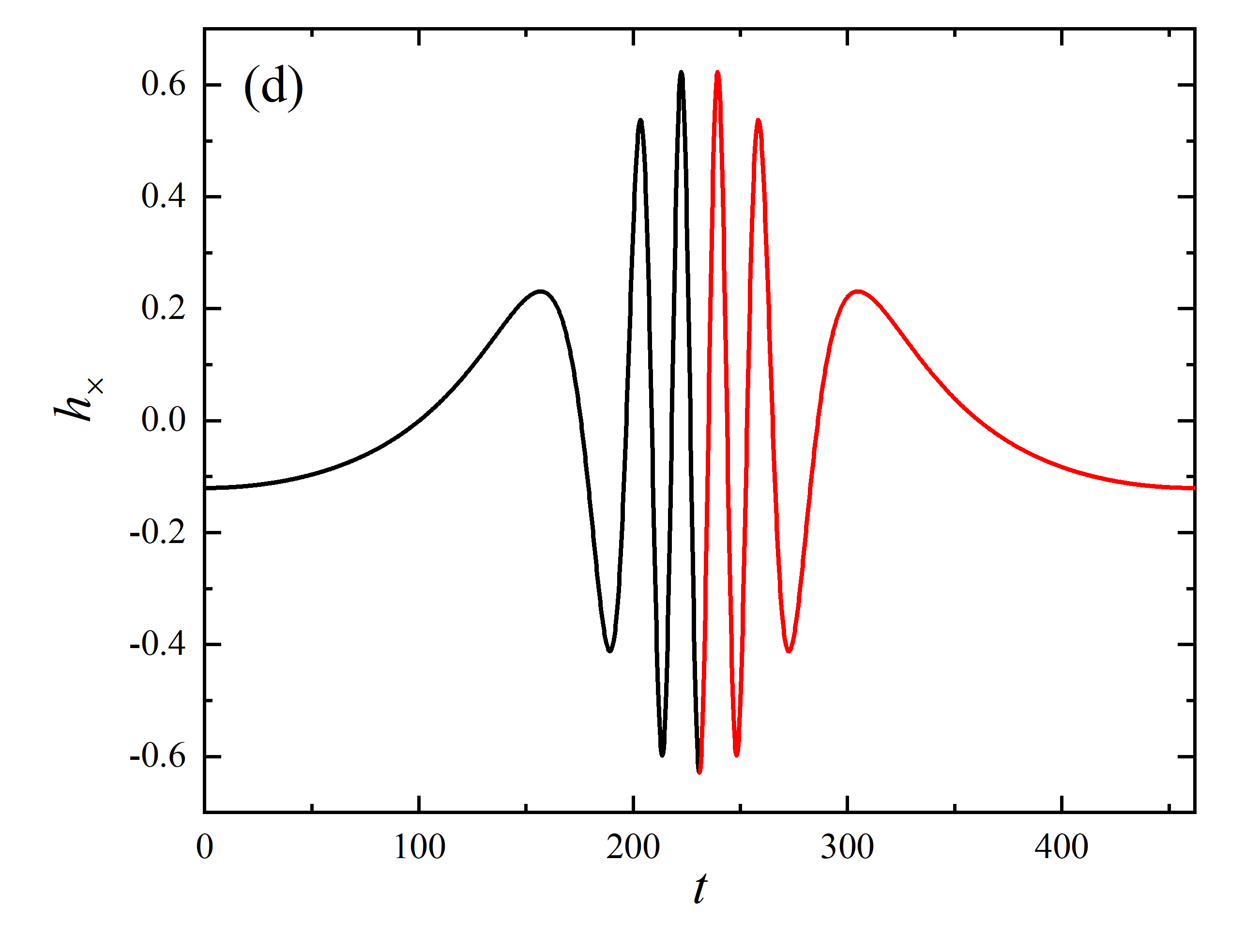}
\includegraphics[height=3.5cm]{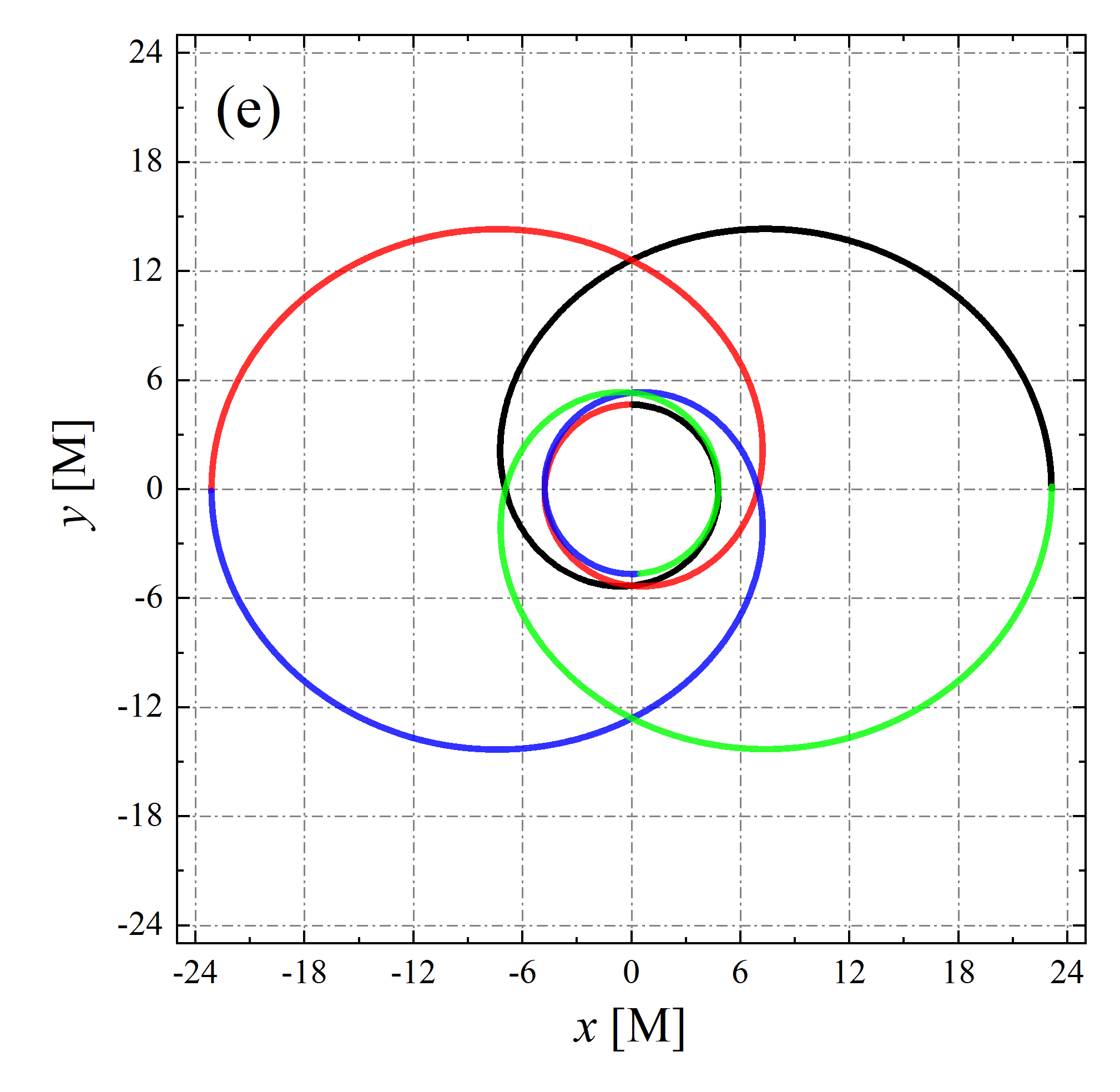}
\includegraphics[height=3.5cm]{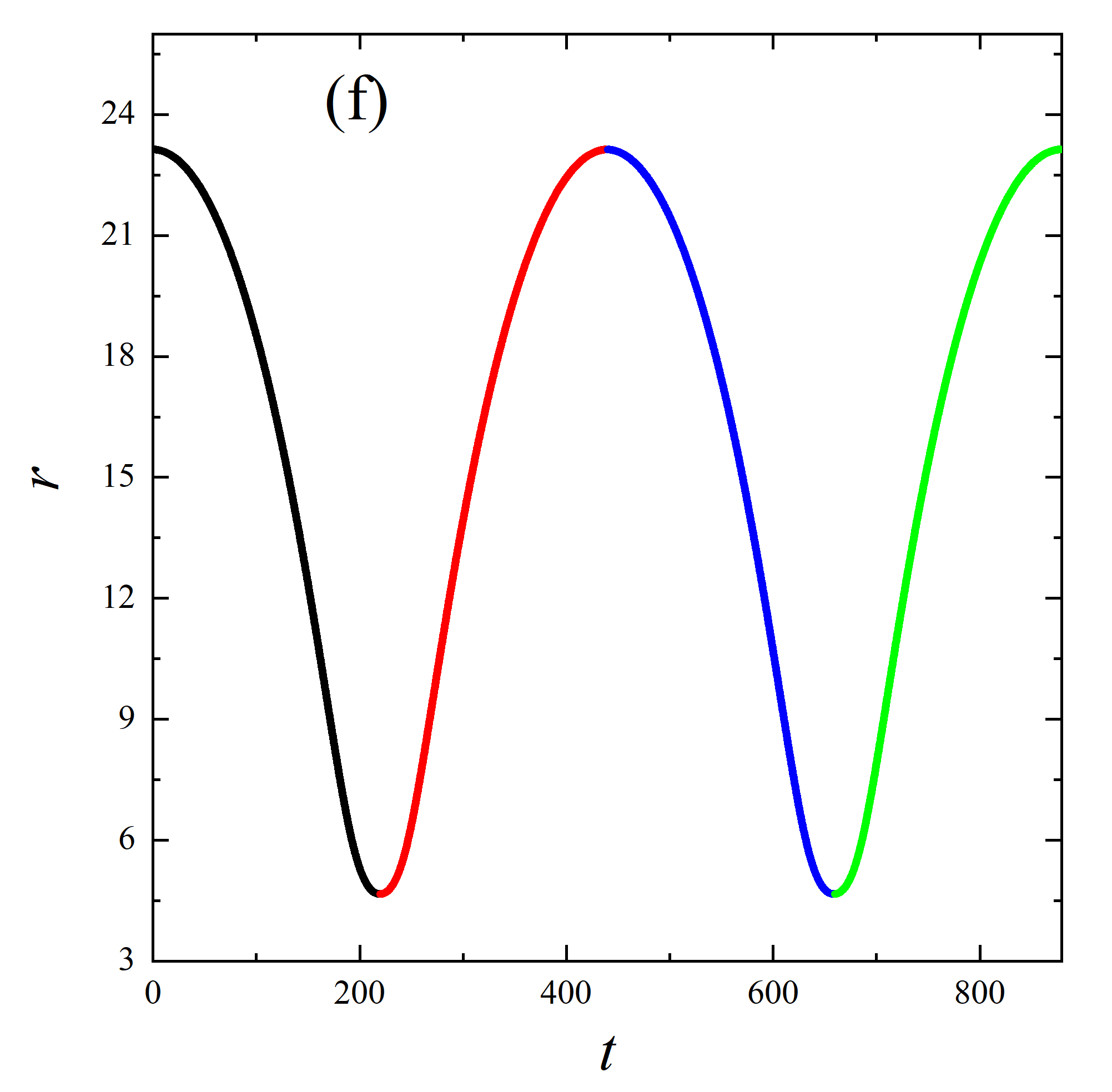}
\includegraphics[height=3.5cm]{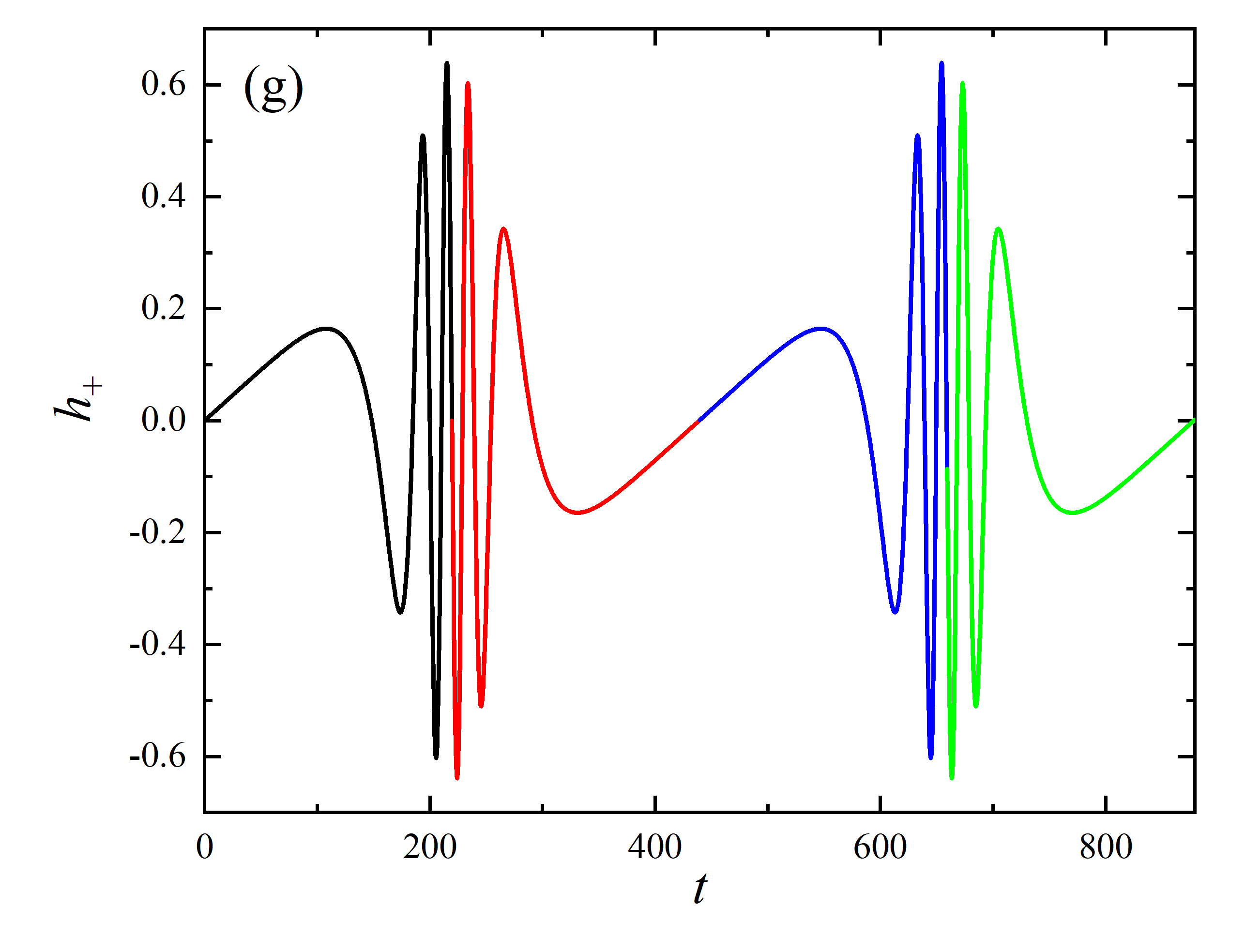}
\includegraphics[height=3.5cm]{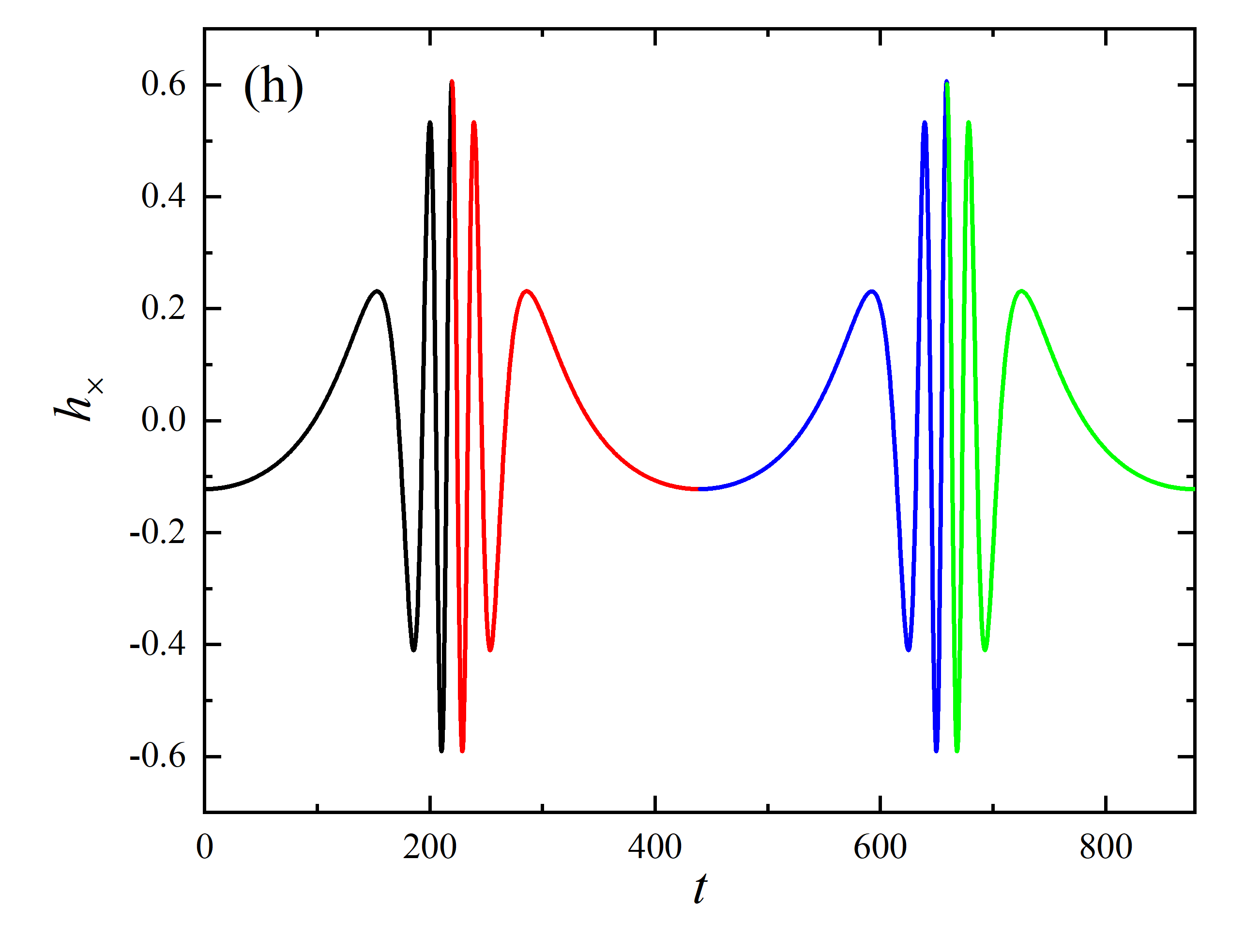}
\includegraphics[height=3.5cm]{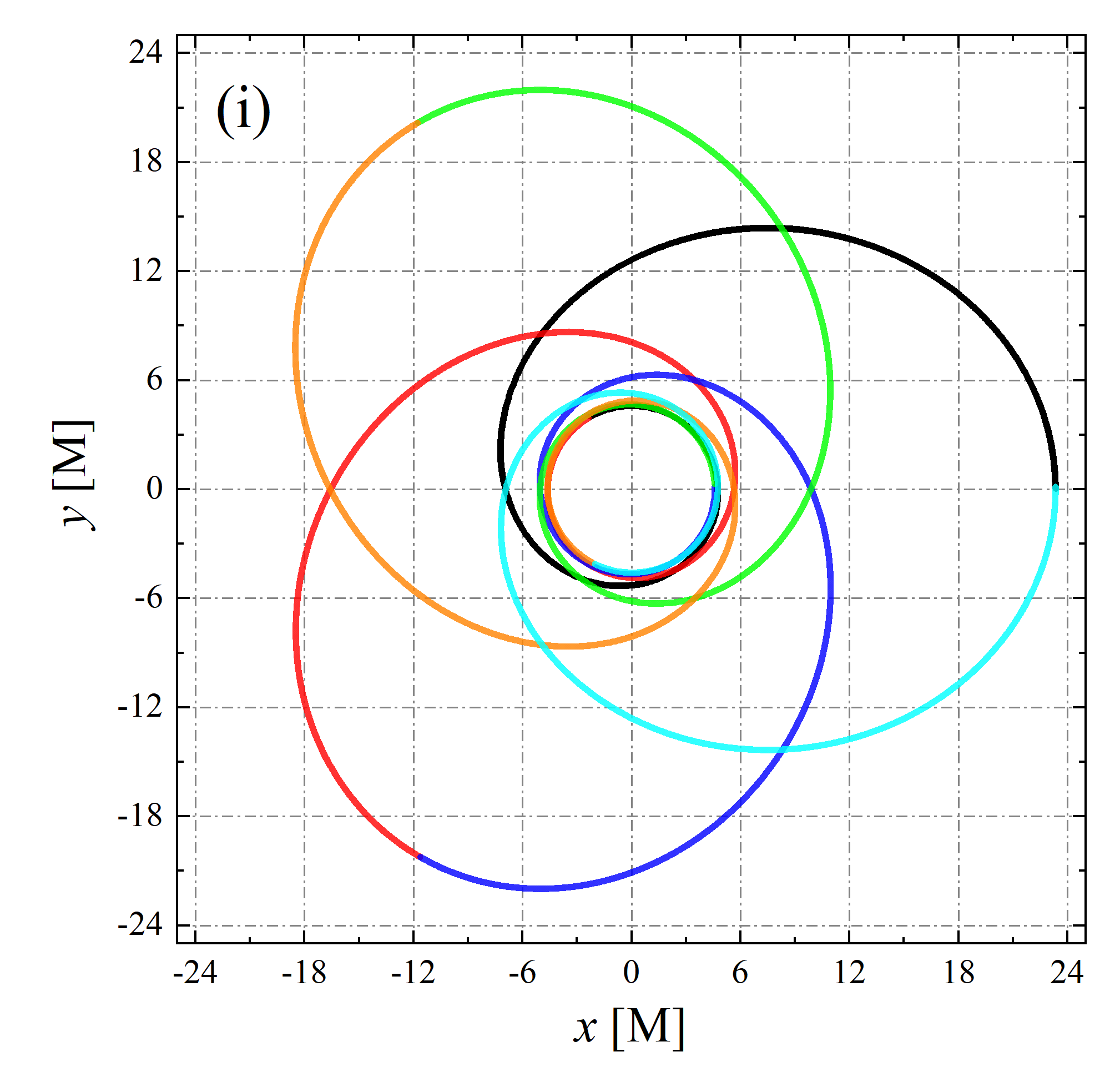}
\includegraphics[height=3.5cm]{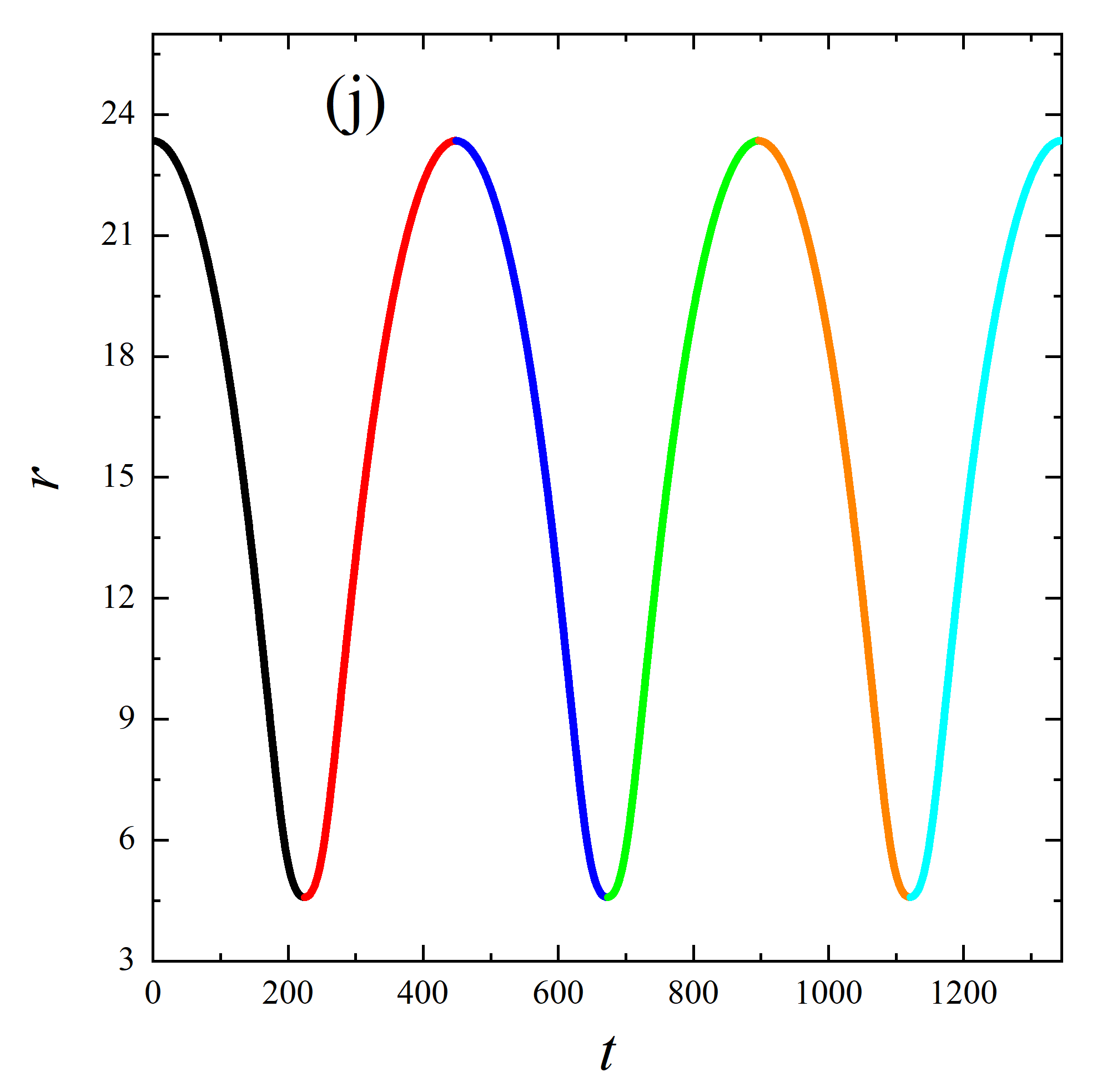}
\includegraphics[height=3.5cm]{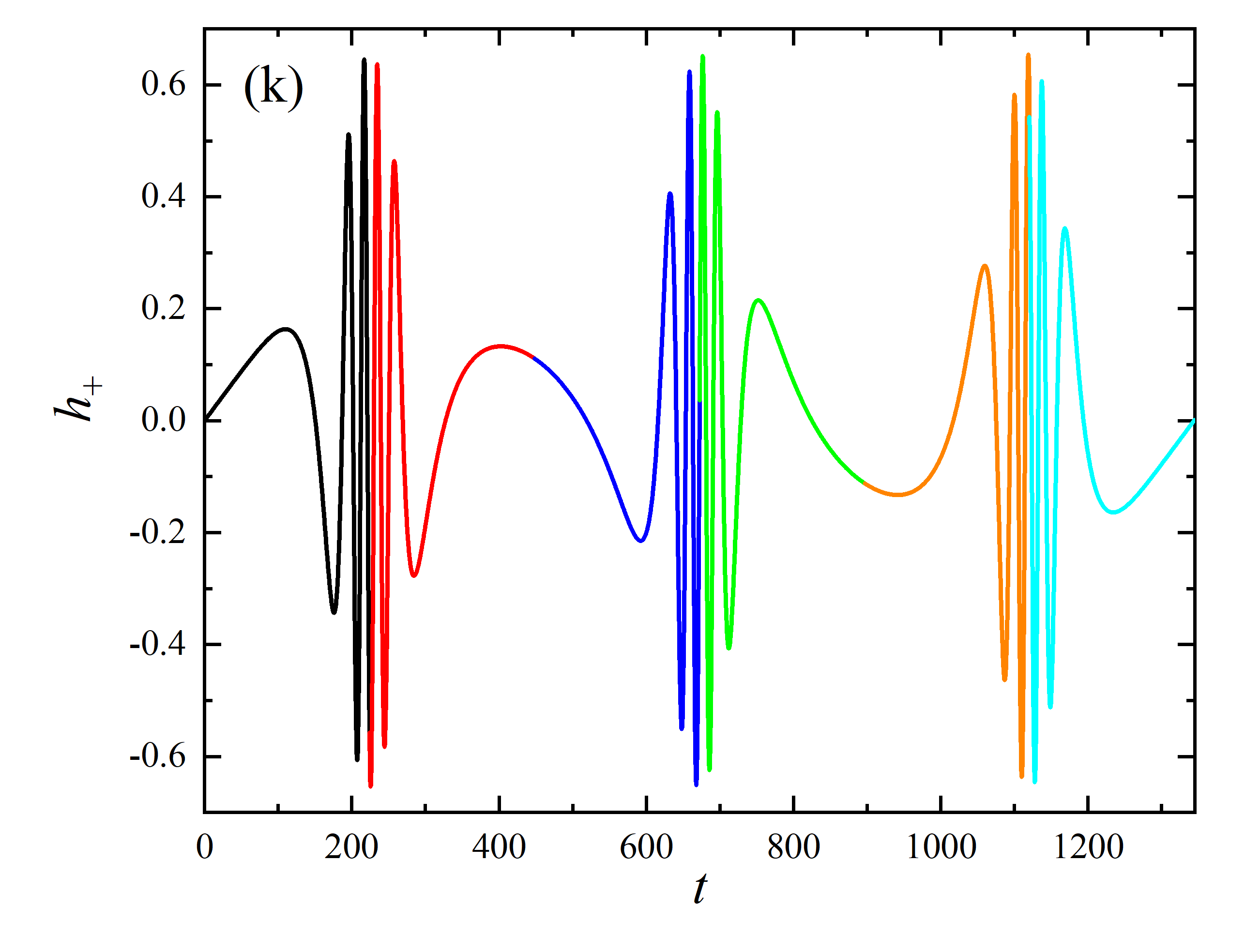}
\includegraphics[height=3.5cm]{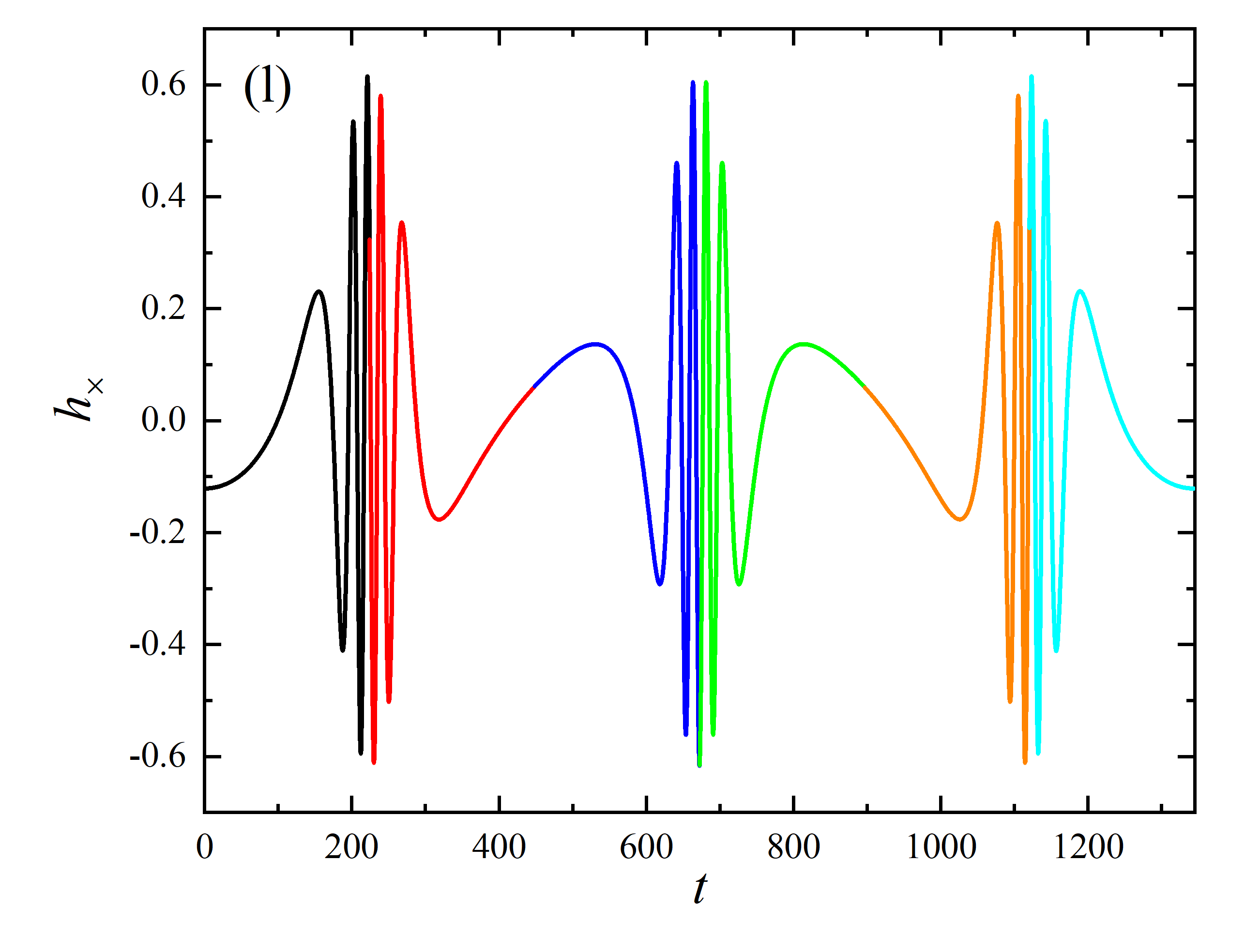}
\includegraphics[height=3.5cm]{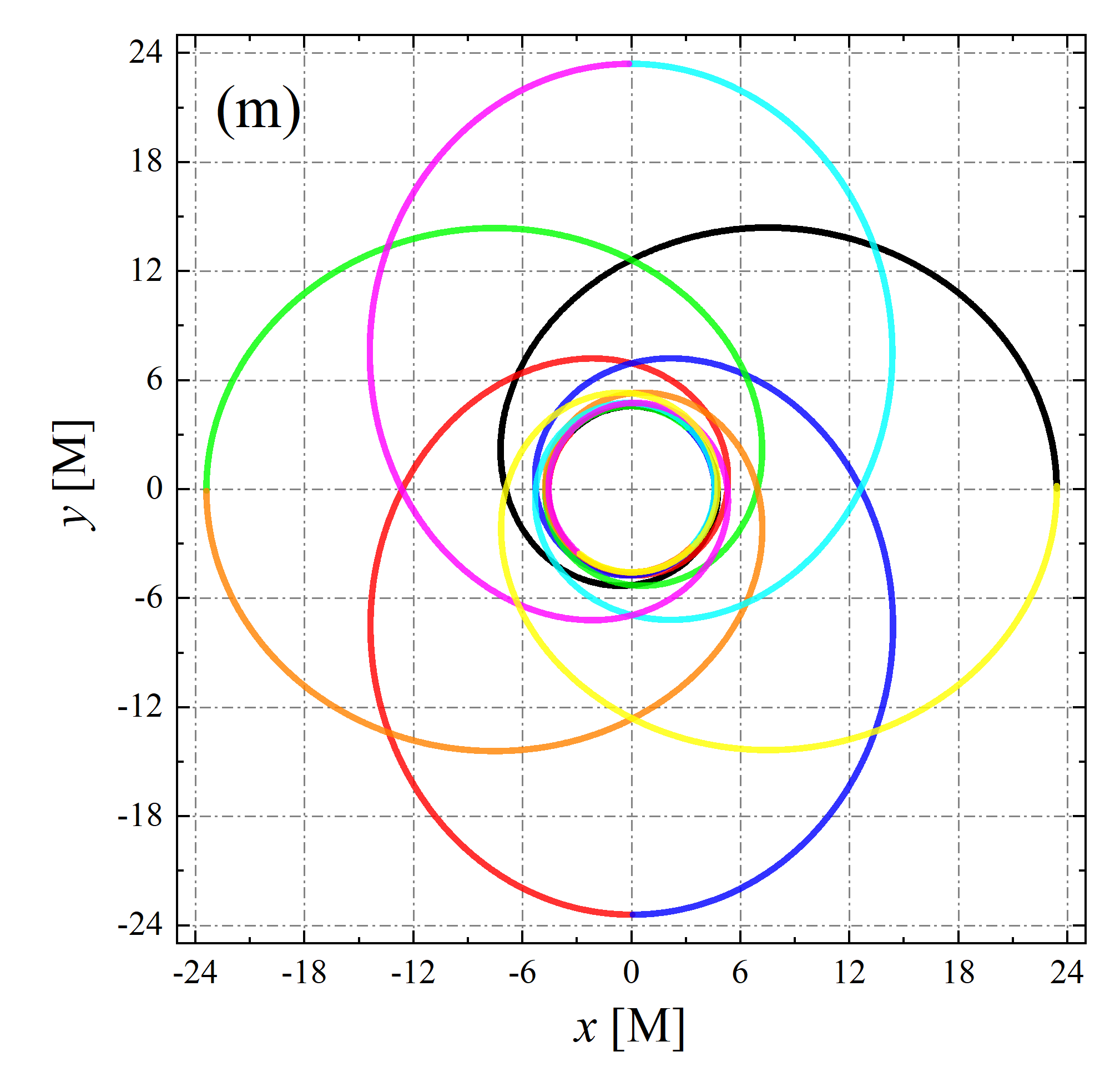}
\includegraphics[height=3.5cm]{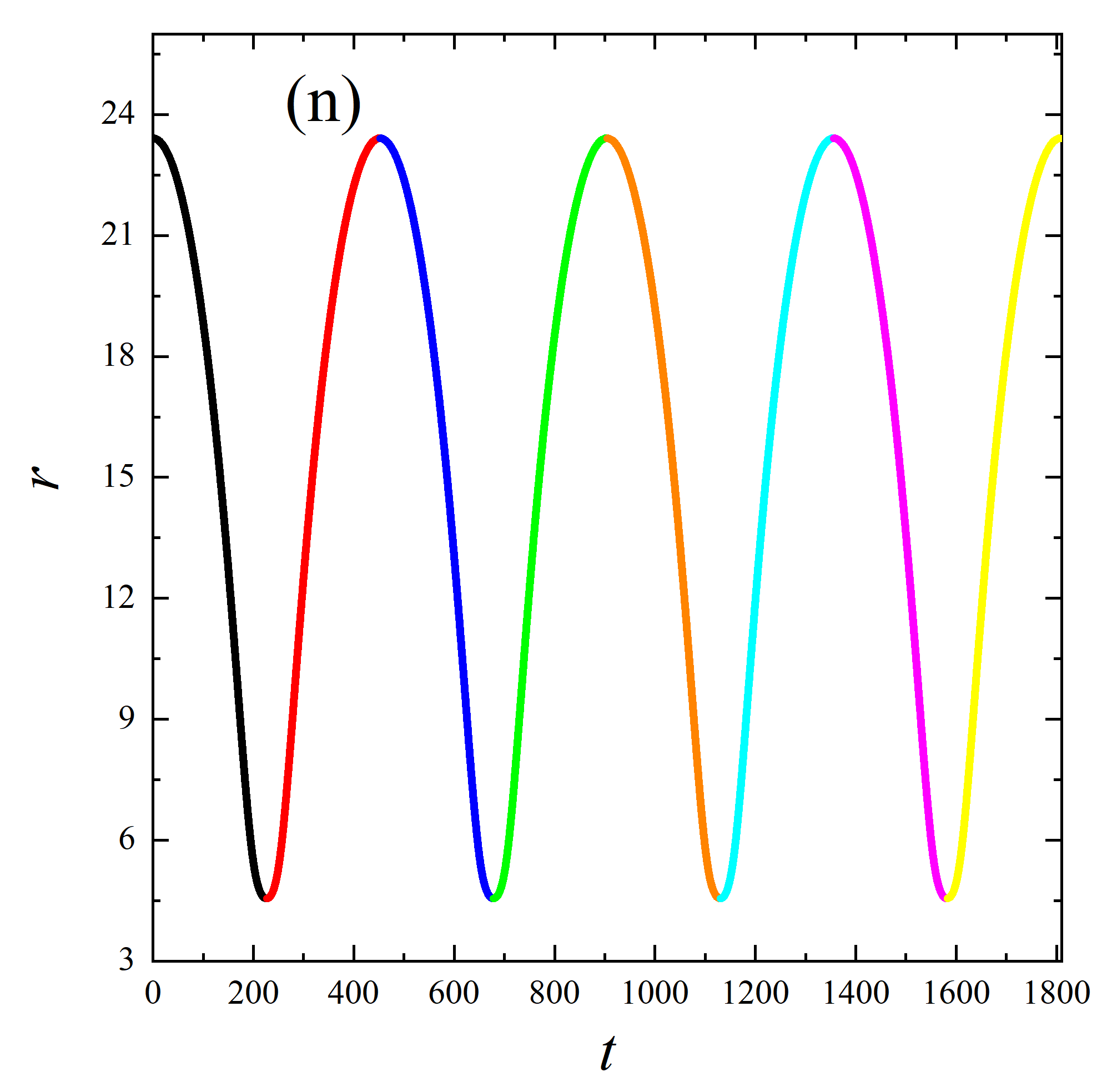}
\includegraphics[height=3.5cm]{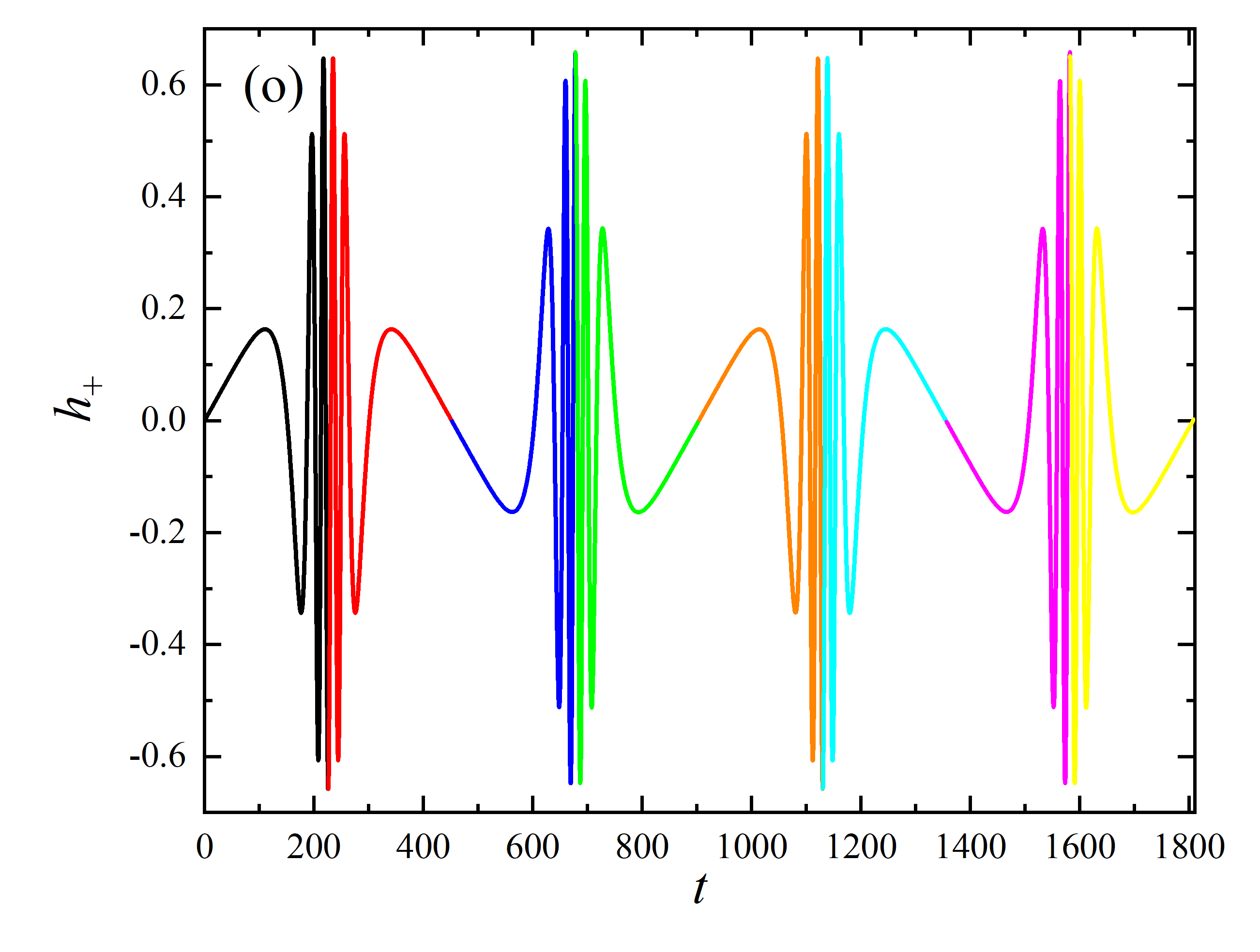}
\includegraphics[height=3.5cm]{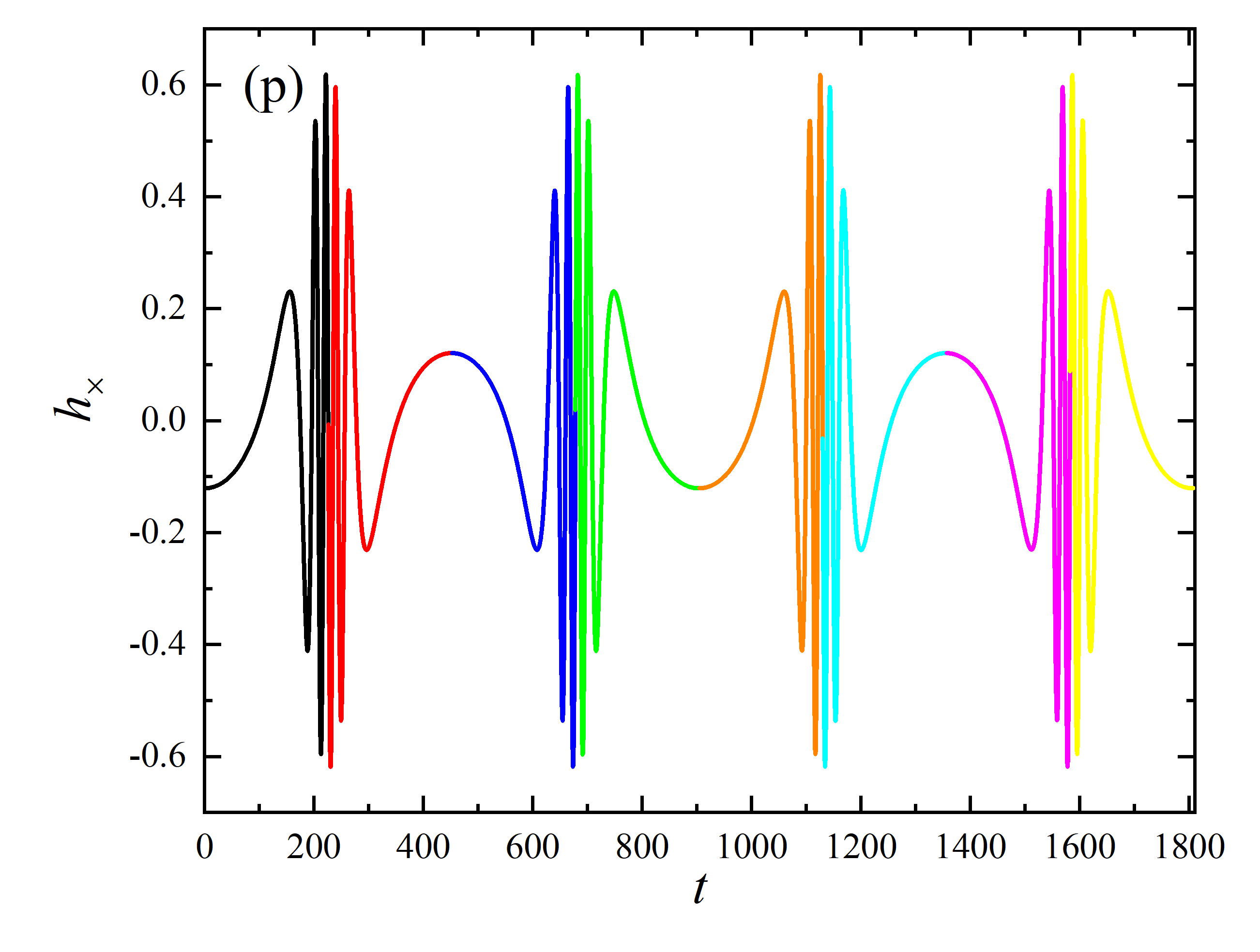}
\caption{From left to right, the columns display the rational orbits, the radial coordinate evolution, and the gravitational wave polarization states within a single complete period, respectively. From top to bottom, the rows correspond to the orbital configurations $(z,w,v)$ of $(1,1,1)$, $(2,1,1)$, $(3,1,2)$, and $(4,1,3)$. For all cases, the dark matter halo parameters are fixed at $(r_{\textrm{s}},\rho_{\textrm{s}})$ $=$ $(0.2,0.2)$, and $\varepsilon=0.5$.}}\label{fig10}
\end{figure*}

To further uncover the underlying sources of the phase delay and the distinct components within the gravitational wave signals, we simulated the orbits with varying numbers of leaves, along with their radial evolution and gravitational wave polarization states, over a single period under two sets of dark matter halo parameters, as depicted in Figs. 10 and 11. In these figures, the colors applied to the radial coordinate evolution and the gravitational wave signals indicate the specific contributions originating from the identically colored segments of the respective orbits. Furthermore, the total orbital period can be calculated as $2z\Delta\tau$, where $\Delta\tau$ is given by \eqref{periodictime}. For $z=1$ (as shown in the first row of Fig. 10), the timelike particle is released from the apastron at $(x,y)=(23.52,0)$ and proceeds toward the periastron along the solid black curve. During this process, the orbit intersects the $x$-axis three times. These intersections consecutively contribute to the peak of the broad bump and the two sharp peaks within the high-frequency oscillation component of the gravitational wave signal. Subsequently, the particle returns to its starting point along the red path, during which the gravitational wave signal transitions from high-frequency oscillations into a smooth, flattened region. These observations conclusively demonstrate that the broad bumps in the gravitational signal originate from the particle's broader transit from the apastron toward the periastron, whereas the high-frequency oscillations are exclusively contributed by the motion in the vicinity of the periastron. Furthermore, we can reasonably anticipate that for different values of the orbital parameter $w$, the number of intersections between the orbit and the $x$-axis during the apastron-to-periastron transit will vary. Consequently, this geometric variation will lead to a differing number of peaks within the high-frequency oscillation component of the resulting gravitational wave signal.

Turning our attention to the cases where $z=2$, $3$, and $4$, it becomes evident that the correspondence between the gravitational wave signal components and the orbital trajectory is identical to that observed for $z=1$. Specifically, as the particle travels from the apastron toward the periastron, the gravitational wave signal transitions from a smooth, broad bump into high-frequency oscillations. Conversely, as the particle returns from the periastron to the apastron, this evolutionary trend is exactly reversed. Furthermore, the variations in the gravitational wave signals induced by an increasing $z$ are highly apparent. This is primarily manifested by the fact that within a single complete period, the number of distinct high-frequency oscillation bursts in the signal perfectly matches the value of $z$.

By comparing Figs. 10 and 11, we find that increasing the dark matter halo parameters does not alter the correspondence between the gravitational wave signal components and the orbital trajectory. However, as illustrated in Fig. 9, it expands the spatial extent of the orbit while preserving its geometric proportions, thereby inducing a phase delay in the gravitational wave signal. For instance, a slight rightward shift of the waveform can be clearly observed when comparing panel (c) in Figs. 10 and 11. The physical mechanism underlying this phenomenon is that, assuming an identical rational orbital structure, an increase in the dark matter halo parameters drives an outward expansion of the orbit. Consequently, this expansion forces the particle to spend a longer transit time traveling between the apastron and the periastron. In other words, in scenarios with smaller dark matter halo parameters, the timelike particle has already reached the periastron and generated the sharp peaks associated with the high-frequency oscillations. Conversely, under larger dark matter halo parameters, the particle has not yet arrived at the periastron, which naturally leads to the delayed appearance of the gravitational wave signal. Notably, this delay effect accumulates and becomes increasingly pronounced as the integration time grows.
\begin{figure*}
\center{
\includegraphics[height=3.5cm]{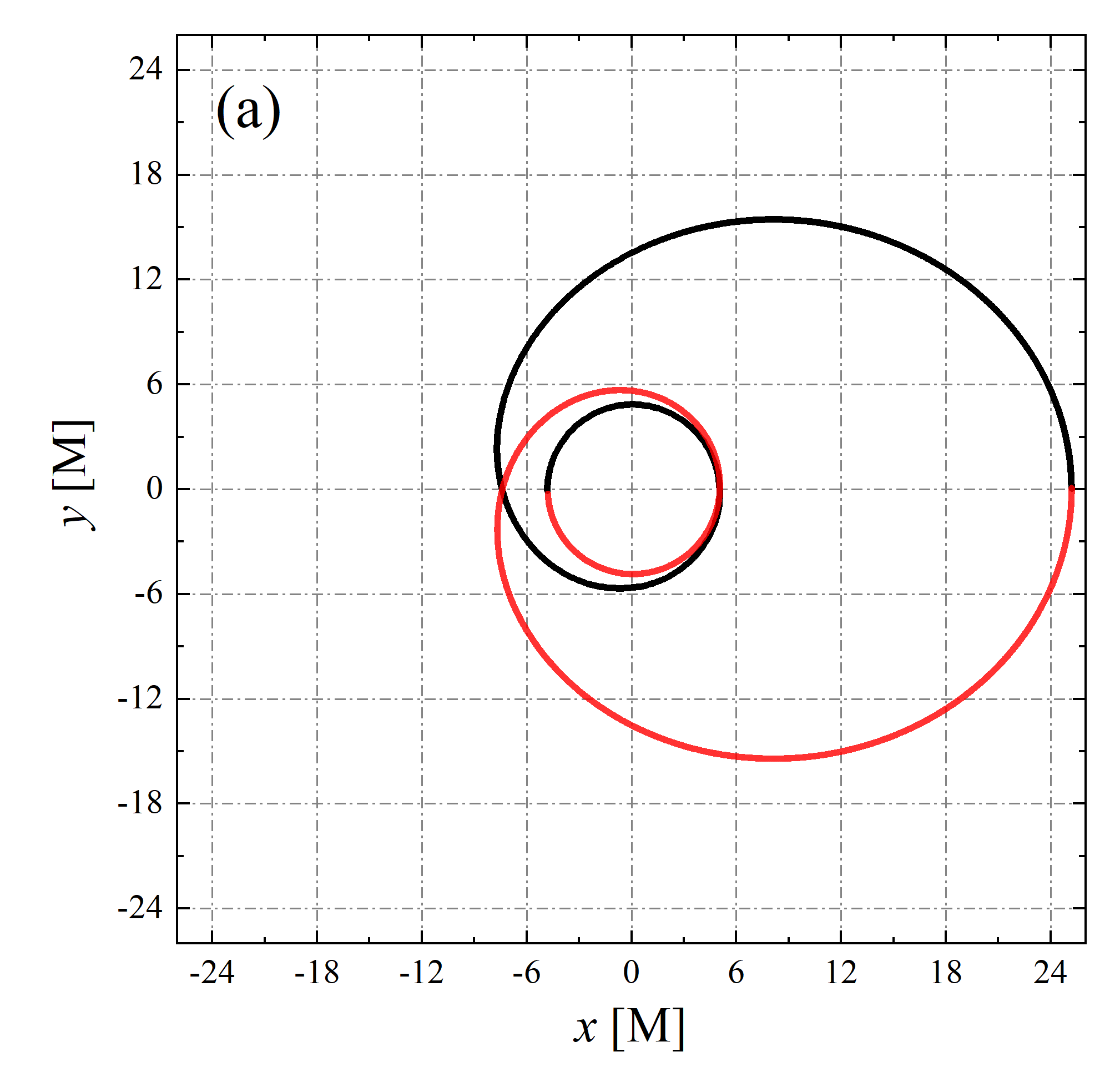}
\includegraphics[height=3.5cm]{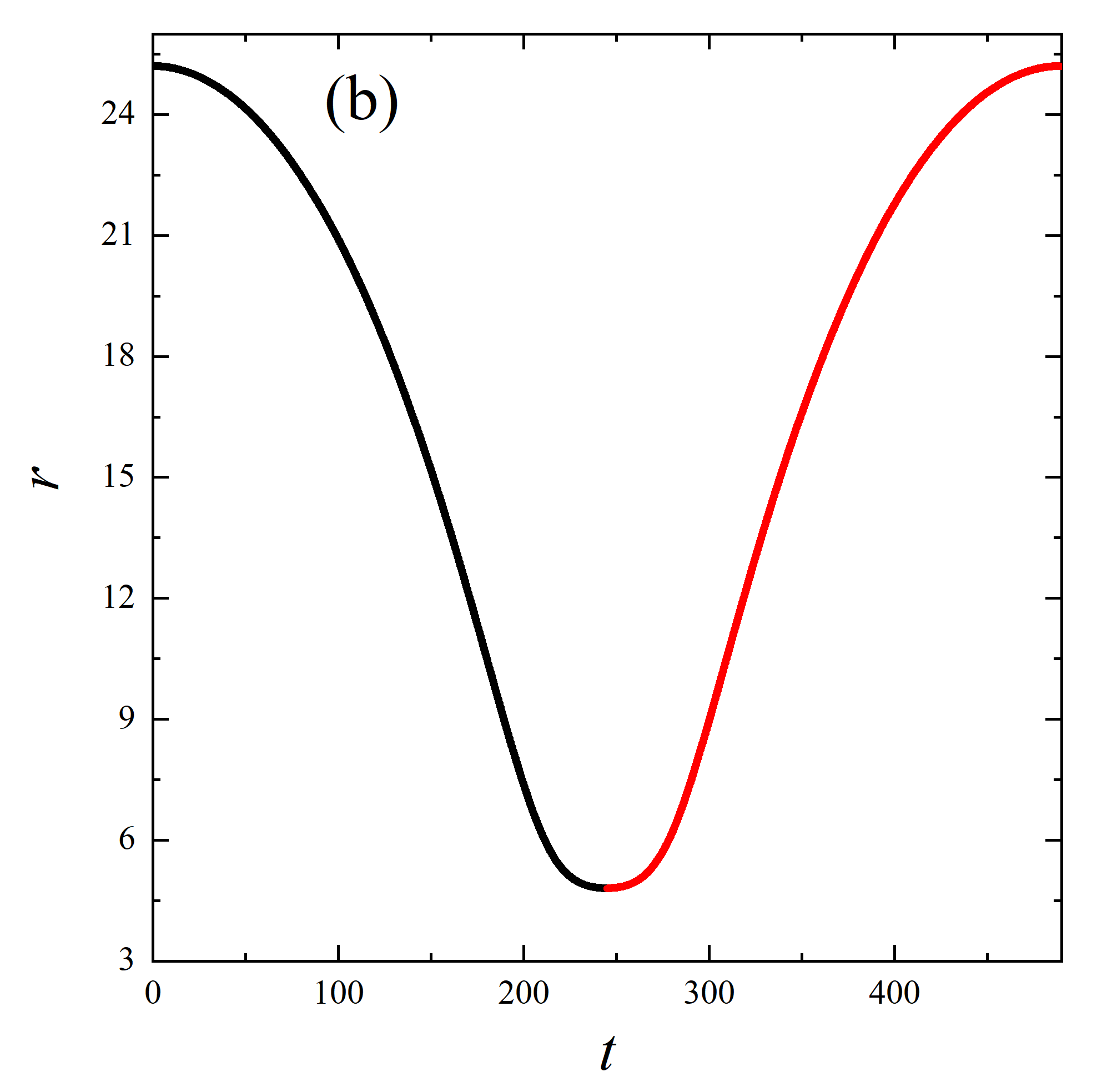}
\includegraphics[height=3.5cm]{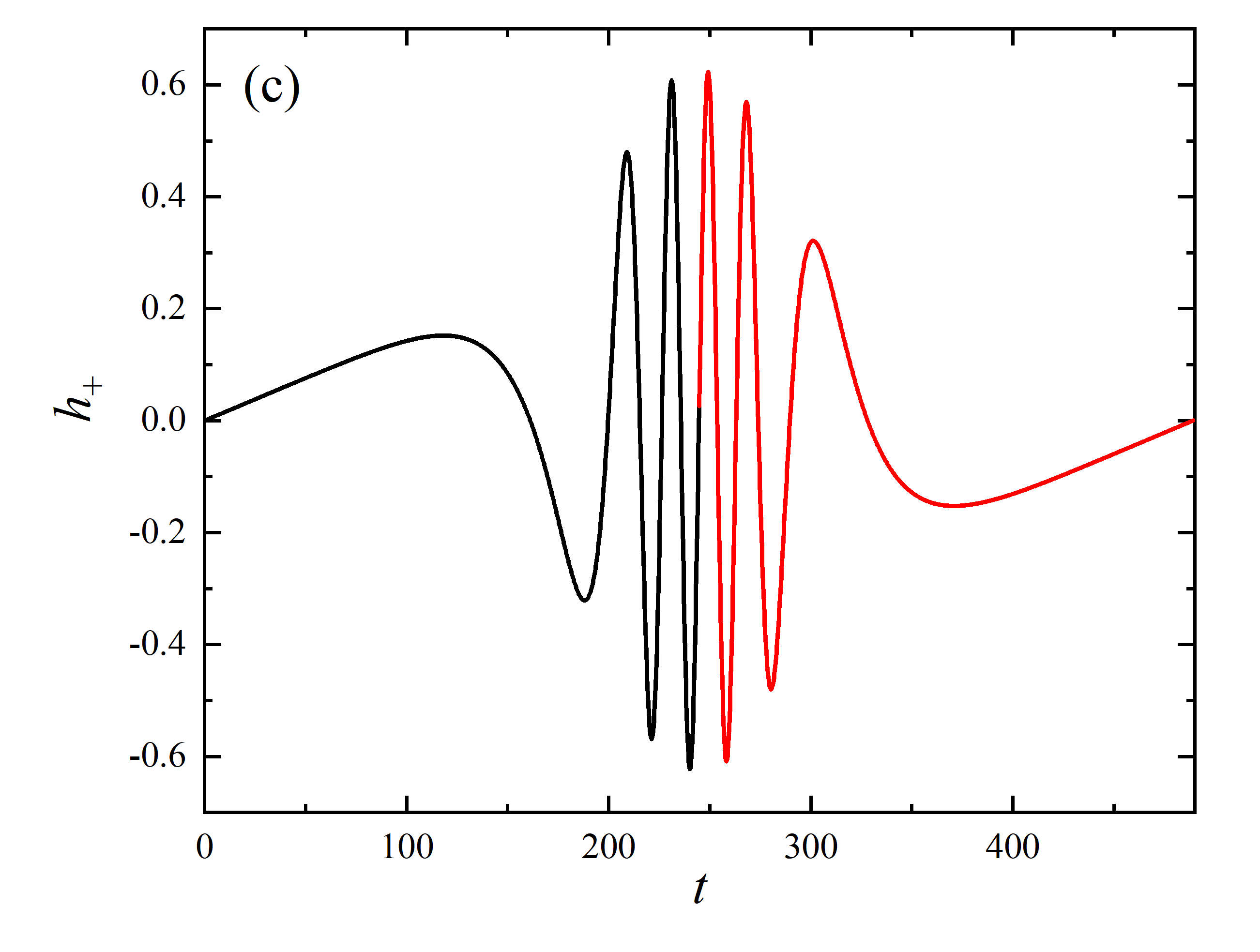}
\includegraphics[height=3.5cm]{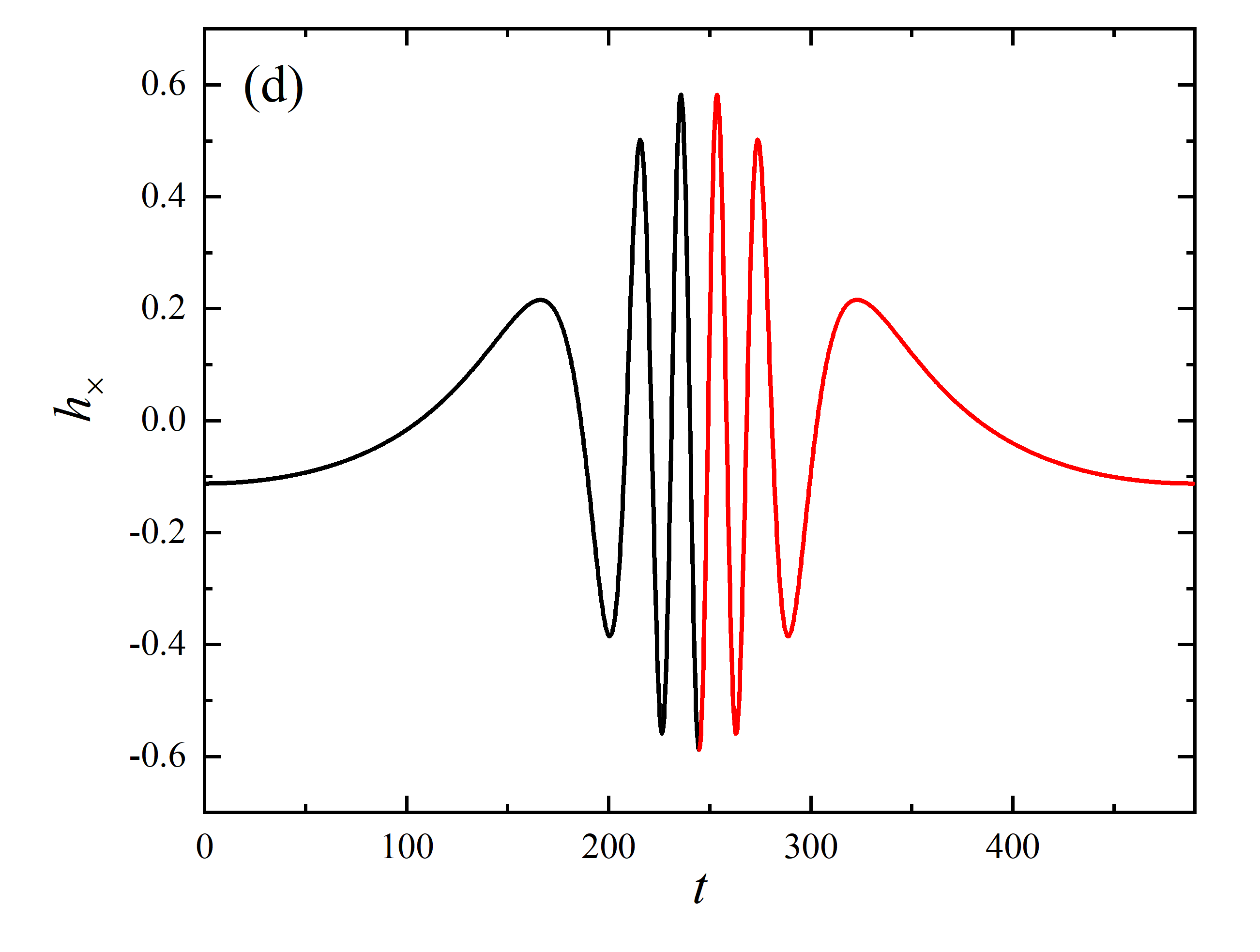}
\includegraphics[height=3.5cm]{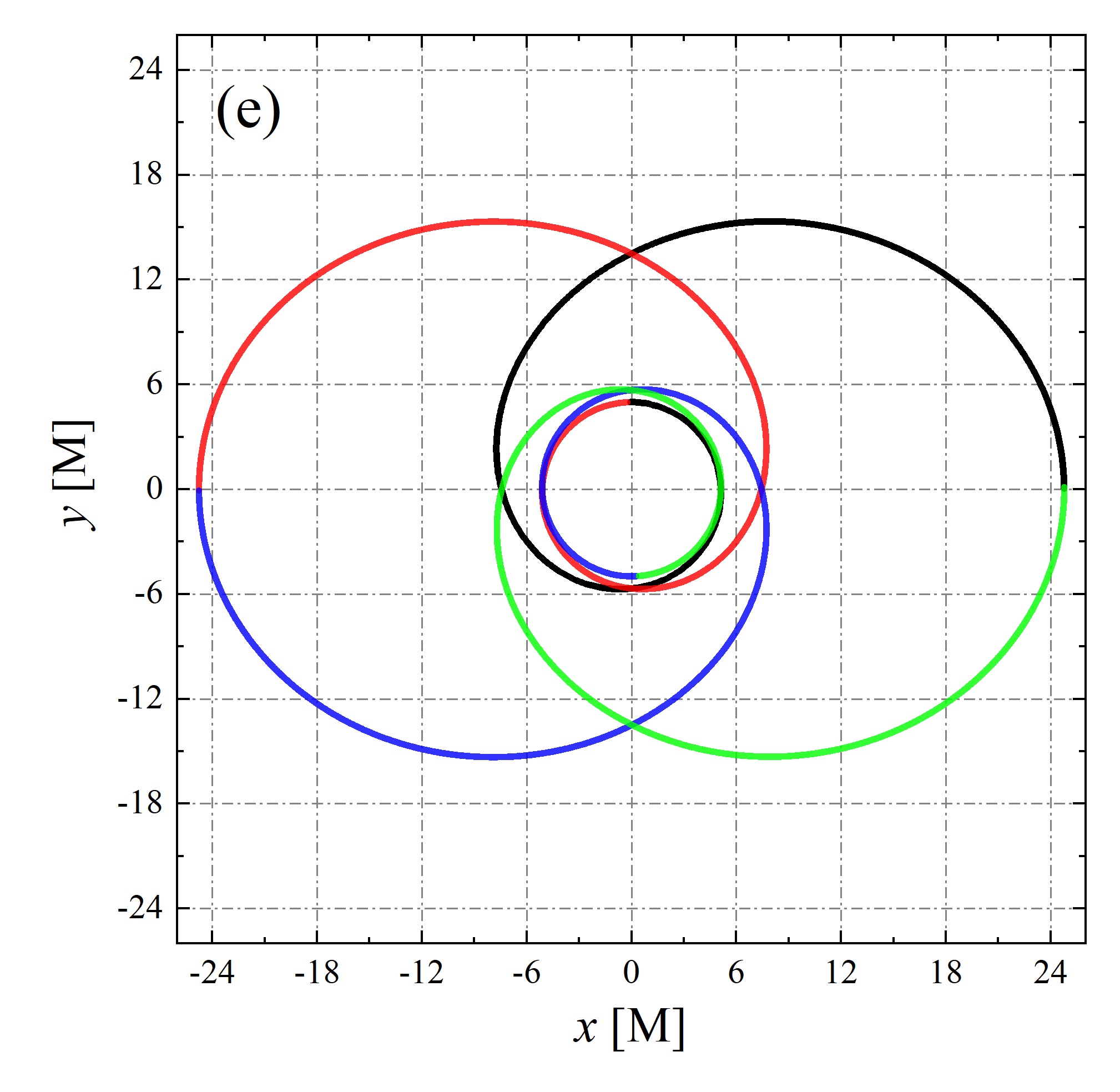}
\includegraphics[height=3.5cm]{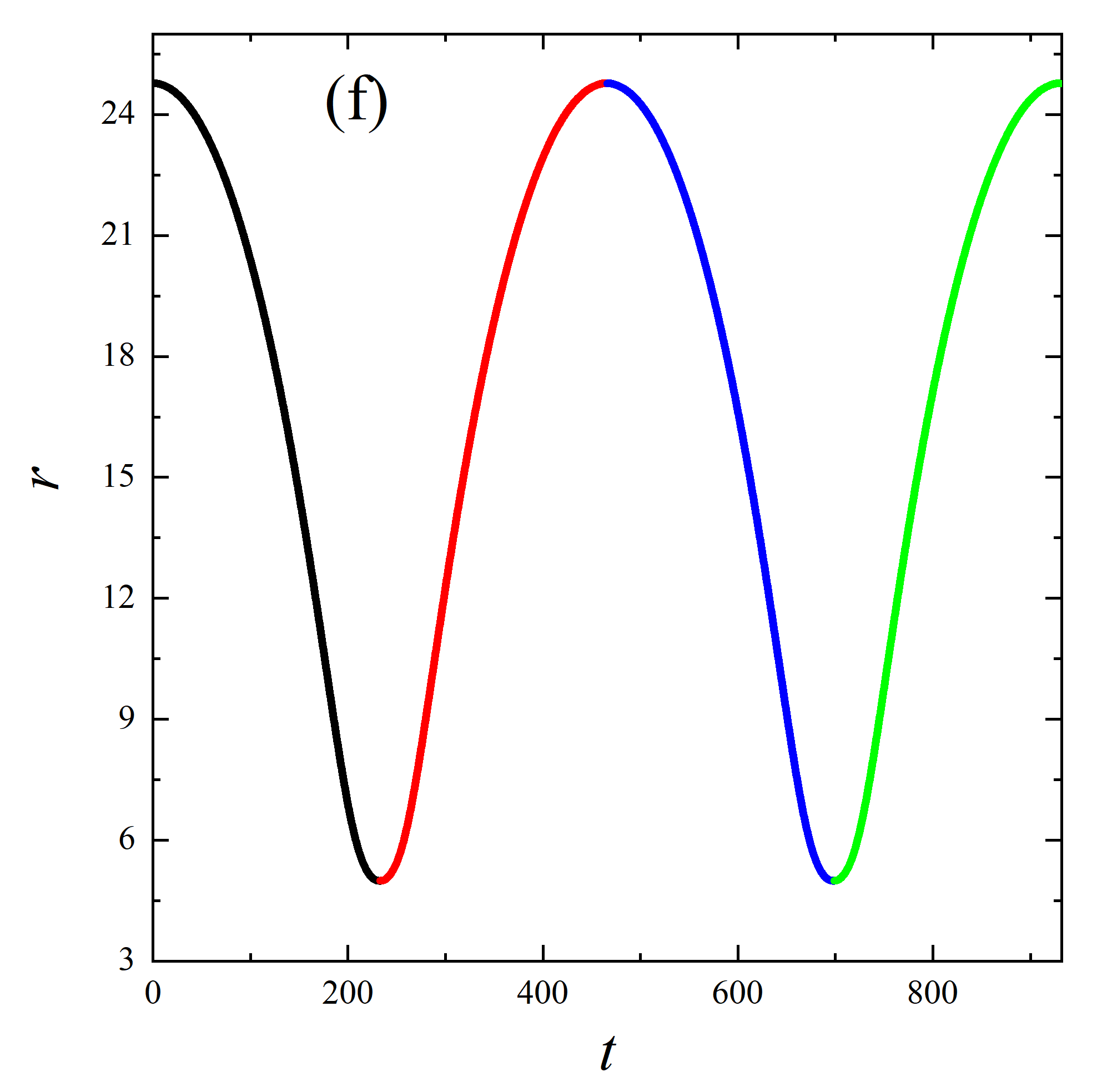}
\includegraphics[height=3.5cm]{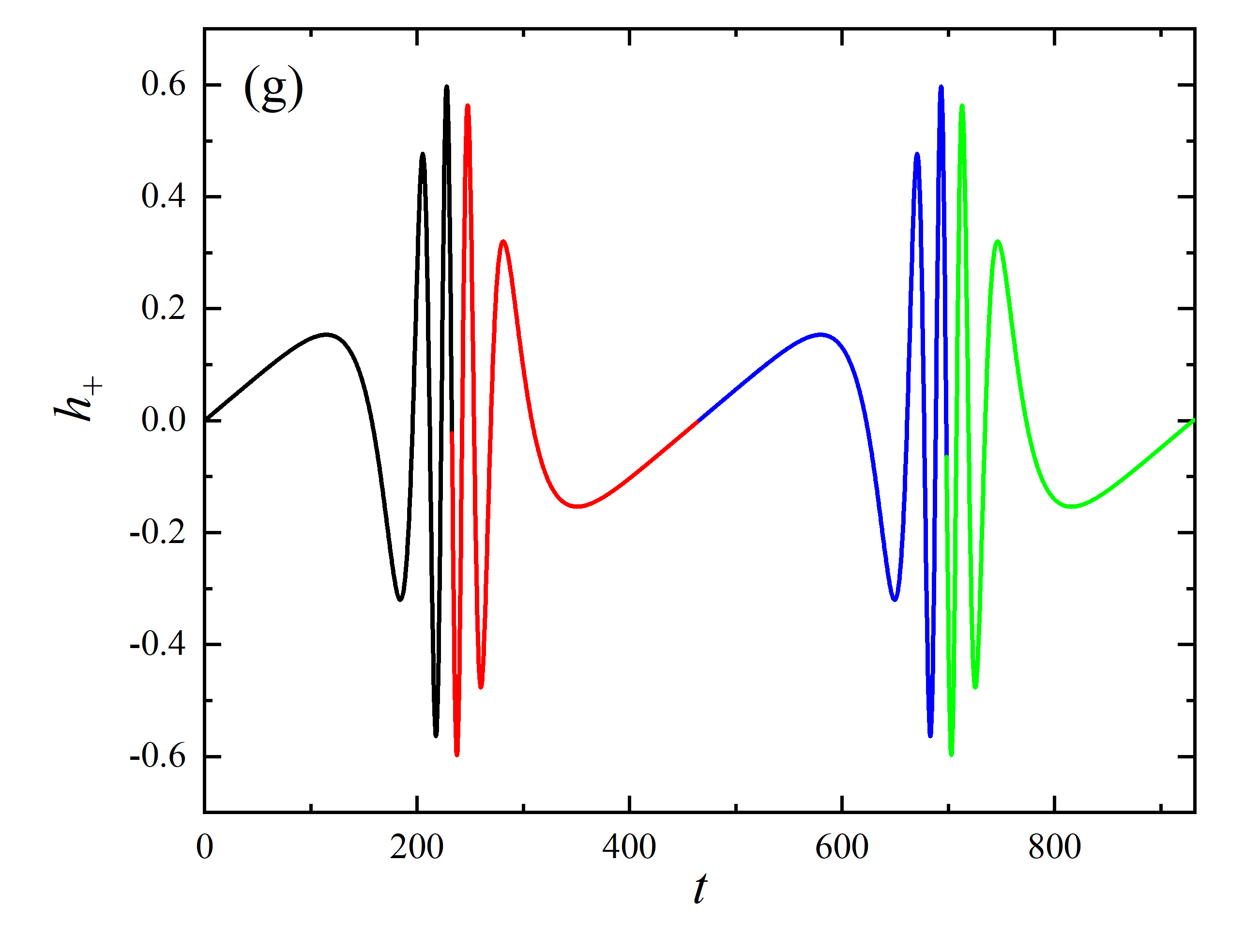}
\includegraphics[height=3.5cm]{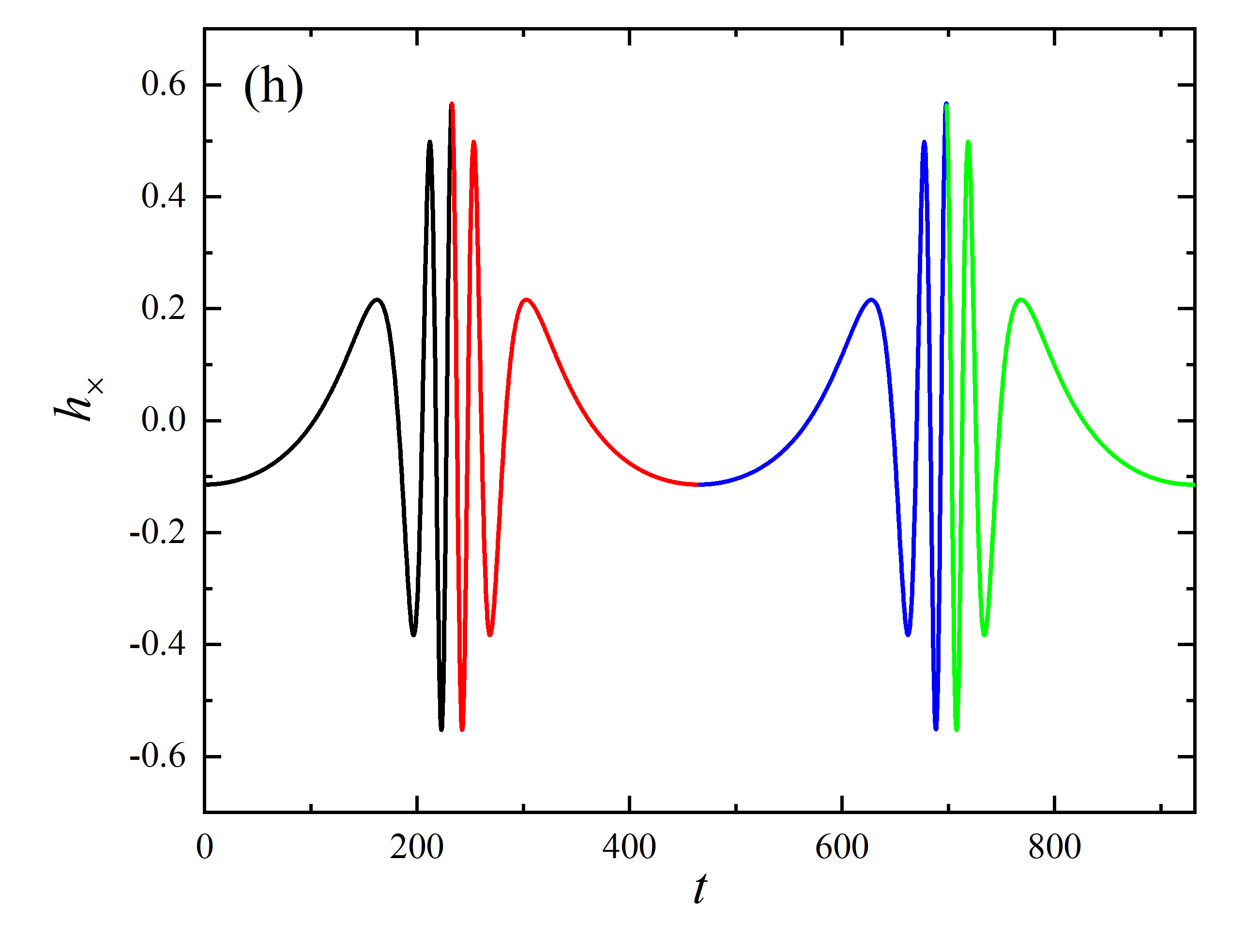}
\includegraphics[height=3.5cm]{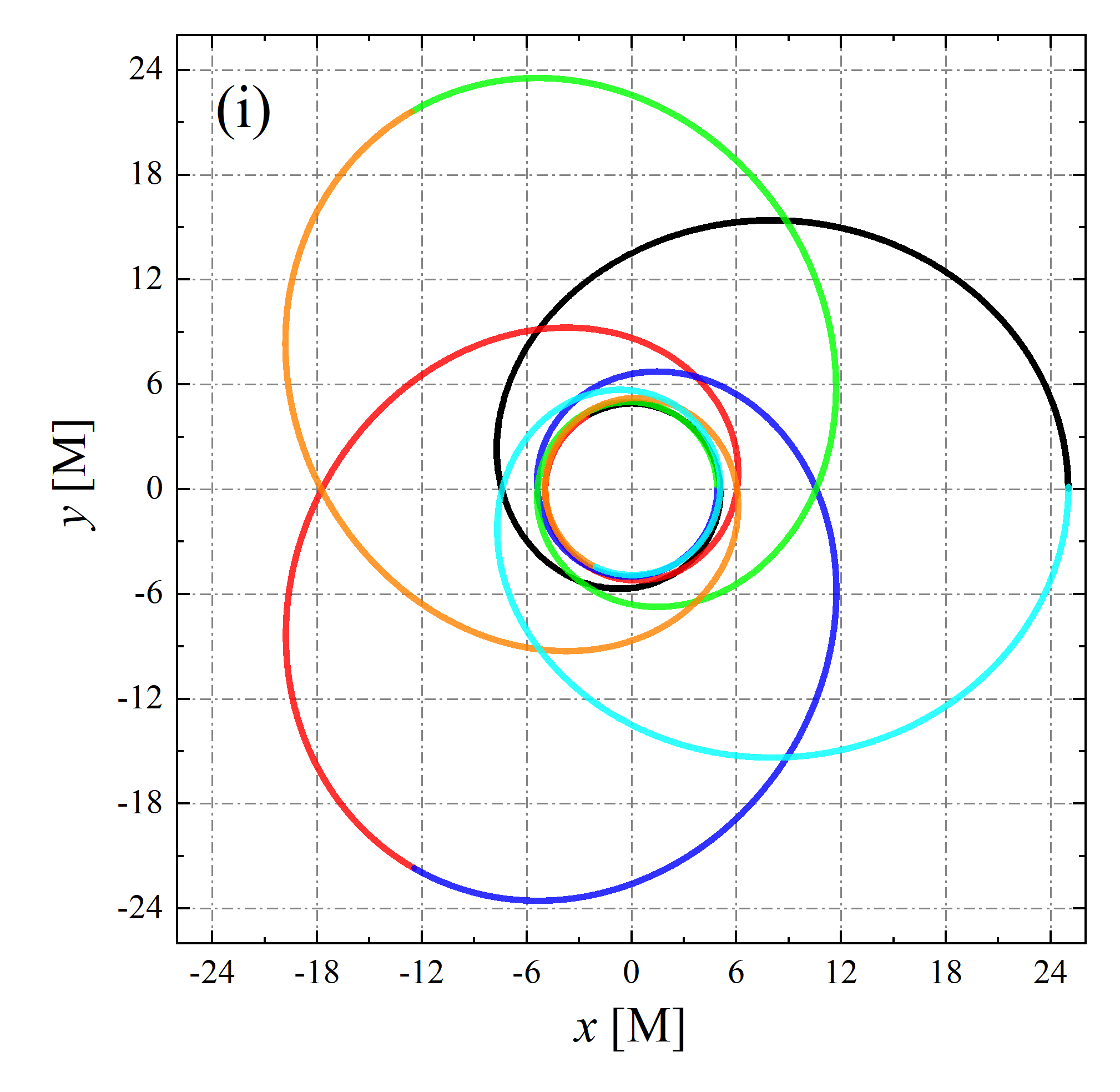}
\includegraphics[height=3.5cm]{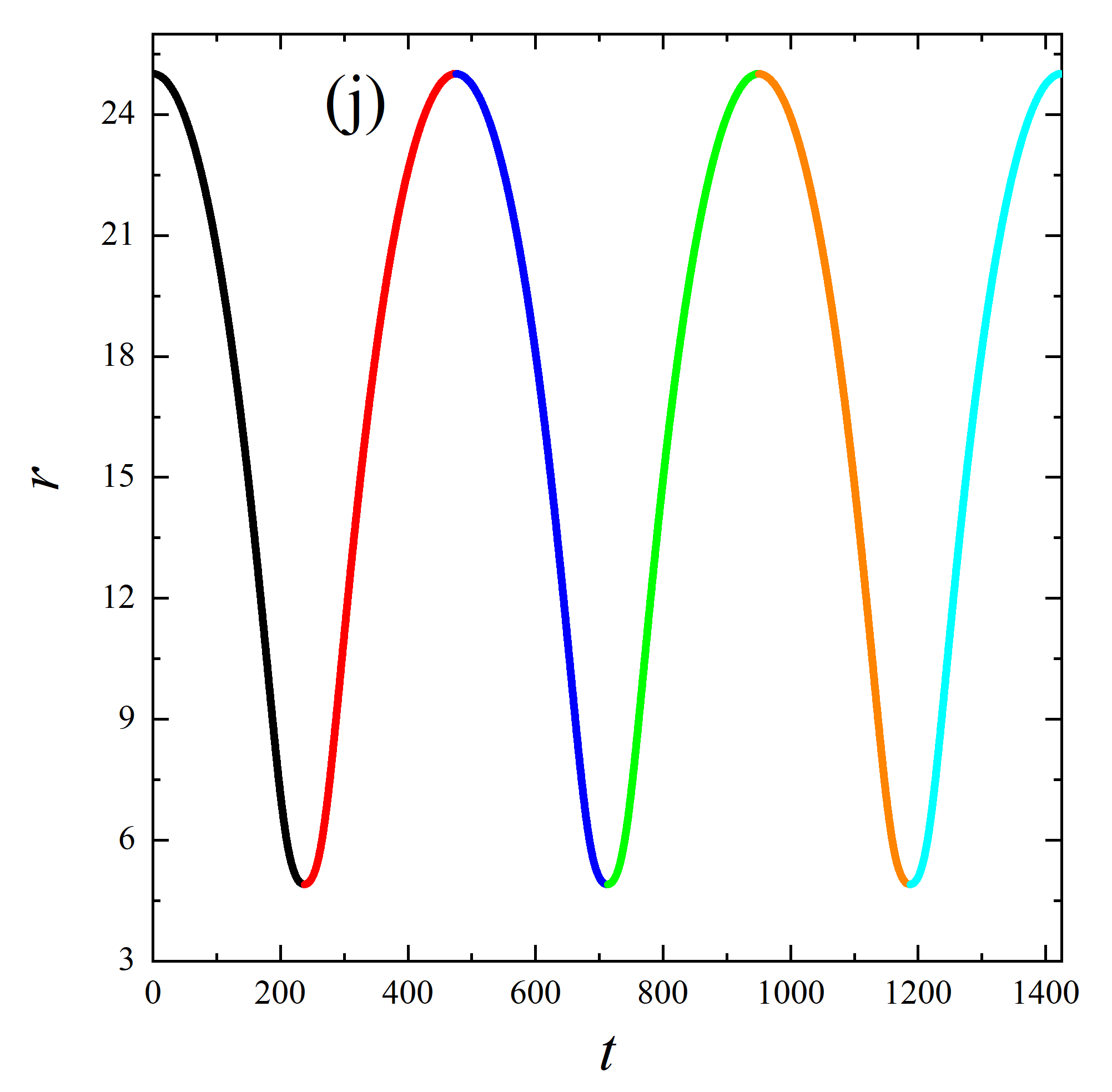}
\includegraphics[height=3.5cm]{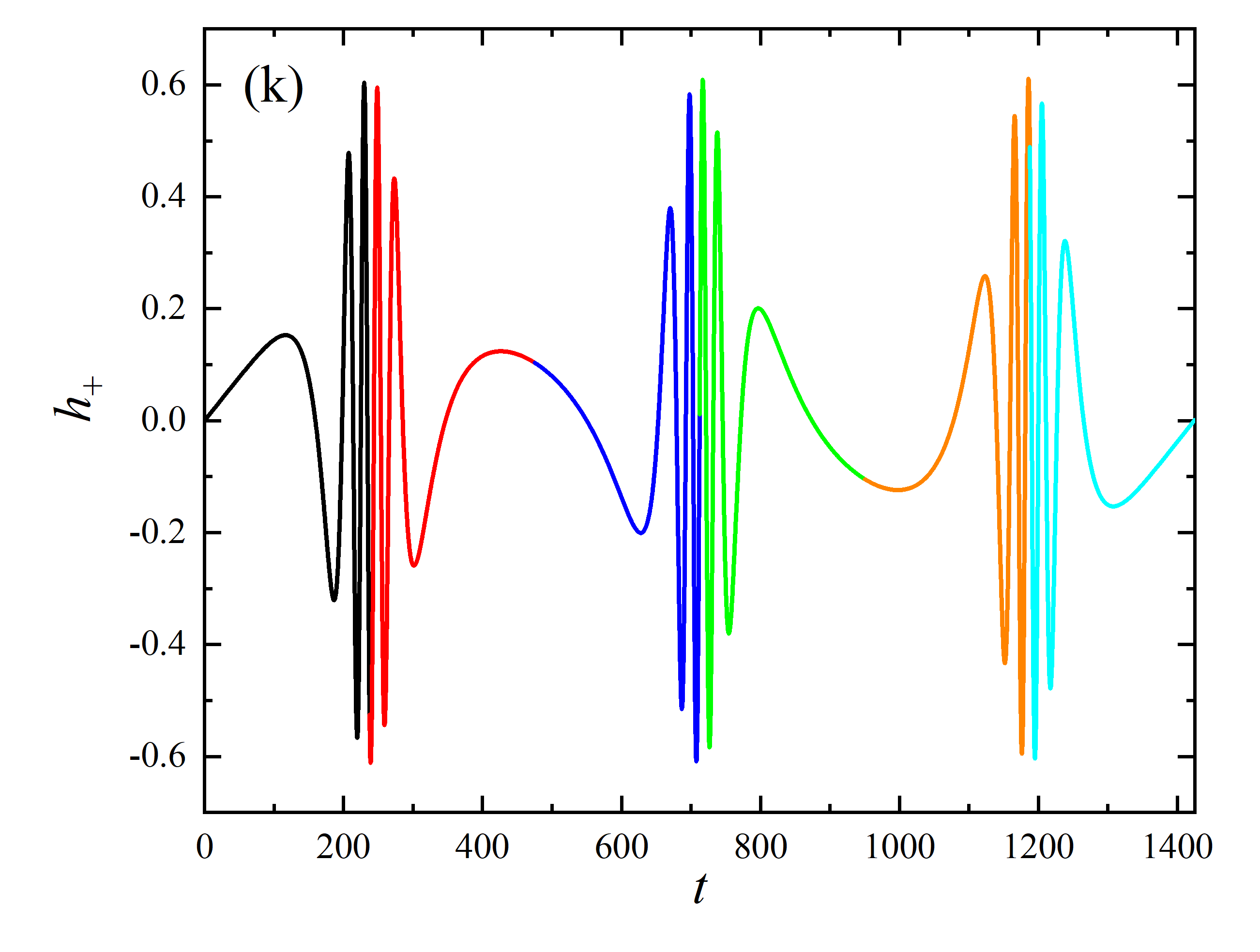}
\includegraphics[height=3.5cm]{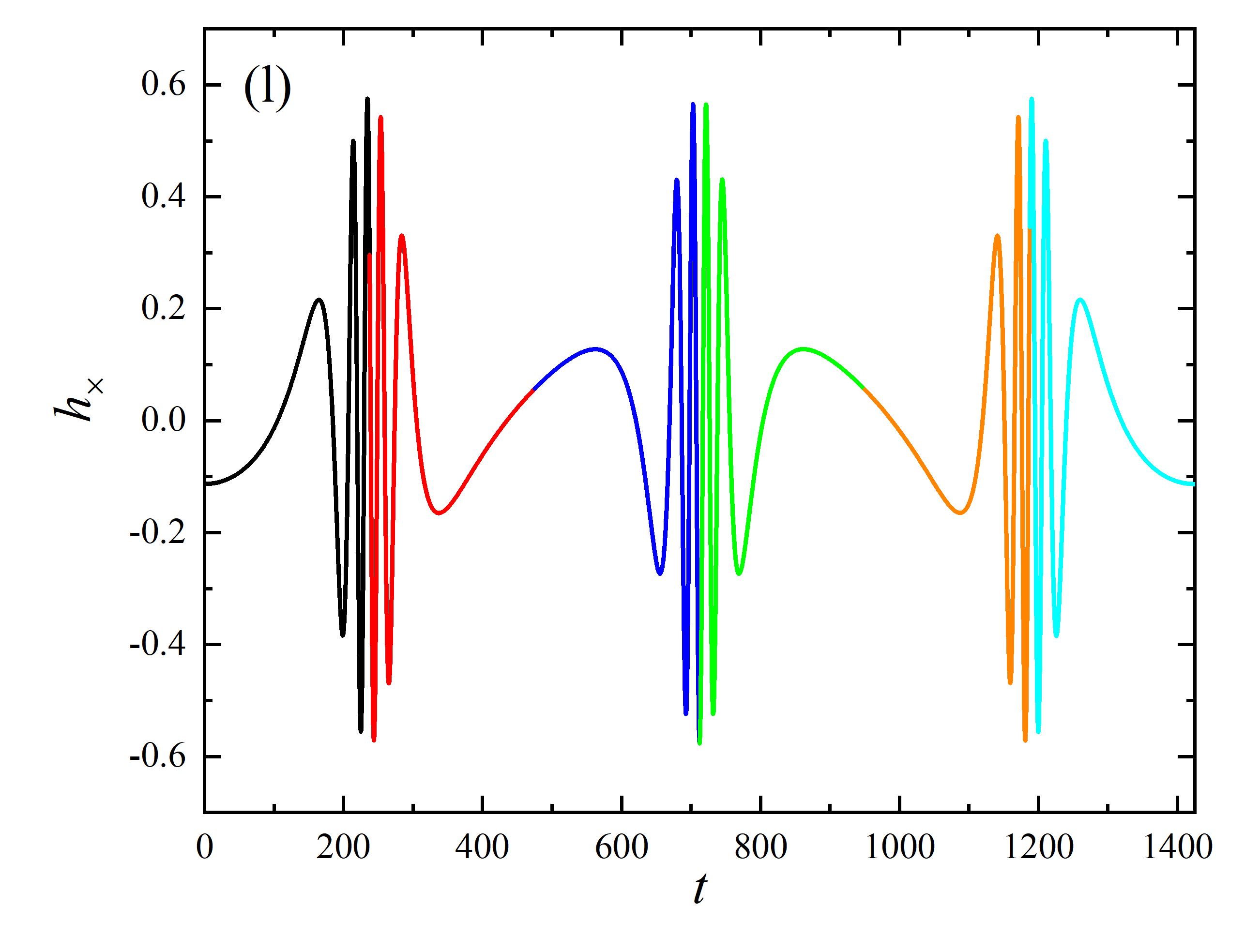}
\includegraphics[height=3.5cm]{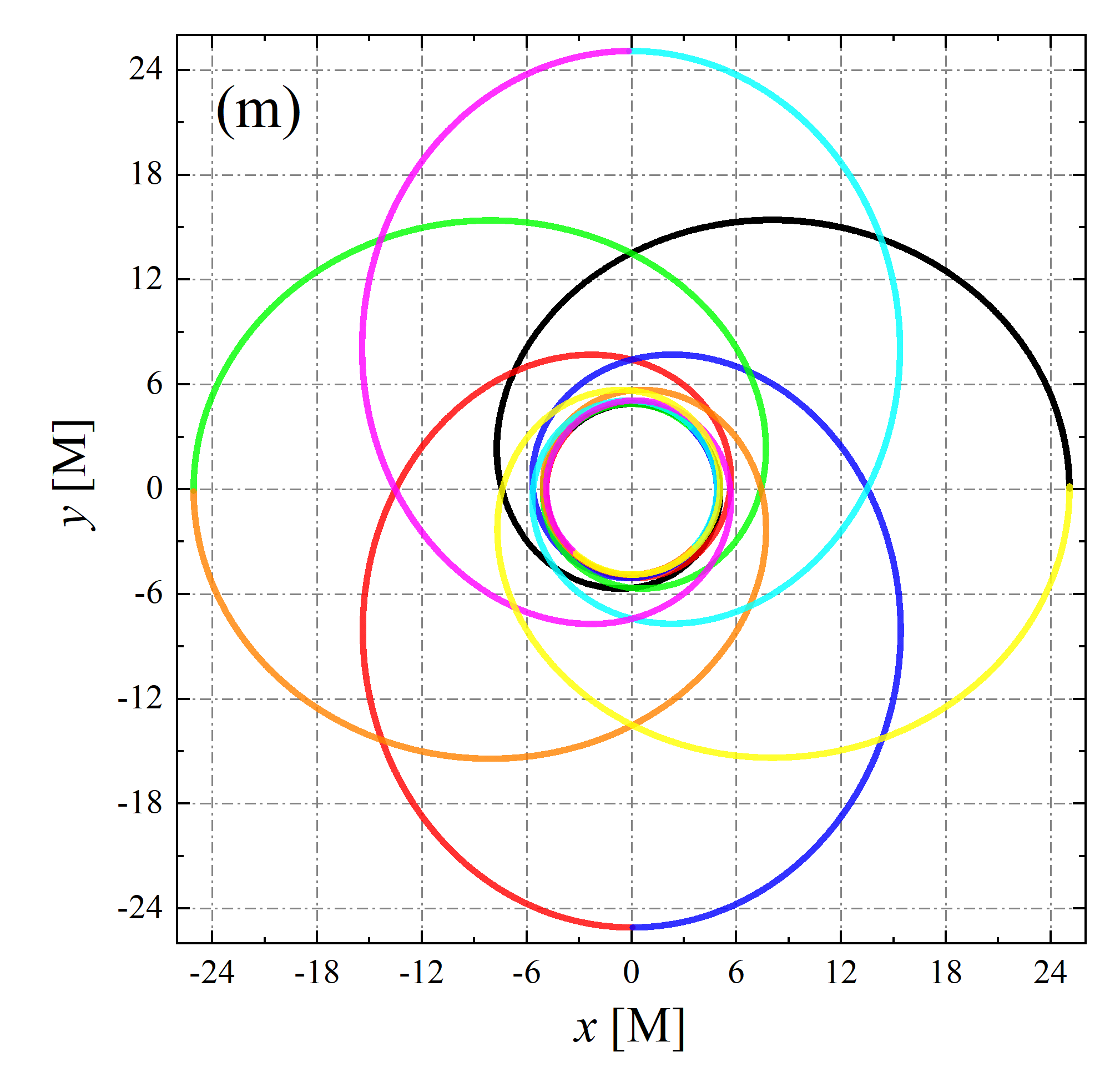}
\includegraphics[height=3.5cm]{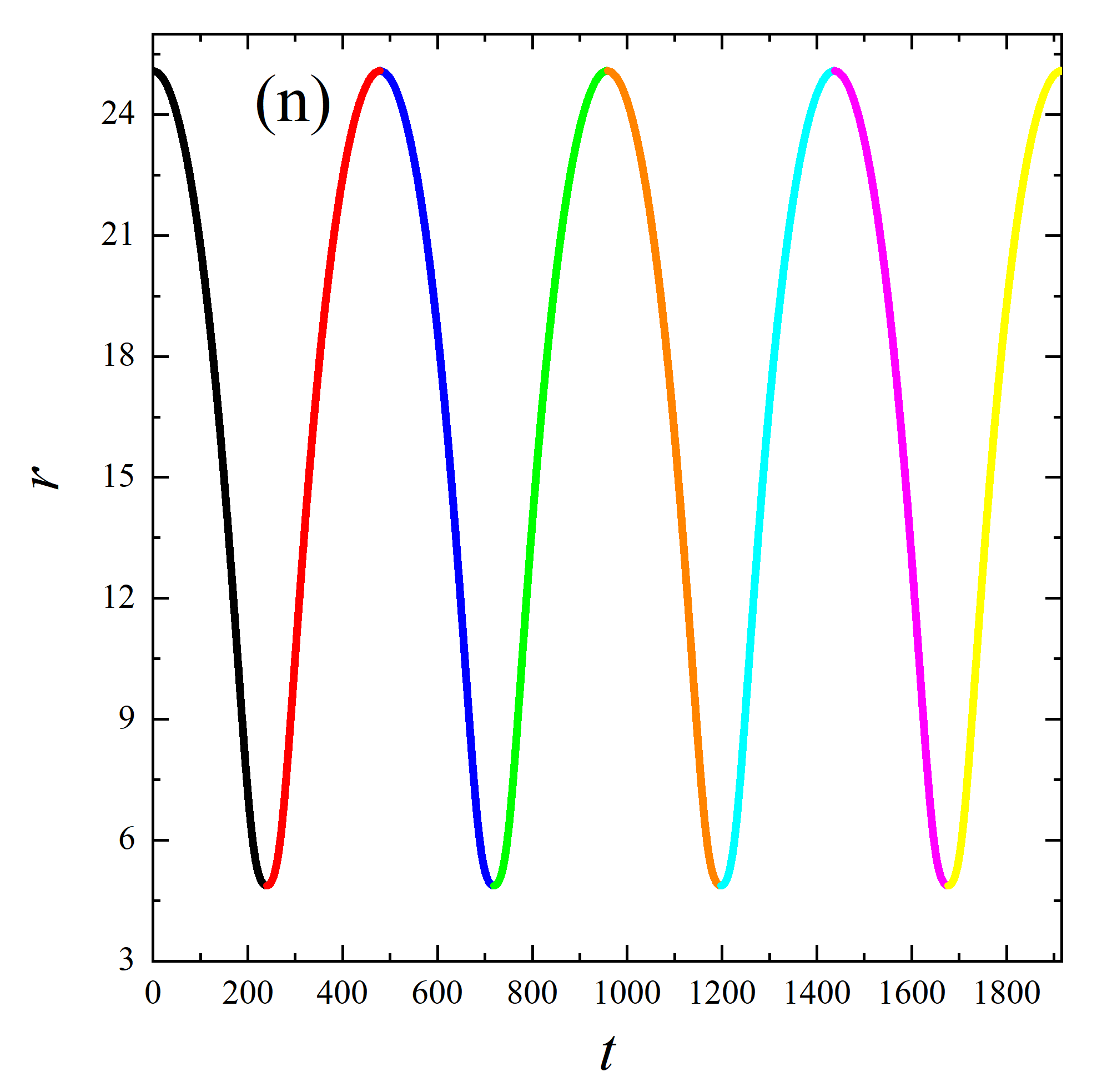}
\includegraphics[height=3.5cm]{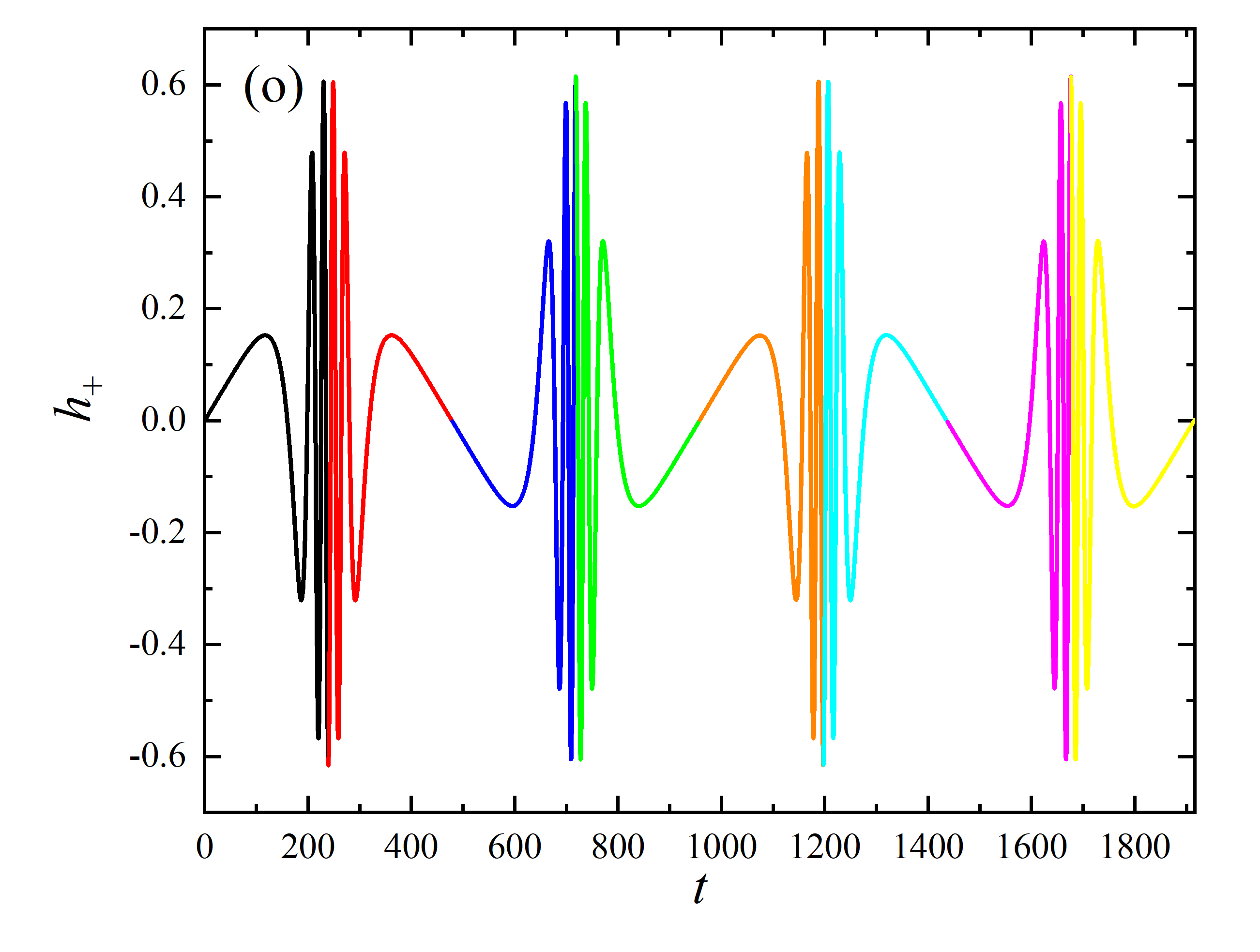}
\includegraphics[height=3.5cm]{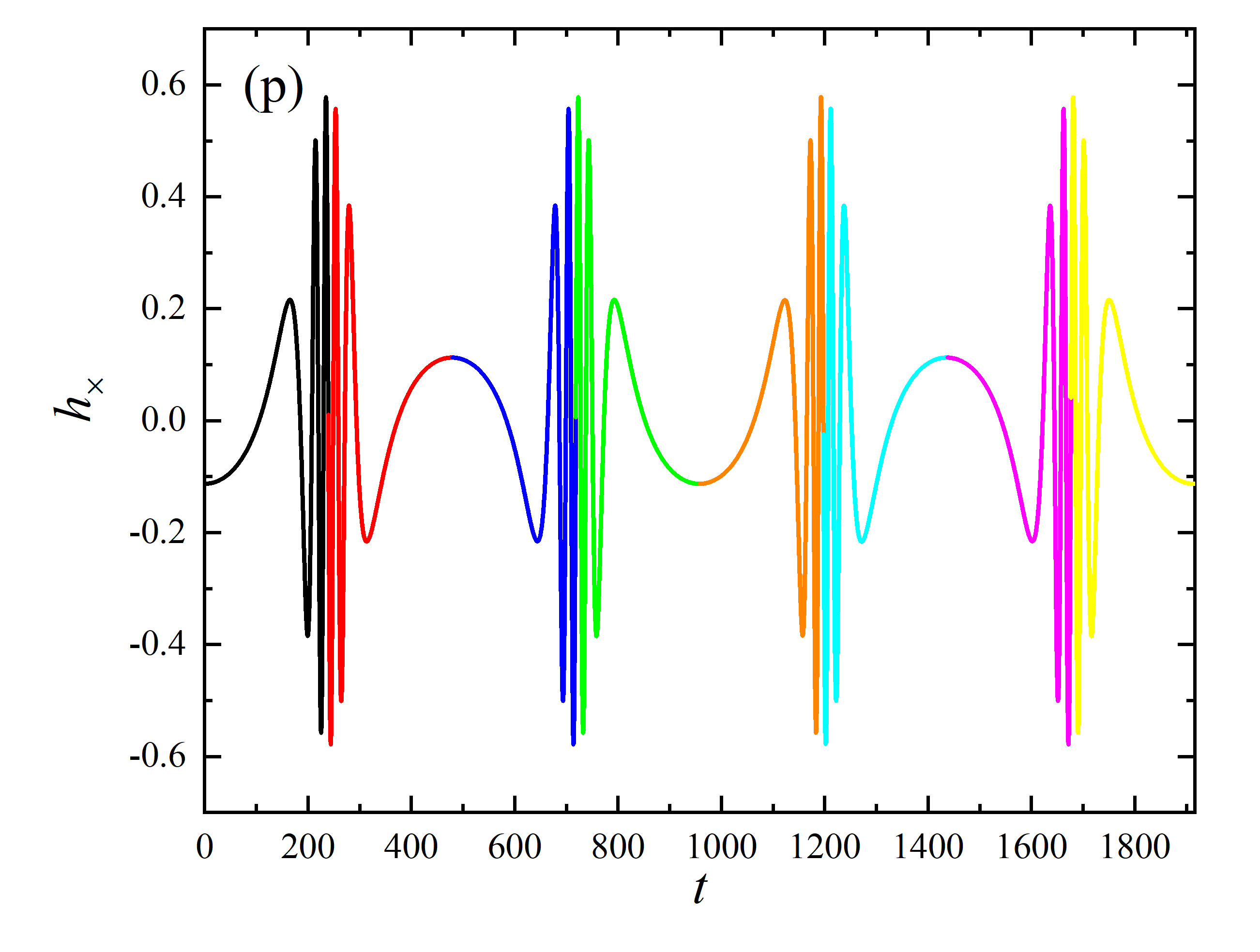}
\caption{Similar to Fig. 10, but with the dark matter halo parameters $(r_{\textrm{s}},\rho_{\textrm{s}})$ set to $(0.5,0.2)$.}}\label{fig11}
\end{figure*}

Next, it is necessary to discuss the potential observability of the modifications to the gravitational wave signals induced by the dark matter halo. We assume the central Schwarzschild black hole has a mass of $M= 4 \times 10^{6}$ $M_{\bigodot}$, where $M_{\bigodot}$ denotes the solar mass, and its luminosity distance from Earth is $D_{\textrm{L}} = 8000 $ pc. The small compact object undergoing the extreme mass ratio inspiral has a mass of $m= 100$ $M_{\bigodot}$. Thus, based on the gravitational wave formulas in Eqs. \eqref{h1} and \eqref{h2}, we obtain the amplitude scale factor $\eta GM/(D_{\textrm{L}}c^{2}) \sim 5.9887 \times 10^{-16}$ and the time scale factor $GM/c^{3} \sim 19.7088$. Using these data, we can convert the gravitational wave signals depicted in geometric units in Figs. 10 and 11 into the International System of Units (SI). After applying a Fast Fourier Transform (FFT), we obtain the frequency spectrum information of the gravitational wave signals, denoted as $\hat{h}_{+}\left(f\right)$ and $\hat{h}_{\times}\left(f\right)$, where $f$ represents the signal frequency. By selecting the four-leaf orbit $(z,w,v)=(4,1,3)$ from the last rows of Figs. 10 and 11 and acquiring the corresponding $\hat{h}_{+}\left(f\right)$ and $\hat{h}_{\times}\left(f\right)$, we apply the relation \cite{2000PhRvD..62l4021F}
\begin{equation}
h_{c}\left(f\right) = 2f\sqrt{|\hat{h}_{+}\left(f\right)|^{2}+|\hat{h}_{\times}\left(f\right)|^{2}},
\end{equation}
to compute the characteristic strain for the gravitational wave observation, as illustrated by the green and black curves in Fig. 12. We find that the frequencies of the gravitational wave signals for both orbits are concentrated around the mHz band, reaching a maximum magnitude of approximately $10^{-18}$. A non-negligible portion of the entire curve lies above the sensitivity curve of the Laser Interferometer Space Antenna (LISA), indicated by the solid red line. This demonstrates that our numerically simulated gravitational wave signals theoretically hold the potential to be captured by LISA. Furthermore, it implies a certain possibility of utilizing the phase delay and the fine details of high frequency oscillations to effectively extract orbital information and infer underlying spacetime parameters. It must be emphasized, however, that actual gravitational wave observations do not merely involve a simple comparison between the theoretical characteristic strain and the sensitivity curve. In practice, multiple critical factors such as the signal to noise ratio, the observation time, the detector response, the astrophysical background confusion, parameter degeneracy, and the accuracy of waveform templates must be thoroughly considered. Therefore, our current conclusions remain largely phenomenological, exploring the observational prospects of these signals solely from a theoretical perspective.
\begin{figure*}
\center{
\includegraphics[width=5cm]{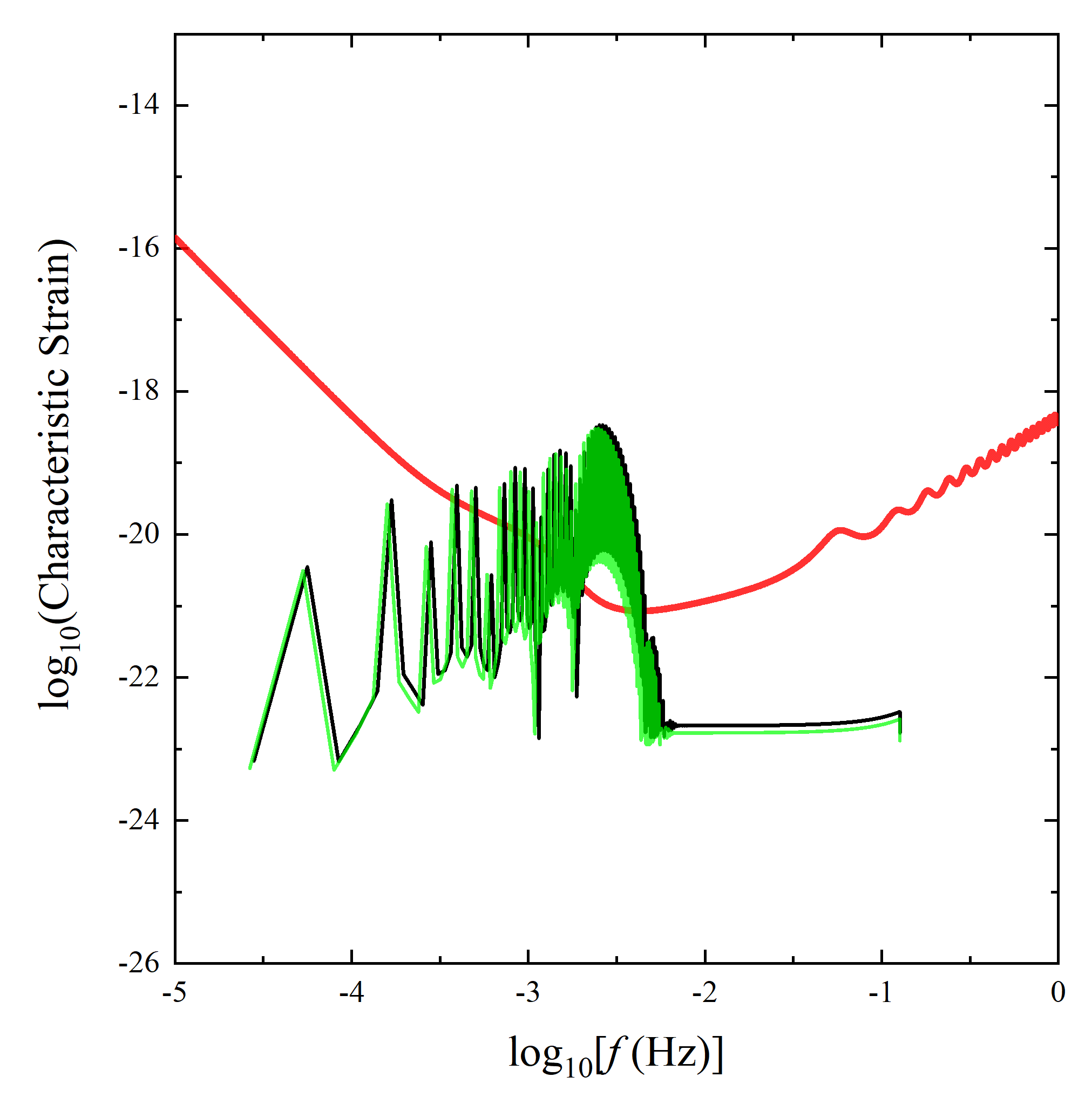}
\caption{The characteristic strain of the four-leaf orbit $(z,w,v)=(4,1,3)$ as a function of frequency, plotted on a logarithmic scale under varying dark matter halo parameters. Here, the parameters are fixed at $\varepsilon=0.5$ and $\rho_{\textrm{s}}=0.2$, with the black and green lines corresponding to $r_{\textrm{s}}=0.2$ and $r_{s}=0.5$, respectively. The red line represents the sensitivity curve of LISA \cite{Robson:2018ifk}. A gravitational wave signal is preliminarily considered to have the potential for detection when its characteristic strain exceeds the sensitivity curve of the given detector.}}\label{fig12}
\end{figure*}

\begin{figure*}
\center{
\includegraphics[width=8cm]{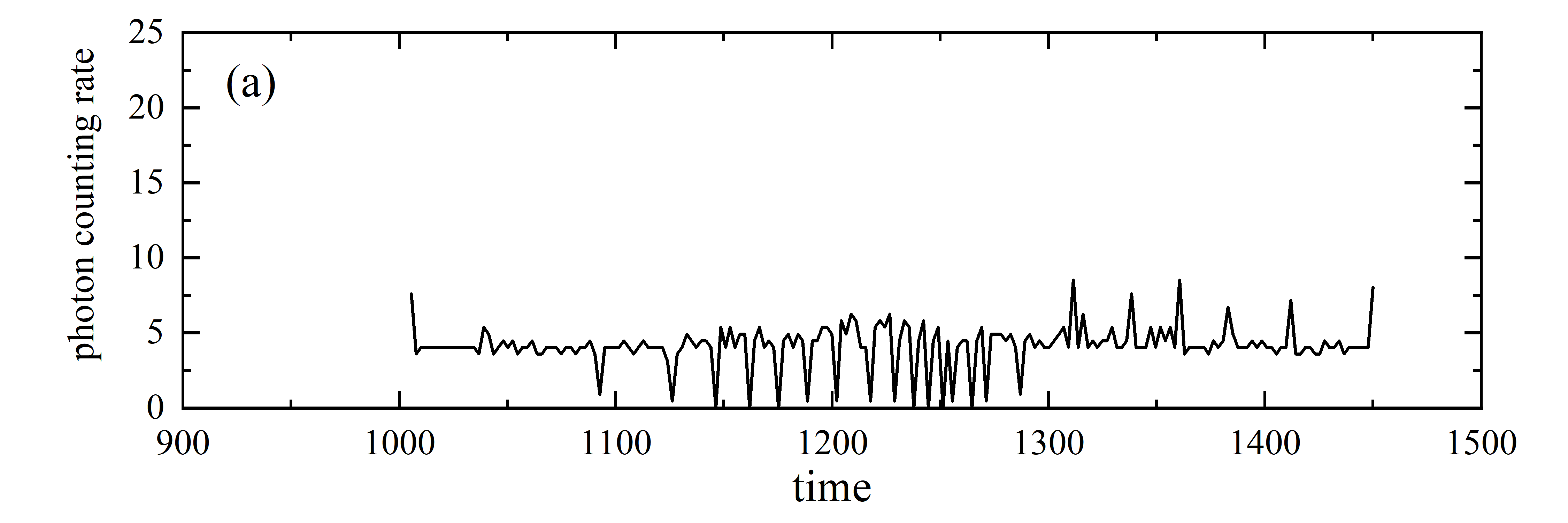}
\includegraphics[width=8cm]{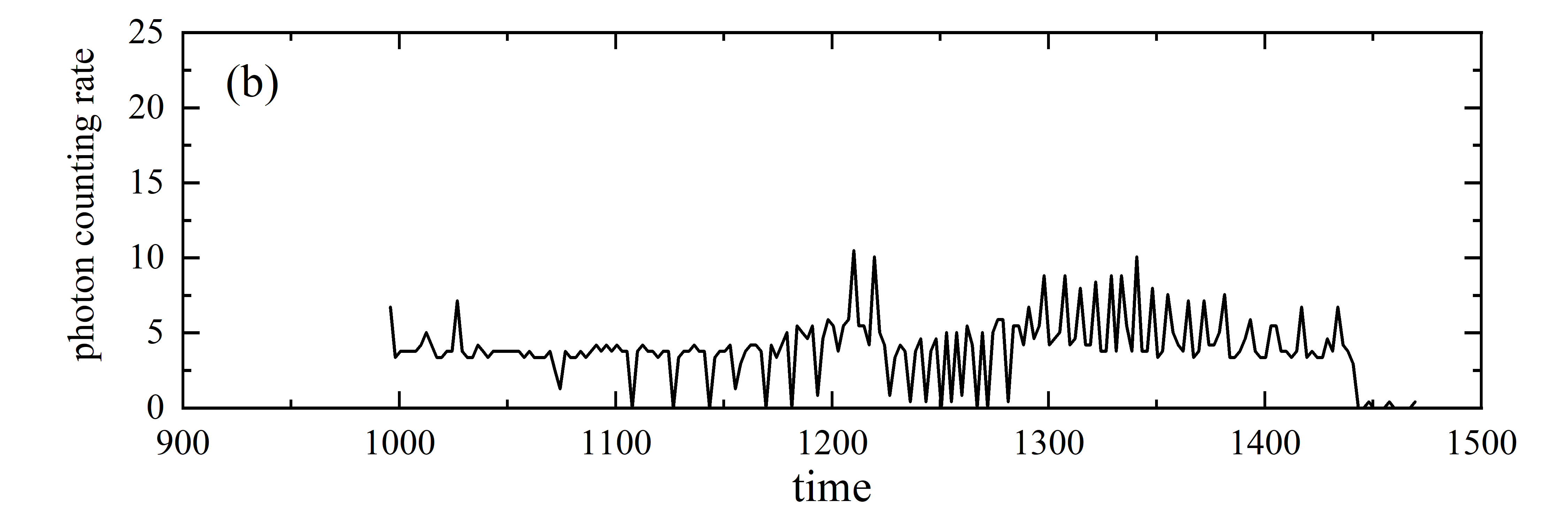}
\includegraphics[width=8cm]{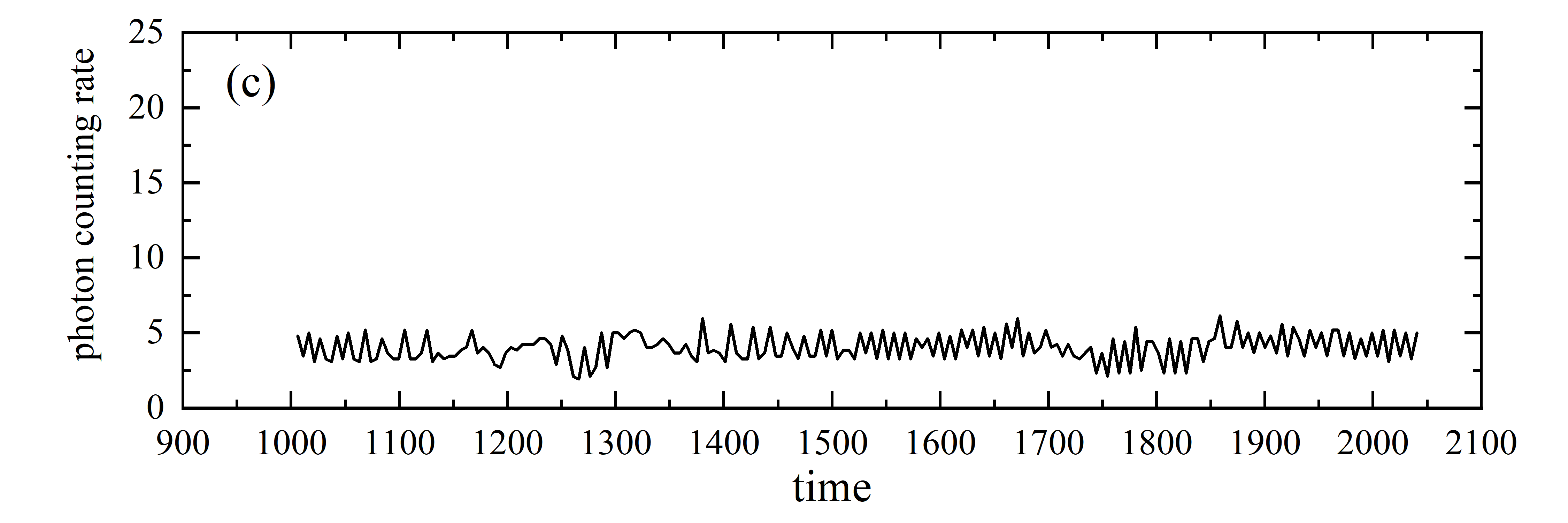}
\includegraphics[width=8cm]{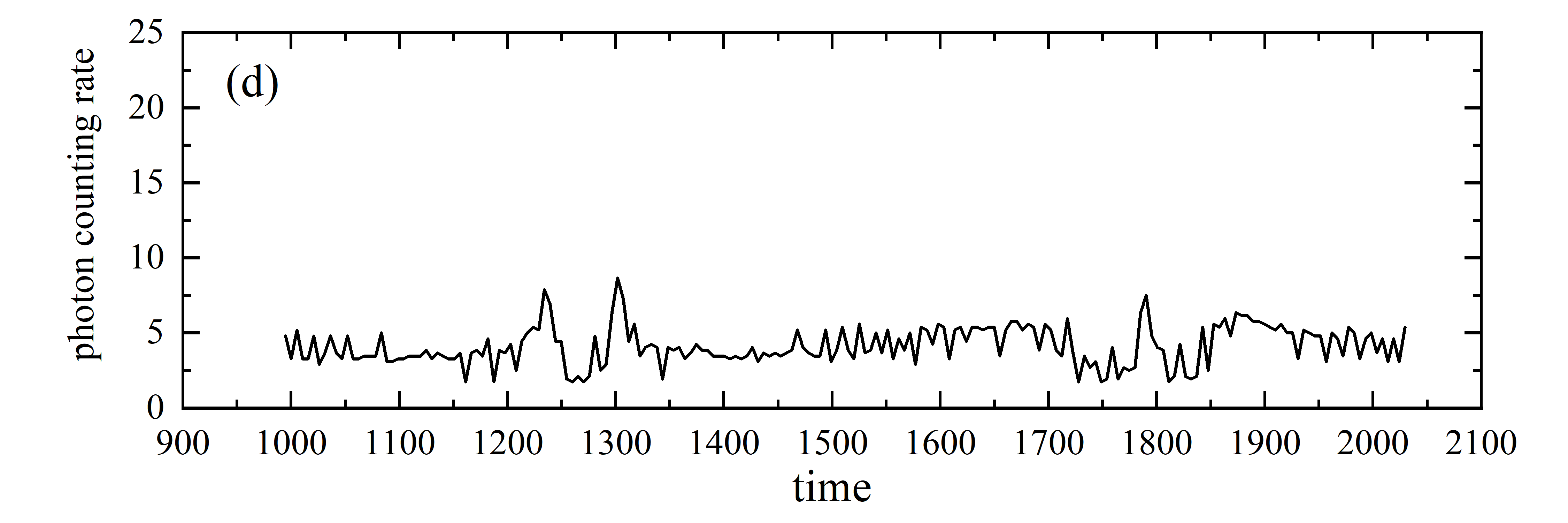}
\includegraphics[width=8cm]{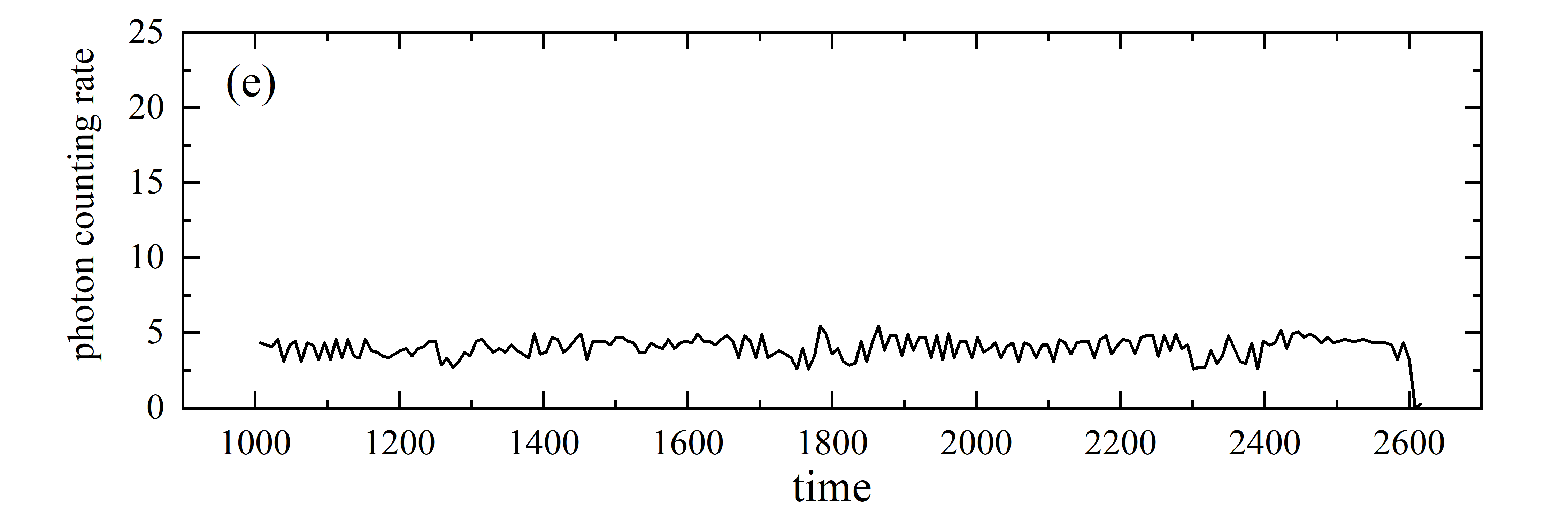}
\includegraphics[width=8cm]{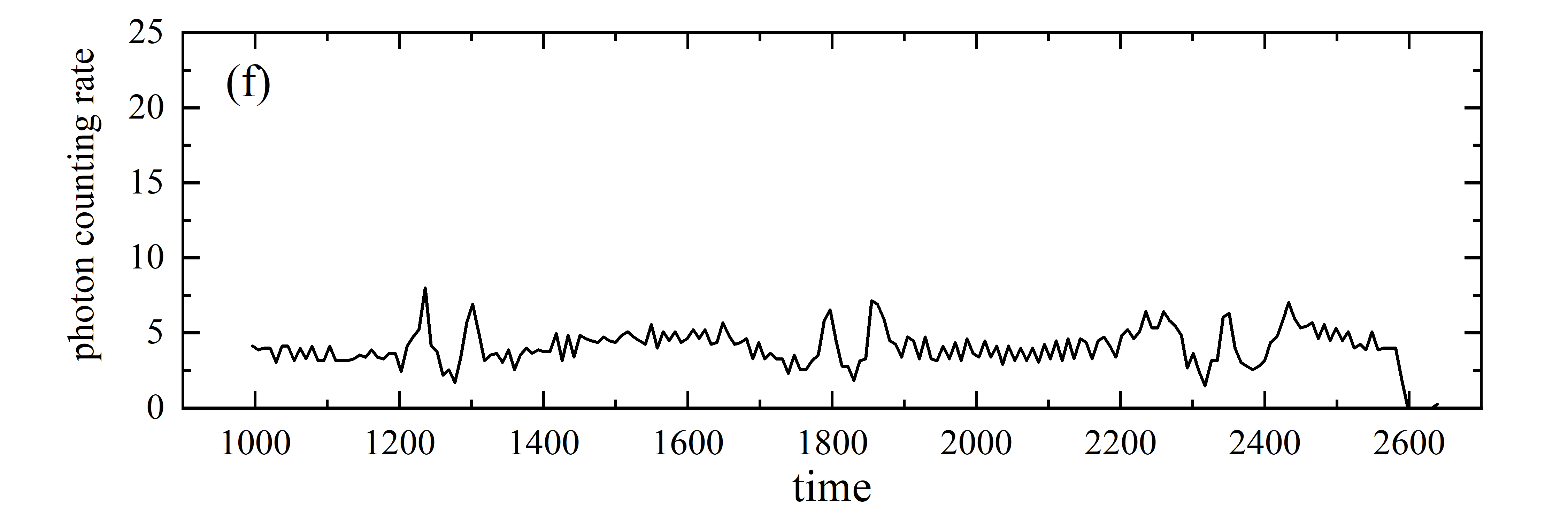}
\includegraphics[width=8cm]{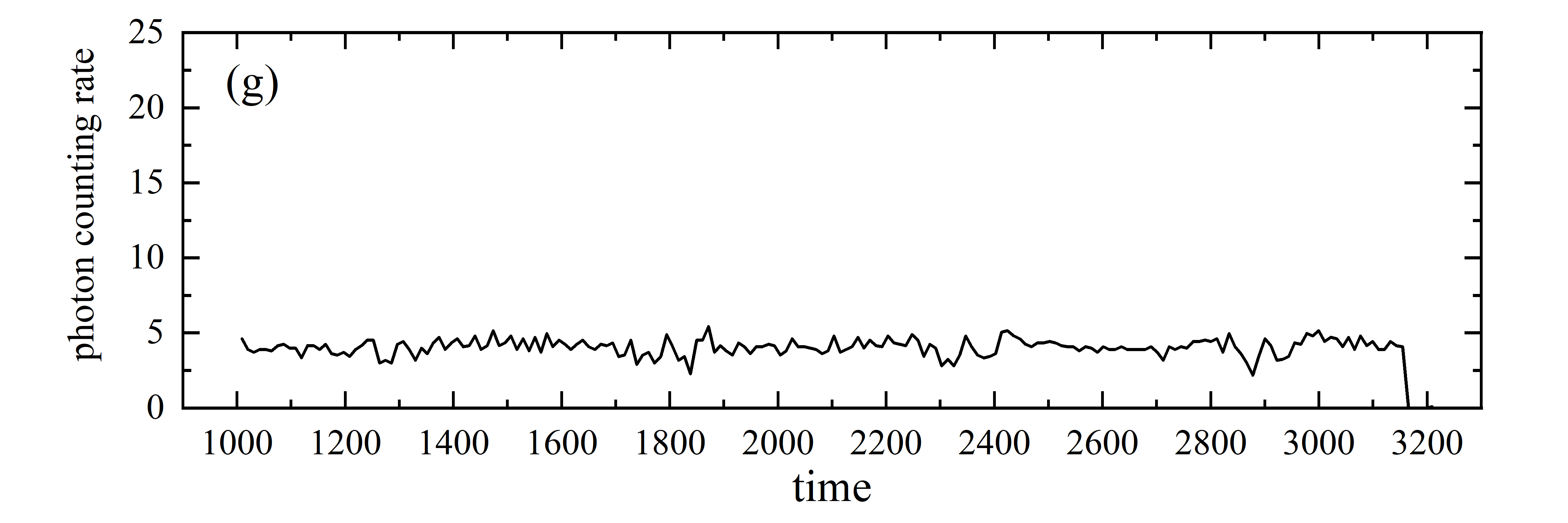}
\includegraphics[width=8cm]{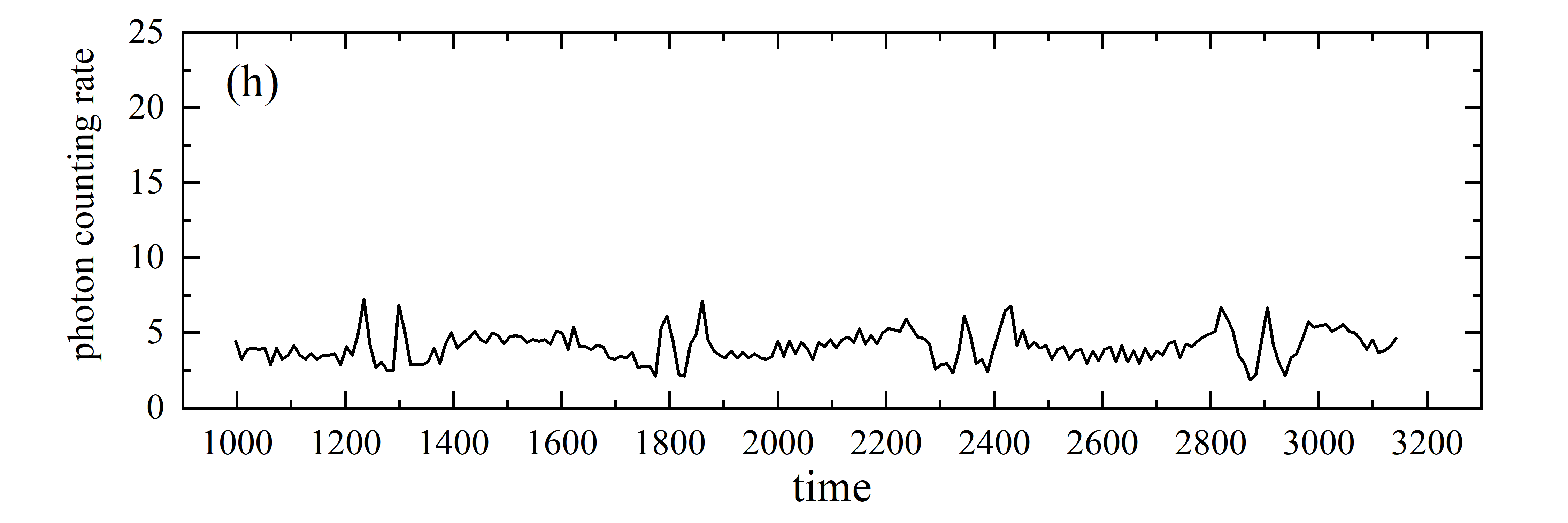}
\includegraphics[width=8cm]{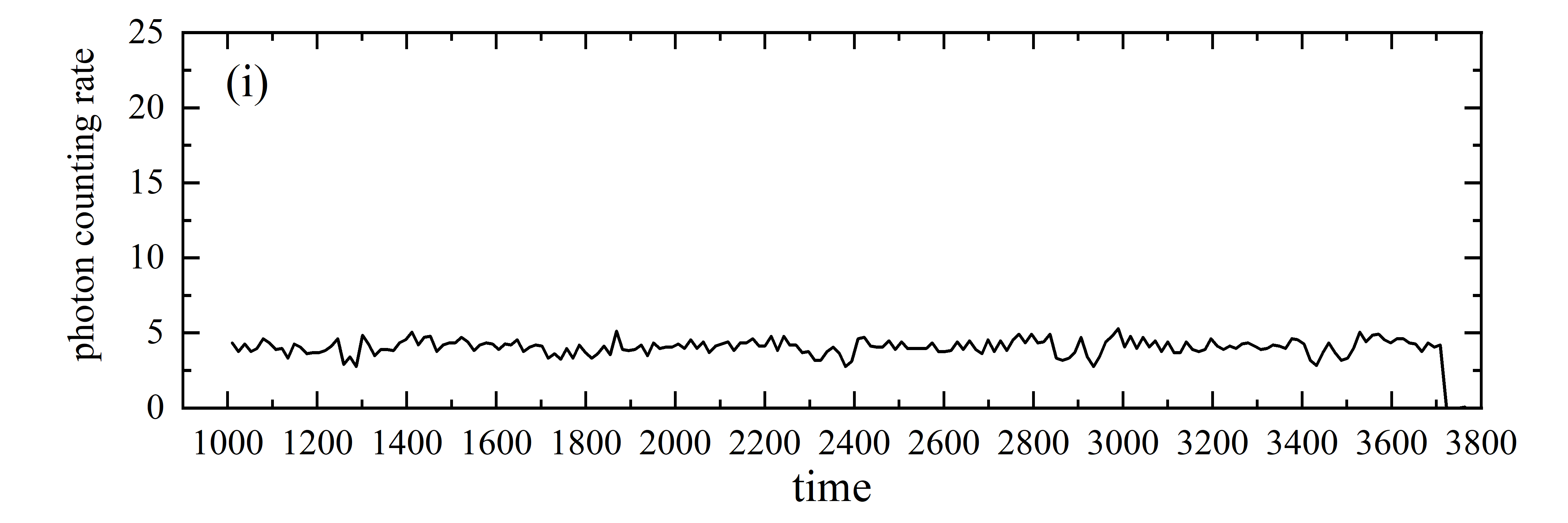}
\includegraphics[width=8cm]{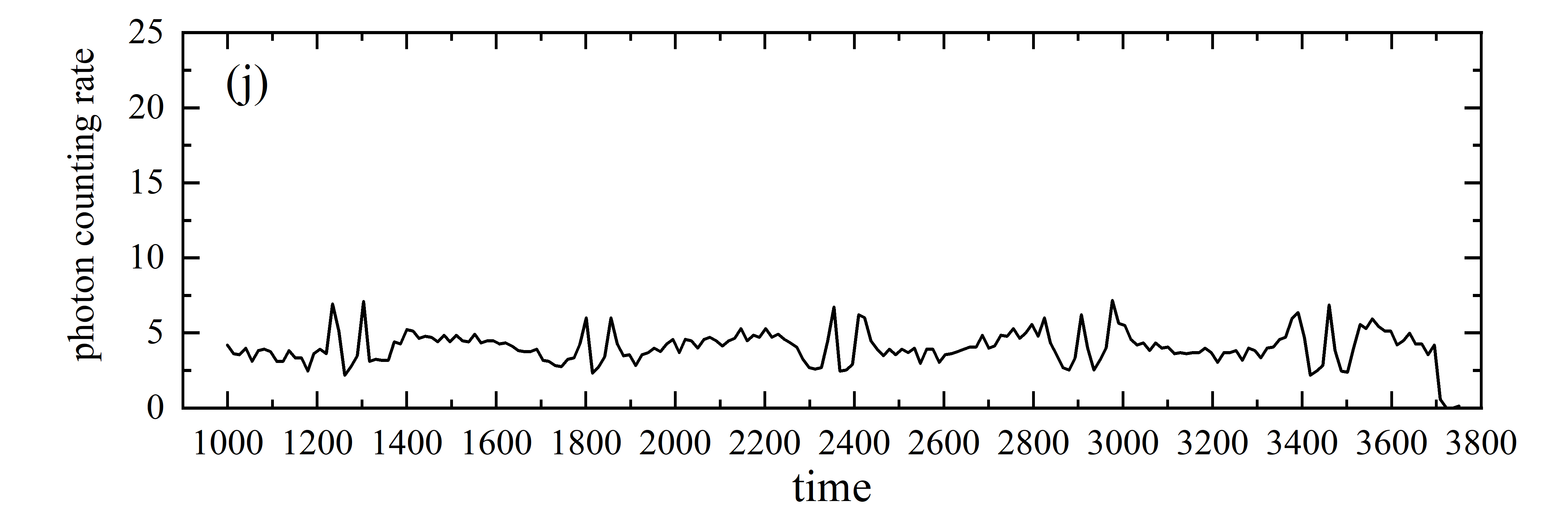}
\includegraphics[width=8cm]{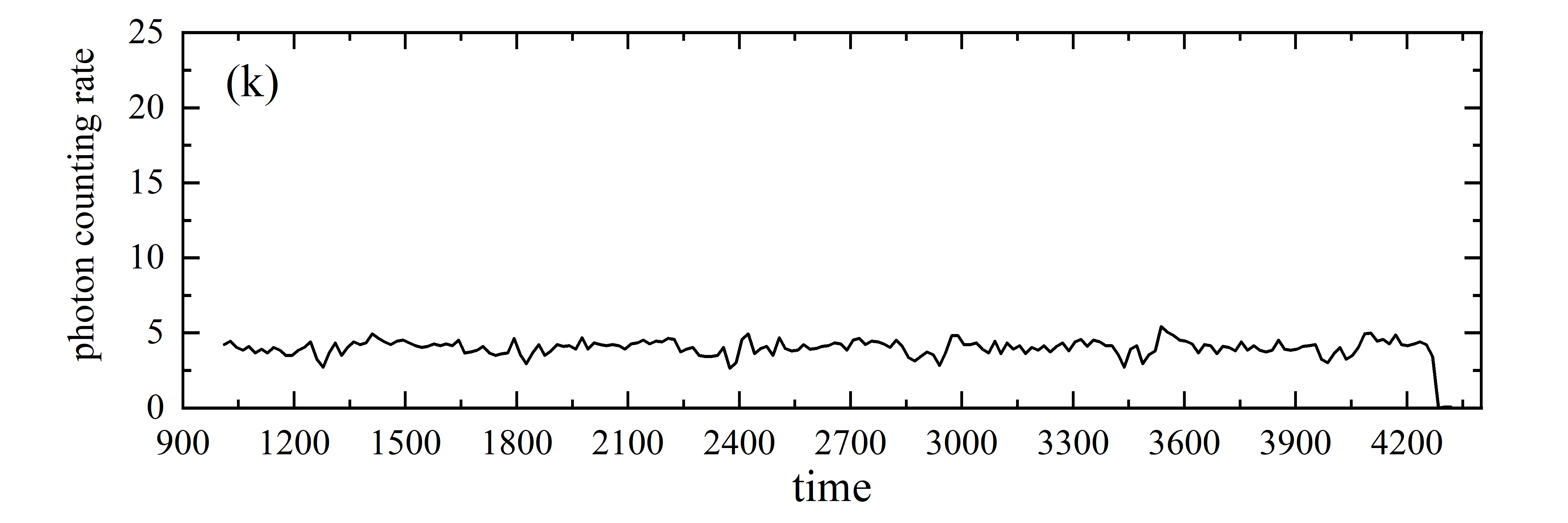}
\includegraphics[width=8cm]{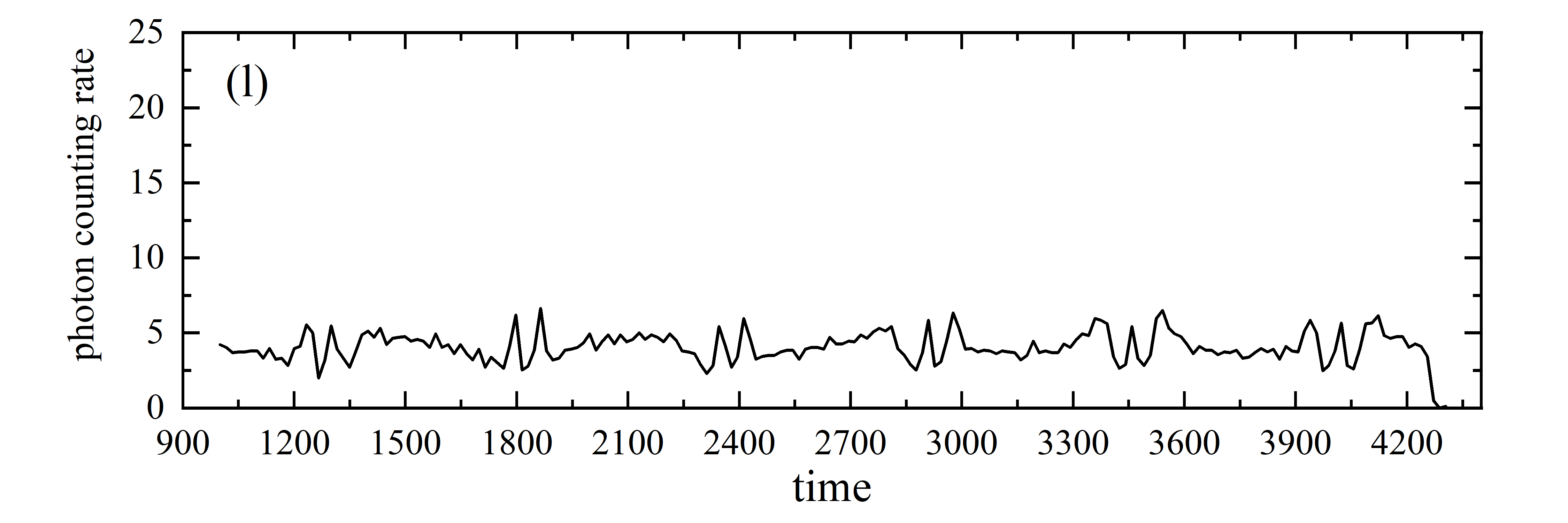}
\caption{Light curves of closed orbits with varying numbers of leaves at observation inclination angles of $17^{\circ}$ (the left column) and $50^{\circ}$ (the right column). From top to bottom, the number of orbital leaves $z$ increases from $1$ to $6$. For all cases, the dark matter halo parameters are fixed at $(r_{\textrm{s}},\rho_{\textrm{s}})$ $=$ $(0.2,0.2)$, $\varepsilon=0.5$.}}\label{fig13}
\end{figure*}

Finally, we examine the imprints of the orbital leaf number on the light curves. Using OCTOPUS, we simulated the light curves of strictly closed orbits with $z=1$, $2$, $3$, $4$, $5$, and $6$ under various observation inclination angles, as illustrated in Fig. 13. It can be observed that at an observation inclination angle of $17^{\circ}$ (the left column), the photon count rate primarily oscillates around a specific baseline value without any distinct periodicity. Conversely, when the inclination angle is increased to $50^{\circ}$ (the right column), the fluctuation amplitude of the light curves expands significantly, allowing for the identification of clear peaks in localized regions. The underlying cause of this phenomenon is that a larger inclination angle aligns the line of sight closer to the orbital plane. Consequently, the time delay induced by gravitational lensing causes a portion of the photons to superimpose onto subsequent peak positions, leading to a surge in the peak values. Simultaneously, as $z$ increases, the light curves begin to exhibit roughly periodic characteristics. Nevertheless, the results presented in Fig. 13 remain insufficient to establish a clear connection between the number of orbital leaves and the specific morphological features of the light curves.
\begin{figure*}
\center{
\includegraphics[width=8cm]{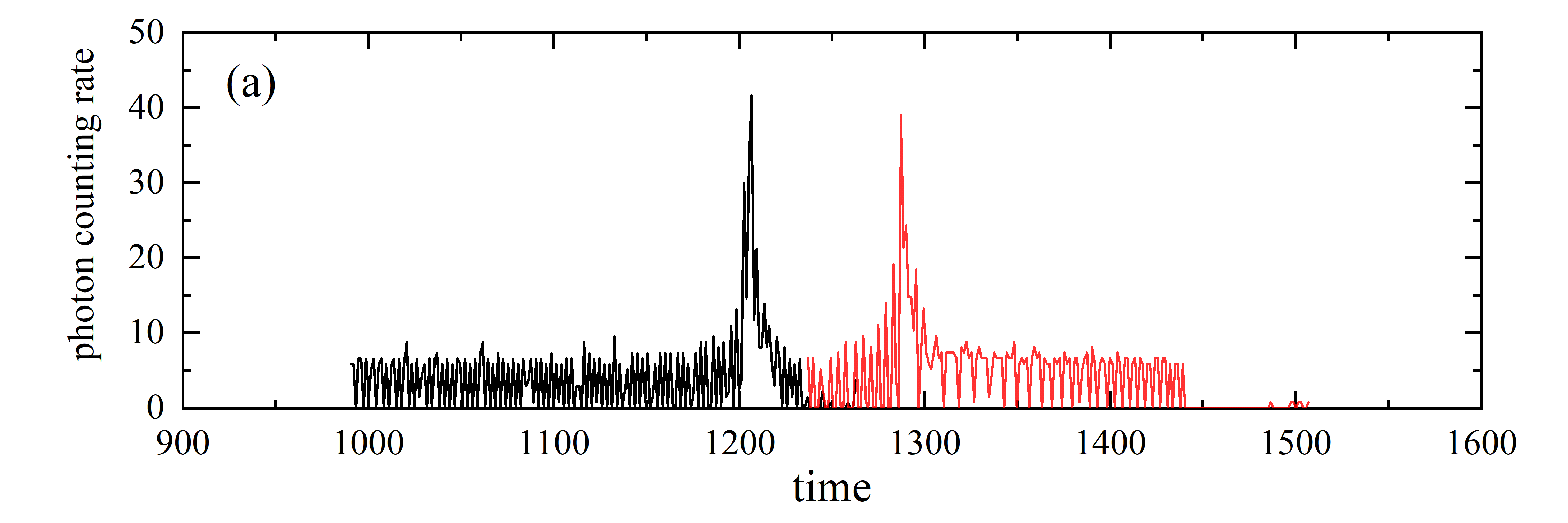}
\includegraphics[width=8cm]{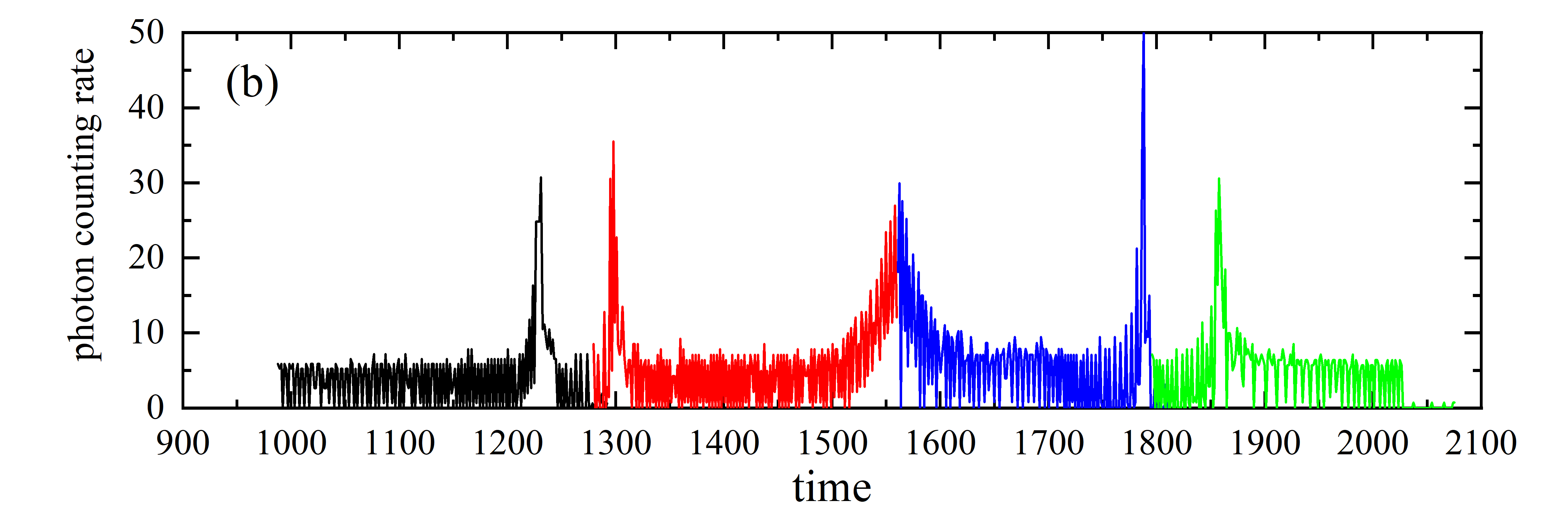}
\includegraphics[width=8cm]{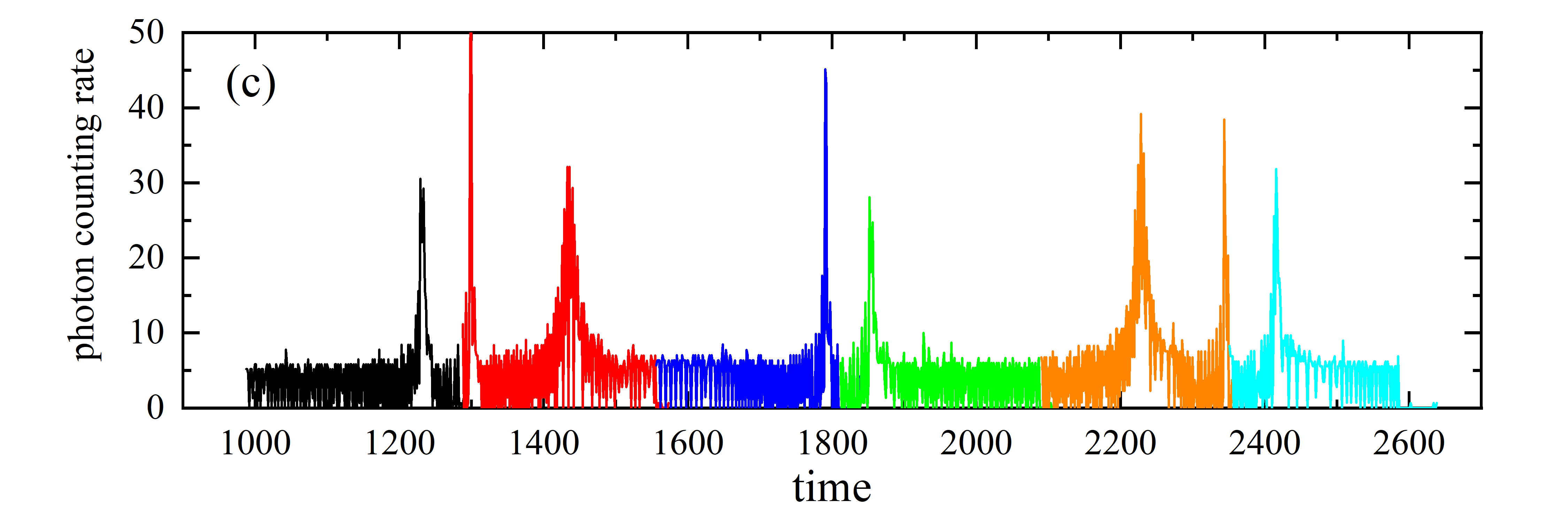}
\includegraphics[width=8cm]{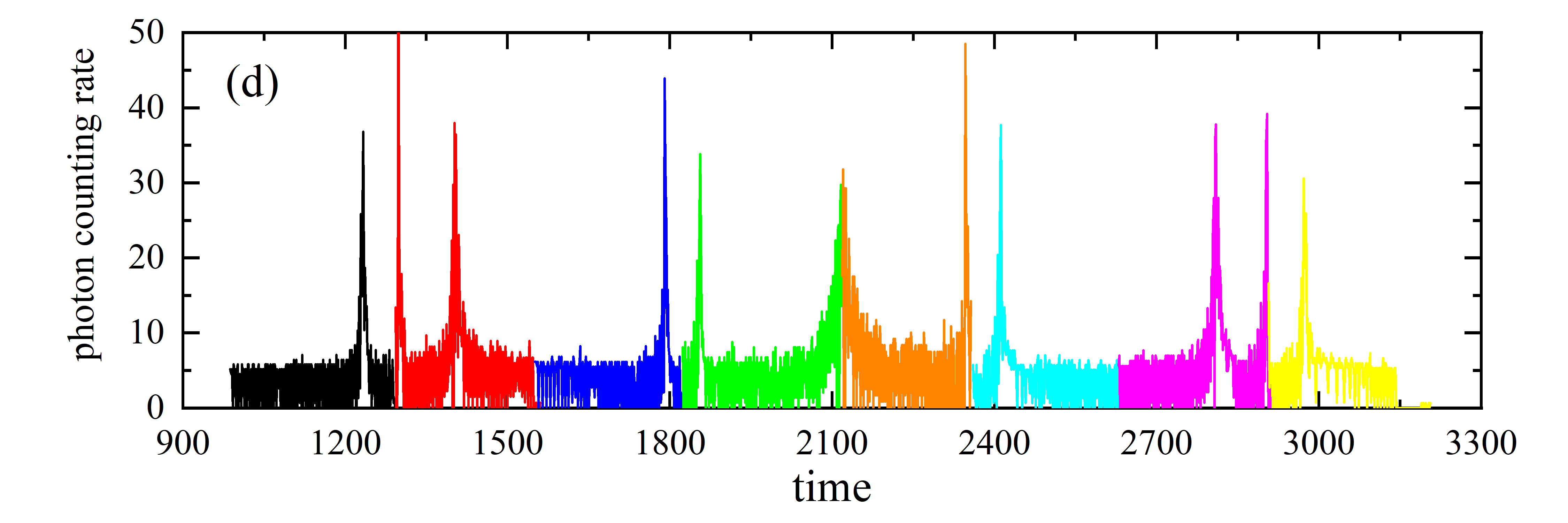}
\includegraphics[width=8cm]{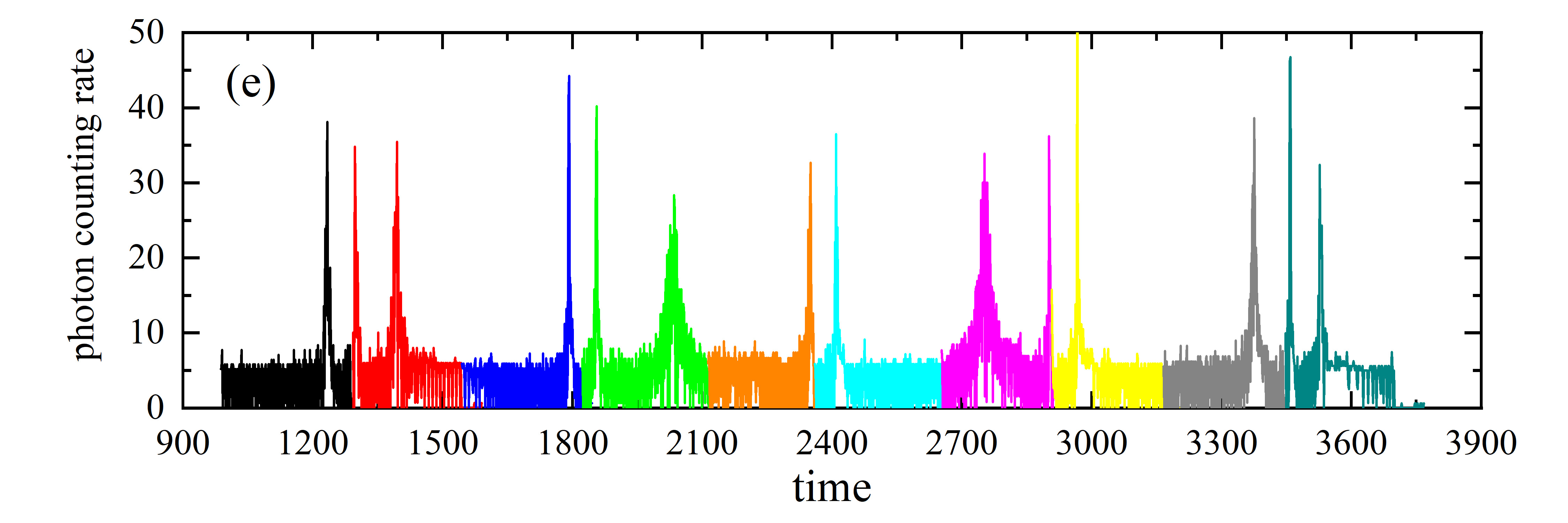}
\includegraphics[width=8cm]{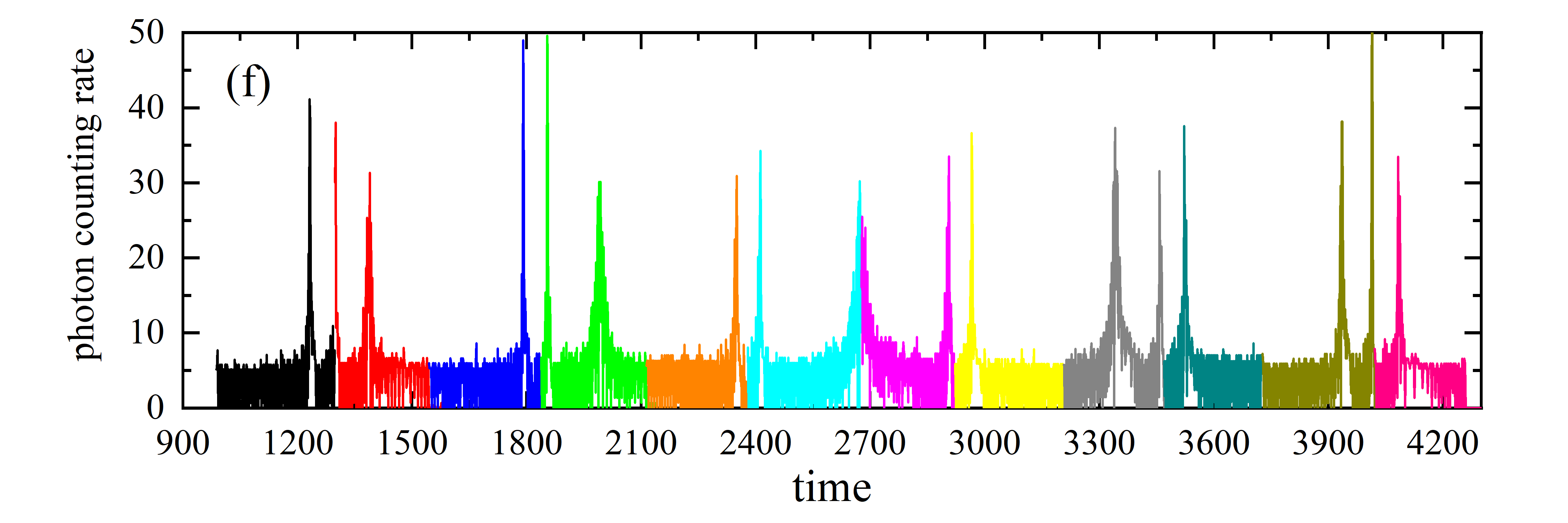}
\caption{Light curves of orbits with varying leaf numbers $z$ within a single complete period at an observation inclination angle of $85^{\circ}$. From panels (a) to (f), the value of $z$ increases from $1$ to $6$. Each distinct color within the curves represents a specific stage of the orbital evolution between the apastron and the periastron.}}\label{fig14}
\end{figure*}

We further increase the observation inclination angle to $85^{\circ}$ and simulate the light curves for orbits with different $z$ values, as illustrated in Fig. 14. In these plots, the light curve segment corresponding to each orbital transit from the apastron to the periastron is depicted in a distinct color. We find that this extreme increase in the observation inclination angle almost perfectly aligns the line of sight with the orbital plane. This configuration renders the gravitational lensing effect highly prominent, consequently generating significantly higher peaks in the light curves. Simultaneously, we emphasize that the peaks in the light curves are predominantly contributed by the orbital motion in the vicinity of the periastron. For instance, in panel (a), the black segment corresponds to the orbital evolution from the apastron to the periastron, while the red segment represents the return journey of the timelike particle from the periastron to the apastron. The two peaks of the light curve are situated precisely in the middle of the total observation window, coinciding exactly with the orbital motion near the periastron. Furthermore, because the periastron is located in close proximity to the black hole, photons in this region experience pronounced focusing and time-delay effects, which directly induce the formation of these prominent peaks.

More intriguingly, we discover a direct correlation between the number of peaks in the light curves and the corresponding orbital leaf number $z$. From panels (a) to (f), we can identify $2$, $5$, $8$, $11$, $14$, and $17$ peaks, respectively, which perfectly satisfies the algebraic relationship of $3z-1$. Furthermore, we examined the light curve corresponding to the $(z,w,v)=(7,1,6)$ orbit and found exactly $20$ peaks within a complete period, which similarly conforms to this established algebraic relation. Fundamentally, this relationship is grounded in a clear physical mechanism. Since the prominent peaks originate from the orbital contributions near the periastron, the number of periastron passages within a complete period directly dictates the resulting number of peaks. Consequently, for orbits possessing a larger number of leaves, the timelike particle inevitably traverses more periastrons, thereby driving the proportional increase in the number of light curve peaks. On the other hand, it must be explicitly clarified that the $3z-1$ algebraic relation is not a universal rule governing the correlation between the number of light curve peaks and the orbital leaf number for all rational orbits. Rather, it is strictly limited to the specific orbital family characterized by $(z,w,v)=(z,1,z-1)$. In fact, it is physically anticipated that both $w$ and $v$ possess the capacity to modify the number of central quasi-circular loops of the orbit. Consequently, they can significantly influence the gravitational lensing effect, which in turn alters the total number of peaks in the light curve. Under such circumstances, the $3z-1$ relationship would naturally be broken. We plan to dedicate our future investigations to exploring a more generalized correlation across broader orbital configurations.

From an observational standpoint, applying the results from Fig. 14 to infer the orbital leaf number fundamentally relies on the accurate enumeration of peaks within the light curve. This prerequisite inherently requires the detector to possess a temporal resolution sufficient for resolving the time interval between adjacent peaks. By analyzing the temporal gaps between the peaks in Fig. 14, we determine that the minimum time interval is approximately $60$ M, which is located between the first two peaks from the left in panel (b). By adopting the mass and distance parameters of the supermassive black hole at the Galactic Center, we establish a time scale factor of approximately $19.7088$. This scaling translates the geometric interval into a physical time difference of roughly $1183$ seconds. Taking the XMM-Newton observatory as an example, its temporal resolution operates on the order of milliseconds to seconds\footnote{For specific details regarding the temporal resolution of XMM-Newton, please refer to the official user handbook at https://xmm-tools.cosmos.esa.int/external/xmm$\_$user$\_$support/documentation/uhb/epicmode.html.}. Such high temporal precision implies that, strictly from a theoretical perspective, the specific number of peaks within our predicted light curves possesses the potential to be extracted in observations.

However, we must explicitly acknowledge that temporal resolution alone does not ensure these predicted structures can be readily identified in real astronomical data. In actual observations, the reliable extraction of such fine signals is inevitably subject to multiple complicating factors, including photon statistics, astrophysical background, intrinsic variability, and the specific emission assumptions. Furthermore, it should be noted that the prominent peaks simulated in our current work are defined strictly from the perspective of the photon count rate. If the observational analysis were to be based on specific intensity, the resulting light curve profiles would be additionally modulated by the Doppler effect, thereby introducing further complexities to the actual observed signatures.
\section{Summary and conclusions}
Strictly closed orbits are a special class of trajectories within the framework of general relativity, emerging only when the azimuthal period is commensurable with the radial period. These closed orbits carry intrinsic spacetime information, serving as excellent probes for investigating spacetime curvature and its internal sources. In this work, we searched for closed orbits of various configurations within a Schwarzschild spacetime embedded in a Dehnen-type dark matter halo. We simulated the gravitational wave signals and light curves associated with these orbits to unveil the impacts of the dark matter halo scale and density parameters on orbital dynamics.

Our investigation reveals that the introduction of a dark matter halo lowers the effective potential of timelike orbits in the equatorial plane, enlarges the ISCO radius, and contracts the specific energy range corresponding to bound orbits. These phenomena stem from the fact that the dark matter halo effectively enhances the gravitational field intensity of the spacetime. Furthermore, we find that the configuration of a closed orbit is primarily determined by the rational number $q$ and the specific angular momentum. Specifically, a sufficiently large angular momentum is required to yield $q \leq 1$; this condition allows for the emergence of multi-leaf orbits with $w=0$, given the relation $q=w+v/z$ and the requirement $v \leq z$. The spatial extent of the closed orbits depends on both the specific angular momentum and the dark matter halo parameters. Notably, increasing either of these parameters expands the orbital range while preserving the underlying orbital configuration. In other words, orbits within a spacetime embedded with a dark matter halo inevitably exhibit a larger spatial extent compared to their counterparts in a pure Schwarzschild spacetime. This characteristic expansion can thus serve as a reliable criterion for detecting the presence of a dark matter halo.

Based on the obtained closed orbits of various configurations in the target spacetime, we also simulated their corresponding gravitational wave signals and light curves. In the gravitational wave signals, the periastron and apastron regions of the closed orbits contribute to distinct components of the waveform. Specifically, the former contributes to the high-frequency oscillations of the signal, whereas the latter accounts for the smooth regions and broad bumps. Particularly, for the gravitational wave signals radiated by rational orbits within a single complete period, the number of high-frequency oscillation bursts equals the number of orbital leaves, while the number of sharp peaks within the high-frequency components correlates with the frequency of periastron passages. Furthermore, the orbital expansion induced by the dark matter halo leads to a noticeable phase delay in the waveforms. This phase delay can be readily identified, and the degree of delay is positively correlated with the dark matter halo parameters. More importantly, based on our preliminary analysis regarding the sensitivity curve, both the calculated characteristic strain and the frequency of the gravitational wave signals appear to fall within the observation window of LISA, suggesting that our findings may extend beyond purely theoretical correlations between physical parameters and orbital phenomena, potentially offering prospects for future observational extraction.

Regarding the light curves, we find that prominent peaks tend to be generated when the orbit evolves into the vicinity of the periastron; simultaneously, we discovered a potential correlation between the number of orbital leaves and the number of peaks. Specifically, within a complete orbital evolution period, the number of peaks in the light curve is exactly equal to $3z-1$. However, this discrepancy is best identified through the gravitational lensing effect, which becomes prominent only when the observation line of sight approaches the orbital plane. Furthermore, it must be emphasized that because the number of peaks in the light curves is intrinsically tied to the periastron passages, it is inevitably governed by the parameters $w$ and $v$. Consequently, the algebraic $3z-1$ correlation discovered in this work is strictly confined to the $(z,w,v)=(z,1,z-1)$ orbital family. The derivation of a more generalized relationship encompassing broader orbital configurations will be a primary focus of our future investigations.

The results of this paper not only unveil the influence of dark matter halos on curved spacetimes but also integrate the two major messengers, gravitational and electromagnetic radiation, into the study of orbital dynamics in black hole spacetimes. This comprehensive approach paves a new avenue for exploring the potential observational properties of orbits and extracting intrinsic spacetime information.

\printcredits

\section*{Declaration of competing interest}
The authors declare that they have no known competing financial interests or personal relationships that could have appeared to influence the work reported in this paper.

\section*{Acknowledgements}
The authors are very grateful to the referee for insightful comments and valuable suggestions. This research has been supported by the National Natural Science Foundation of China [Grant Nos. 12403081 and 12475057].

\section*{Data availability}
Data will be made available on request.

\bibliographystyle{model1-num-names}
\bibliography{references}
\end{document}